\documentclass[prd,english,preprintnumbers,amsmath,amssymb,nofootinbib,superscriptaddress,twocolumn]{revtex4-1}

\pdfoutput=1

\usepackage{mathtools}
\usepackage{amsfonts}
\usepackage{mathrsfs}
\usepackage{bbm}
\usepackage{slashed}

\usepackage{graphicx}
\usepackage[dvipsnames]{xcolor}
\usepackage{array}

\usepackage{xspace}
\usepackage{siunitx}
\usepackage{hyperref}

\usepackage{xifthen}
\usepackage{dsfont}
\usepackage{appendix}
\usepackage{enumitem}
\usepackage{booktabs}
\usepackage{units}

\usepackage[latin1]{inputenc}
\usepackage[dvipsnames]{xcolor}
\usepackage{array}

\setkeys{Gin}{width=0.48\textwidth}

\graphicspath{
	{./}
}

\def\0#1#2{\frac{#1}{#2}}

\def\s0#1#2{\mbox{\small{$ \frac{#1}{#2} $}}}

\def\CO{{\mathcal O}}

\newcommand{\be}{\begin{eqnarray}}
\newcommand{\ee}{\end{eqnarray}}

\newcommand{\nn}{\nonumber }

\newcommand{\beq}{\begin{equation}}
\newcommand{\eeq}{\end{equation}}
\newcommand{\bea}{\begin{eqnarray}}
\newcommand{\eea}{\end{eqnarray}}

\newcommand{\tinytext}[1]{\text{\tiny{#1}}}
\def\0#1#2{\frac{#1}{#2}}

\def\eq#1{\eqref{#1}}
\def\Eq#1{Eq.~\eqref{#1}}

\newcommand{\gettitle}{Renormalization group consistency and low-energy effective theories}

\hypersetup{
	colorlinks,
	linkcolor={red!75!black},
	citecolor={blue!75!black},
	urlcolor={blue!75!black},
	pdftitle={\gettitle},
	pdfauthor={Braun, Leonhardt, Pawlowski},
	pdfkeywords={Renormalization group} {Cutoff}
	{Correlations functions} {Initial conditions},
	bookmarksopen=true,
	bookmarksopenlevel=2,
	bookmarksnumbered=true
}

\usepackage{babel}
\makeatother
\begin{document}

\title{{Renormalization group consistency and low-energy effective theories}}

\author{Jens Braun}
\affiliation{Institut f\"ur Kernphysik (Theoriezentrum), Technische Universit\"at Darmstadt, 
D-64289 Darmstadt, Germany}
\affiliation{ExtreMe Matter Institute EMMI, GSI, Planckstra{\ss}e 1, D-64291 Darmstadt, Germany}
\author{Marc Leonhardt} 
\affiliation{Institut f\"ur Kernphysik (Theoriezentrum), Technische Universit\"at Darmstadt, 
D-64289 Darmstadt, Germany}
\author{Jan M. Pawlowski} 
\affiliation{Institut f\"ur Theoretische Physik, Universit\"at Heidelberg, Philosophenweg 16, 
D-69120 Heidelberg, Germany}
\affiliation{ExtreMe Matter Institute EMMI, GSI, Planckstra{\ss}e 1, D-64291 Darmstadt, Germany}

\begin{abstract}
  Low-energy effective theories have been used very
  successfully to study the low-energy limit of QCD,
  providing us with results for a plethora of phenomena, ranging from bound-state formation to phase transitions
  in QCD. These theories are consistent
  quantum field theories by themselves and can be embedded in QCD, but typically have a physical ultraviolet
  cutoff that restricts their range of validity. 
   Here, we provide a discussion of the concept of
  renormalization group consistency, aiming at an analysis of cutoff
  effects and regularization-scheme dependences in general studies of low-energy effective theories. 
 For illustration, our findings are applied to low-energy effective models of QCD in
  different approximations including the mean-field approximation. 
  More specifically, we 
  consider hot and dense as well as finite systems and 
  demonstrate that violations of renormalization group consistency affect
 significantly the predictive power of the corresponding model calculations.

\end{abstract}

\maketitle

%
\section{Introduction}
The computation of quantum corrections in field theories in general
requires a regularization and renormalization procedure. In
perturbation theory, the regularization procedure allows us to compute
the perturbative loop diagrams in a well-defined fashion, e.g., by
introducing a momentum cutoff $\Lambda$ for the momentum
integrals. This cutoff dependence can be absorbed in counter terms in
the underlying bare action, here called $\Gamma_{\Lambda}$. The latter
then consists of all ultraviolet (UV) relevant terms allowed by the
symmetry of the classical theory or classical action
  $S$. Potentially, additional terms have to be introduced in
$\Gamma_{\Lambda}$, if the momentum cutoff breaks symmetries present
in the classical action. Here, a prominent example is provided by
gauge symmetries that are explicitly broken by a momentum
cutoff. Then, symmetry-breaking counter terms such as a mass term for
the gauge field in $\Gamma_{\Lambda}$ indeed restores the gauge
symmetry for the full quantum effective action.

This perturbative reasoning extends to the general non-perturbative
case. In the past decade non-perturbative functional methods, such as
the functional renormalization group (FRG), Dyson-Schwinger equations
(DSE) and $n$PI methods, have made rapid progress and improved our
understanding of strongly-correlated systems, ranging from condensed
matter over heavy-ion physics and high energy-physics to quantum
gravity. Inherent to all these functional approaches is their
formulation in terms of non-perturbative loop equations being 
structurally very similar to the perturbative setting briefly
discussed above.
In most cases, and in particular in numerical applications, these
approaches feature momentum cutoffs for the non-perturbative loops
involved as well as respective counter terms in the bare action
$\Gamma_\Lambda$. The explicit cutoff dependence of $\Gamma_\Lambda$ 
ensures the cutoff independence of the full quantum effective action $\Gamma$, 
\begin{align}
\Lambda \0{d\Gamma}{d{\Lambda}} =0\,.
\label{eq:RGconsistency}
\end{align}
{This is the requirement of a consistent regularization and
renormalization of a given} theory, and is called {\it RG con\-sistency}. As
a central ingredient in a non-perturbative functional setup, RG consistency will be
discussed in detail in Sec.~\ref{sec:RGconsistency}. 
Evidently, these considerations are very general and are not bound to perturbatively
renormalizable theories or to a specific class of field theories. 

In this work, we discuss how cutoff artefacts can be removed
consistently within a given low-energy effective theory in order to
ensure the important property of RG consistency
\eq{eq:RGconsistency}. Irrespective of possible fundamental UV
completions, the discussion of cutoff artefacts is required for a
meaningful application of a given model and a test of its range of
applicability in terms of external parameters. In
Sec.~\ref{sec:RGconsistency}, we therefore discuss this issue on very
general grounds.  In Sec.~\ref{sec:rgcex}, we then demonstrate the
application of these general considerations to specific model
calculations. This includes a quark-meson model in the vacuum limit, a
diquark model at finite density, a quark-meson-diquark model at finite
temperature and density, and the quark-meson model confined in a
finite box. Our conclusions can be found in~Sec.~\ref{sec:conc}.

\section{RG consistency}\label{sec:RGconsistency}
In this section we focus on general aspects of RG consistency which
includes both discussions of formal as well as phenomenological
aspects. The reader may skip this section in a first reading and
readily start with Sec.~\ref{sec:rgcex} where we exemplify the meaning
of RG consistency in the context of specific QCD low-energy effective
theories.

\subsection{RG consistency and low-energy phenomenology}

The computation of quantum corrections in field theories in general
requires the introduction of a UV cutoff~$\Lambda$.  This
scale is said to be asymptotically large when 
\begin{align}
\label{eq:asUV}
\0{s_{i}}{\Lambda}\ll 1 \quad {\rm with} \quad s=\{m_\text{phys},m_\text{ext}\}\,,
\end{align}
where the set $s$ stands for all mass scales in the theory, including
dimensionful couplings. In particular, this set consists of the
intrinsic fundamental parameters of the theory, $m_\text{phys}$
(e.g. masses of particles), as well as the external
scales $m_\text{ext}$.  For example, for the low-energy effective
theories (LEFT) of QCD discussed below, we have
$m_\text{ext}=\{T,V^{-1/3},\mu\}$, where $T$ is the temperature, $V$
is the volume of the system, and $\mu$ is the quark chemical
potential.

When \Eq{eq:asUV} is ensured, the bare action $\Gamma_\Lambda$ only
encodes the microphysics of the situation at hand, and changes of the
intrinsic parameters are simply triggered by changing the respective
bare parameters in the action. In particular, Eq.~\eqref{eq:asUV}
entails that a change of the external parameters of the theory does
not change the regularization and renormalization of the theory
encoded in the $\Lambda$-dependence of
$\Gamma_\Lambda$. For $m_{\text{ext},i}/\Lambda\to 0$, the
corresponding property then reads
\begin{align}
\0{d}{d m_{\text{ext},i}}\,\left[
\Lambda \0{d\Gamma_\Lambda}{d{\Lambda}} \right]= 0\,, 
\label{eq:dm_ext0}
\end{align}
which highlights the similarity of this condition to the RG-consistency 
condition given in Eq.~\eqref{eq:RGconsistency}. In turn, if
Eq.~\eqref{eq:asUV} does not hold, $\Gamma_\Lambda$ has to vary with a
change of $m_\text{ext}$ to ensure that the RG-consistency
condition~\eqref{eq:RGconsistency} holds. However, a dependence
of $\Gamma_\Lambda$ on the external parameter then implies
\begin{align}
\0{d}{d m_{\text{ext},i}}\,\left[
\Lambda \0{d\Gamma_\Lambda}{d{\Lambda}} \right]\neq 0\,.
\label{eq:dm_ext}
\end{align}
This is elaborated below. Note that,
if Eq.~\eqref{eq:dm_ext0} is violated, (part of) the physics related
to the fluctuation physics of the respective external parameters is
already carried by the bare action $\Gamma_\Lambda$. It has to be
computed separately, which necessitates an explicit expression for the
right-hand side of Eq.~\eqref{eq:dm_ext}. This computation is of
eminent importance. As we shall exemplify in this work, violations of
RG consistency may indeed significantly spoil predictions for physical
observables.

For a plethora of physically interesting theories, the cutoff
$\Lambda$ may be limited by a validity bound.  A strict bound is
present, if the effective theory at hand cannot be extended beyond a
certain UV scale. For example, a Landau pole at the
scale $\Lambda_{\text{UV}}$ is such a strict bound.  Then, we have to
choose $\Lambda\leq\Lambda_{\text{UV}}$.
This situation applies to most effective theories for the low-energy regime of QCD,
such as Nambu--Jona-Lasinio-type models (NJL) and quark-meson-type
models (QM) with or without Polyakov-loop 
extensions~\cite{Hatsuda:1985eb,*Asakawa:1989bq,*Klevansky:1992qe,Jungnickel:1995fp,*Berges:1997eu,*Berges:2000ew,Fukushima:2011jc,*Kamikado:2012bt,*Tripolt:2013jra,*Andersen:2014xxa,*Yokota:2017uzu,*vonSmekal:2012vx,%
Rajagopal:2000wf,*Buballa:2003qv,*Shovkovy:2004me,*Alford:2007xm,Strodthoff:2011tz,%
Floerchinger:2012xd,*Drews:2014wba,*Drews:2014spa,*Weyrich:2015hha,%
Meisinger:1997jt,*Pisarski:2000eq,*Ratti:2005jh,*Fukushima:2003fw,*Roessner:2006xn,*Schaefer:2007pw,*Skokov:2010wb,*Skokov:2010uh,*Herbst:2010rf,*Strodthoff:2013cua,*Haas:2013qwp,*Pisarski:2016ixt,*Fukushima:2017csk}, and it also applies
to quantum electrodynamics and a variety of condensed-matter models.

A further, qualitatively different, validity bound of LEFTs 
is related to the fact, that they
typically lack some of the microscopic degrees of freedom that are
relevant at momentum scales $\Lambda>\Lambda_\text{phys}$. Then,
Eq.~\eqref{eq:asUV} may hold for a given LEFT but, beyond the
scale $\Lambda_\text{phys}$, the LEFT lacks the dynamics associated
with the fundamental microscopic degrees of freedom. Consequently,
such a LEFT cannot describe the physics at hand beyond~$\Lambda_\text{phys}$.
For example, in conventional QCD low-energy effective theories, the
gluon dynamics is missing. These LEFTs describe QCD solely in terms
of hadronic degrees of freedom which can only hold true for low
momentum scales.  

Of course, by definition, a determination of the
scale $\Lambda_\text{phys}$ is involved as it requires an actual study
of the fundamental dynamics at all momentum scales. Within the FRG
approach to fundamental
QCD~\cite{Gies:2002hq,Gies:2006wv,Braun:2009gm,Mitter:2014wpa,Braun:2014ata,Rennecke:2015eba,Cyrol:2017ewj},
however, it has been shown in various studies that the gluonic sector
of QCD at low baryon-density decouples from the matter sector at
scales $\Lambda_\text{phys} \sim 0.4 \dots 1\,\text{GeV}$, see e.g.\
\cite{Pawlowski:2010ht,*Pawlowski:2014aha,Herbst:2010rf,Haas:2013qwp,Herbst:2013ufa,Springer:2016cji}.
In this context, it should be noted that the scales $\Lambda$,
$\Lambda_{\text{UV}}$, and $\Lambda_{\text{phys}}$ depend on the chosen
regularization scheme, and are only related to physical momentum
scales by the renormalization procedure.
 
In our discussion of specific models in Sec.~\ref{sec:rgcex}, we do
not aim at a determination of their values of $\Lambda_{\text{phys}}$
but rather aim at a discussion of how cutoff artefacts can be removed
consistently within a given model study. Irrespective of possible
fundamental UV completions, such a discussion of cutoff artefacts is
required for a meaningful application of a specific model within given ranges 
for the external parameters, in
particular in the absence of an accurate knowledge of the
scale~$\Lambda_{\text{phys}}$.

\subsection{Quantum effective action and regularization}\label{sec:qefa}
The central object of our general discussion is the quantum effective
action $\Gamma[\Phi]$ of a given theory with field content
$\Phi=(\Phi_1,\Phi_2,\dots)^{T}$ in $d$ Euclidean space-time
dimensions. It is the quantum analogue of the classical action and its
saddle points are solutions of the quantum equations of motion
(EoM). Its $n$th field derivatives, evaluated at the minimal
  quantum EoM, are the one-particle-irreducible (1PI) parts of the
  $n$-point correlation functions of the theory,
  $\Gamma^{(n)}[\Phi]=\langle \Phi_{i_1} \cdots
  \Phi_{i_n}\rangle_{\text{1PI}}$ for $n>2$. For the two-point
function we
have $\Gamma^{(2)}[\Phi]\cdot\langle \Phi_{i_1}
\Phi_{i_2}\rangle_{\text{1PI}}=1$. In a functional approach, the
effective action has the generic representation
\begin{align}\label{eq:GammaDiagrams}
\Gamma[\Phi]= {\mathcal D}_{\Lambda}[\Phi] +\Gamma_\Lambda[\Phi]\,, 
\end{align}
where ${\mathcal D}_{\Lambda}$ stands for all momentum-loop diagrams
evaluated in the presence of the momentum cutoff $\Lambda$. This
cutoff leads to finite diagrams, as momentum fluctuations with
$p^2\gtrsim \Lambda^2$ are suppressed in $ {\mathcal
  D}_{\Lambda}$. Hence, these fluctuations must reside in $\Gamma_\Lambda$.

The relation~\eq{eq:GammaDiagrams} together with
the RG-consistency condition~\eq{eq:RGconsistency} is simply the requirement that
the $\Lambda$-dependence of the loops is cancelled by that in the bare
action $\Gamma_\Lambda$. Moreover, we can shift the fluctuation
information contained in ${\mathcal D}_{\Lambda}[\Phi]$
to $\Gamma_\Lambda$ by lowering the scale $\Lambda$.  Indeed, we
have $\lim_{\Lambda\to 0}{\mathcal D}_{\Lambda}=0$ and
$\lim_{\Lambda\to 0}\Gamma_{\Lambda}= \Gamma$.

For this interpretation, $\Gamma_\Lambda$ has to be seen as an
effective action that misses the infrared dynamics of the theory
carried by the diagrams ${\mathcal D}_{\Lambda}[\Phi]$. Hence, the UV-cutoff
$\Lambda$ in the diagrams serves as an infrared (IR) cutoff
$k=\Lambda$ for the scale-dependent effective action $\Gamma_k$:
\begin{align}
k\partial_k \Gamma_k[\Phi] = {\mathcal F}_k[\Phi]\,,
\label{eq:genflow}
\end{align}
where ${\mathcal F}_k= -k \partial_k {\mathcal D}_k$. This fruitful
block-spinning perspective is taken for FRG approaches. In its modern form, Eq.~\eqref{eq:genflow} is a simple
one-loop equation, the Wetterich equation~\cite{Wetterich:1992yh} with
\begin{align}
{\mathcal F}_k[\Phi]=\frac{1}{2}{\text{Tr}}\, \0{1}{\Gamma_k^{(2)}[\Phi]+R_k}\, k\partial_k R_k\,.
\label{eq:wetterich}
\end{align}
The trace in \eq{eq:wetterich} sums over momentum, space-time and
internal indices as well as species of fields.  The latter includes a
minus sign for fermionic degrees of freedom as known from perturbation
theory. The regulator function $R_k$ depends on the IR cutoff
scale $k$ and defines the regularization scheme. It adds to the full
two-point function of the regularized theory and changes the
dispersion. In the IR limit, it acts as a mass and thus suppresses the
IR momentum fluctuations in $\Gamma_k$. For UV momenta
$p^2\gtrsim k^2$, it decays sufficiently fast in order to keep the
UV physics unchanged.  Let us discuss this here at the
example of a scalar field. To this end, we parameterize the regulator
as
\begin{align}\label{eq:Rk}
R_k(p^2) = p^2 r(x)\,,
\end{align}
with $x=p^2/k^2$ and a dimensionless shape function~$r$ 
determining the IR and UV asymptotics. The prefactor $p^2$ carries the
classical dispersion of the scalar field. More elaborated choices
substitute the latter with the momentum-dependent part of the full
inverse two-point function $p^2\to \Gamma^{(2)}_k$, known as RG- or
spectrally adjusted
regulators~\cite{Reuter:1993kw,*Pawlowski:2001df,*Litim:2002hj,*Gies:2002af,Pawlowski:2005xe}.

In general, an admissible regulator $R_k$ has to obey certain
conditions regarding its behavior in the low- and large-momentum
limit. For example, it has to render the momentum part of the trace
in~\eqref{eq:wetterich} finite in the UV limit and, by providing a
mass gap for the fields, also in the IR limit. Hence, from an RG point
of view, the regulator function specifies the Wilsonian momentum-shell
integration, such that the right-hand side of the differential
equation~\eqref{eq:genflow} is dominated by fluctuations with
momenta $|p|\sim k$. It should be added that fast decays of $r(x)$
improve the convergence of the approximation scheme used, for details
see Refs.~\cite{Pawlowski:2005xe,Pawlowski:2015mlf}. A common
approximation scheme that is also used in the present work, the
derivative expansion, is based on the expansion in powers of momenta.
The applicability of this scheme to any order requires shape functions
that decay faster than any polynomial in $x$. In summary, exponential
or even compact support regulators are best suited for common
systematic approximation schemes, ranging from the derivative
expansion to vertex expansions as used in QCD.

In the set of diagrams ${\mathcal D}_k[\Phi]$, the cutoff $k$ acts as a
UV cutoff. UV suppression is achieved by the occurrence of
\begin{align}
\0{1}{\Gamma^{(2)}_{k=0}[\Phi]}- \0{1}{\Gamma^{(2)}_k[\Phi]+R_k}
\end{align}
for internal lines. This pattern is easily seen by integrating the
flow equation~\eqref{eq:genflow} with~\eqref{eq:wetterich} within one-loop perturbation theory. On the
right-hand side of the flow equation the bare action enters with
$\Gamma_\Lambda =S$ with~$S$ being the classical action. For a sharp cutoff and
constant background fields, we find
\begin{align}
\0{1}{S^{(2)}}- \0{1}{S^{(2)}+R_k}= 
\0{1}{S^{(2)}}\theta(k^2-p^2)\,.
\end{align}
Additional subtractions occur in
perturbation theory by iteratively generating higher loop orders by
re-inserting the result on the right-hand side of the flow equation,
see e.g. Refs.~\cite{Litim:2001ky,Litim:2002xm,Pawlowski:2005xe}. In
total, the flow equation~\eqref{eq:genflow} with
Eq.~\eqref{eq:wetterich} leads to a generalized
Bogoliubov-Parasiuk-Hepp-Zimmermann (BPHZ)-type regularization scheme:
the regularization is achieved by subtraction.

\subsection{RG consistency -- formal discussion}\label{sec:RGconsistencyII}

The effective action $\Gamma$ is obtained from Eq.~\eqref{eq:GammaDiagrams} by 
integrating Eq.~\eqref{eq:genflow} from the initial UV scale $k=\Lambda$ to
$k=0$. For finite $k$, we find
\begin{subequations} \label{eq:IntegratedFlow}
\begin{align}\
  \Gamma_k[\Phi] = \Gamma_{\Lambda}[\Phi] + \int_{\Lambda}^{k}\frac{{\rm d}
  k^{\prime}}{k^{\prime}}{\mathcal F}_{k^{\prime}}[\Phi]\,, 
\label{eq:IntegratedFlow1}
\end{align}
which leads to \eq{eq:GammaDiagrams} for $k\to 0$. The RG consistency
condition~\eq{eq:RGconsistency} follows immediately for any
$k\neq \Lambda$ from Eq.~\eq{eq:IntegratedFlow1} by taking the
$\Lambda$-derivative.\footnote{Note that, within the standard convention of the FRG approach, 
the partial derivative with respect to $\Lambda$ 
corresponds to the total derivative in Eq.~\eq{eq:RGconsistency}.}
 We have
\begin{align}
 \Lambda  \partial_\Lambda \Gamma_k[\Phi] = \Lambda\partial_\Lambda 
\Gamma_{\Lambda}[\Phi] - {\mathcal F}_{\Lambda}[\Phi]=0\,, 
\label{eq:IntegratedFlow2}
\end{align}
where we have used \eq{eq:genflow} in the last
step. Note that in Eqs.~\eqref{eq:IntegratedFlow1} and~\eqref{eq:IntegratedFlow2}  
the scale $\Lambda$ is not necessarily
the largest scale possible in the theory at hand,
i.e. $\Lambda_\text{UV}$. It is only some scale at which we fix the couplings of the theory.
\end{subequations}

The seemingly simple relations~\eqref{eq:IntegratedFlow1}-\eqref{eq:IntegratedFlow2} 
offer a lot of information that is in general
more difficult to access in other approaches. First of all,
Eqs.~\eqref{eq:IntegratedFlow1}-\eqref{eq:IntegratedFlow2} entail that
RG consistency and hence cutoff independence of a theory, fundamental
or effective, follows trivially in the FRG approach: the effective
action at the initial scale $k=\Lambda$ has to obey the flow equation,
if we vary the initial scale. Moreover, its $\Lambda$-dependence is
easily extracted for large initial cutoff scales $\Lambda$ with the
aid of Eq.~\eqref{eq:asUV}. In this case, $\Gamma_\Lambda$ can be
expanded in powers of $\Lambda$ with
\begin{subequations}\label{eq:Intialinfty}
\begin{align}\label{eq:Gammainfty}
\Gamma_\Lambda[\Phi] = \sum_{n\leq N_\text{max}}  \tilde{\gamma}_n[\Phi] \,\Lambda^n 
+ \tilde{\gamma}_\text{log}[\Phi] \,\ln \0{\Lambda}{s_0}\,, 
\end{align}
where the term with $n=0$ carries the
physics part of the initial condition at the scale $\Lambda$. Here, we
have normalized the logarithmic term with some physical scale
$s_0\in s$, e.g.\ the physical mass gap of the theory at
hand. Choosing a different reference scale shifts terms from
$\tilde{\gamma}_0$ to $\tilde{\gamma}_\text{log}$. Note that the
$\tilde{\gamma}_n$'s can be a collection of different field-dependent
terms with the same $\Lambda$-behavior.  The right-hand side of the
flow equation can also be expanded in powers of $\Lambda$, and the
expansion coefficients only depend on the shape function $r(x)$
and $\tilde{\gamma}=\{\tilde{\gamma}_{N_\text{max}},
\tilde{\gamma}_{N_\text{max}-1},...,\tilde{\gamma}_\text{log},\tilde{\gamma}_0,\tilde{\gamma}_{-1},\dots\}$,
\begin{align}\label{eq:Finfty}
  {\mathcal F}_{\Lambda}[\Phi] = \sum_{n\leq N_\text{max}}  f_n[\Phi;\tilde{\gamma},r] \,\Lambda^n\,. 
\end{align} 
Inserting Eqs.~\eqref{eq:Gammainfty} and~\eqref{eq:Finfty} into the flow equation~\eqref{eq:genflow} leads to
\begin{align}\label{eq:dGammainfty}
 \tilde{\gamma}_{n\neq 0}   = \0{1}{n} f_n[\Phi,\tilde{\gamma},r] \,,\quad 
\quad  \tilde{\gamma}_\text{log}  =  f_0[\Phi,\tilde{\gamma},r]\,,
\end{align}
\end{subequations}
where we have used $\Lambda\partial_{\Lambda}\Lambda^n = n\Lambda^n$, $\Lambda\partial_{\Lambda}\ln \Lambda=1$. 
Note that there is no relation for $\tilde{\gamma}_0$ as it contains the physics input. Nonetheless, $\tilde{\gamma}_0$
appears on the right-hand side of the relations for the
$\tilde{\gamma}_{n\neq 0}$ and $\tilde{\gamma}_\tinytext{log}$. 

The set of relations \eq{eq:dGammainfty} can be solved recursively and
provides the intial effective action in a well-defined and practically
applicable way. Note that only a finite number of terms matter due to
the $\Lambda$-suppression of the rest. 
The relations~\eqref{eq:Gammainfty}-\eqref{eq:dGammainfty} also make
apparent that, for asymptotically large values of $\Lambda$, the initial effective
action is nothing but the bare action for the given FRG scheme. As
such, it depends explicitly on the cutoff $\Lambda$, see e.g.\ \cite{Pawlowski:2005xe,Rosten:2010vm}. 

The setting above is the standard one for perturbatively
renormalizable theories in the absence of Landau poles. Strictly
speaking, the formulation above only applies to asymptotically free
theories. In QCD, for example, we are in the fortunate situation that this simple setting
applies. In general LEFTs, we typically have to deal with the existence of an actual finite UV extent 
given in form of a maximal UV cutoff scale $\Lambda_\text{UV}$ due to an instability of the theory, or
a phenomenologically existing UV extent $\Lambda_\text{phys}$ above which a given LEFT does no longer
provide a valid description of a more fundamental theory.
A priori, a safe choice is then 
\begin{align}\label{eq:Lambdamax} 
  \Lambda\leq \Lambda_\text{max}\,,
\end{align}
where $\Lambda_{\text{max}}\in\{\Lambda_{\text{phys}},\Lambda_{\text{UV}}\}$.
For such a choice, $\Lambda$ may not be sufficiently large compared to
the external parameters $m_\text{ext}$ of interest and we are left
with the situation as described by Eq.~\eq{eq:dm_ext}.  Moreover, the
intrinsic scales may not even be small compared
to $\Lambda$.\footnote{In the following we focus on the external
  parameters for clarity. However, the discussion can be
  straightforwardly generalized to the case of intrinsic scales.}
Then, the determination of the initial effective
action $\Gamma_{\Lambda}$ with
Eqs.~\eqref{eq:Gammainfty}-\eqref{eq:dGammainfty} is no longer
possible: for low initial scales $\Lambda$, the initial effective
action is a complicated object itself. 

In LEFTs of QCD, for example, this issue may potentially be surmounted by
computing $\Gamma_{\Lambda}$ with the aid of RG studies of the
fundamental theory, see
e.g. Refs.~\cite{Braun:2003ii,Haas:2013qwp,Herbst:2013ufa,Springer:2016cji}.
However, if a sufficiently accurate determination
of $\Gamma_{\Lambda}$ from a more fundamental theory is not available,
we still have to ensure that cutoff artefacts associated with a
specific choice for the scale $\Lambda$ are suppressed or even removed
in our model study. Otherwise, over a wide range of the external
parameters, such a model study may only resolve peculiarities of the
underlying regularization schmeme. In this case, we have to make use
of ``pre-initial'' flows that provide a systematic determination of the
effects of the violations described by Eq.~\eq{eq:dm_ext} by an
RG-consistent UV completion of the LEFT at hand.

To illustrate this, let us assume that we know the effective action at
some scale. In case of QCD models, for example, the
effective action is often chosen to assume a simple quadratic form at
some scale. In the following, this scale is denoted as $\Lambda^{\prime}$, see also
Sec.~\ref{sec:rgcex} for a discussion of specific models. 
The UV completion $\Gamma_{\Lambda}$ of the LEFT is then obtained by
following the RG flow from $k=\Lambda^{\prime}<\Lambda$
to $k=\Lambda\leq\Lambda_{\text{UV}}$, such that we
have $\Lambda\partial_{\Lambda}\Gamma_{\Lambda}\to 0$
for $m_{\text{ext},i}/\Lambda \ll 1$, i.e.  RG consistency is ensured
in the presence of finite external parameters.  This is demonstrated
in Sec.~\ref{sec:rgcex} for specific models and exploits the fact that
the effective action $\Gamma_{\Lambda}$ can be determined from
Eq.~\eq{eq:IntegratedFlow} as
\begin{align}\nn
&\Gamma_{\Lambda}[\Phi;m_{\text{ext}}^{(0)}] 
\\[1ex] &\hspace{0.5cm}= \Gamma_{\Lambda^{\prime}}[\Phi;m_{\text{ext}}^{(0)}] 
-  \int_{\Lambda}^{\Lambda^{\prime}}\frac{{\rm d}k^{\prime}}{k^{\prime}}{\mathcal F}_{k^{\prime}}[\Phi;m_{\text{ext}}^{(0)}]
\,,
\label{eq:gammainiL}
\end{align}
with $\Lambda$ chosen such
that $m_{\text{ext},i}/\Lambda \ll 1$ for all parameters of
interest. Here, ${\mathcal F}_{k^{\prime}}$ depends on $\Phi$
and $m_{\text{ext}}^{(0)}$, the latter denoting a given set of
``benchmark values'' for the external parameters at
which $\Gamma_\Lambda$ has been fixed with the aid of some set of
physical low-energy observables.  Typical benchmark values are the
vacuum values of the external parameters. For QCD, this is vanishing
temperature, infinite volume, and vanishing quark chemical
potential, see also our examples in Sec.~\ref{sec:rgcex}.

If not indicated otherwise, we shall assume from now on
that $\Gamma_{\Lambda}$ has been fixed in the limit of vanishing
external parameters. From our choice~\eqref{eq:gammainiL}, we then
deduce that the effective action $\Gamma_k$ remains unchanged in this
limit:
\begin{align}\nn
\Gamma_{k}[\Phi;m_{\text{ext}}^{(0)}] &= 
\Gamma_{\Lambda^{\prime}}[\Phi;m_{\text{ext}}^{(0)}] + 
\int_{\Lambda^{\prime}}^{k}\frac{{\rm d}k^{\prime}}{k^{\prime}}{\mathcal F}_{k^{\prime}}[
\Phi;m_{\text{ext}}^{(0)}]\\[2ex] 
& =
\Gamma_{\Lambda}[\Phi;m_{\text{ext}}^{(0)}] + \int_{\Lambda}^{k}
\frac{{\rm d}k^{\prime}}{k^{\prime}}{\mathcal F}_{k^{\prime}}[\Phi;m_{\text{ext}}^{(0)}]\,,
\label{eq:Gamma}
\end{align}
where $k<\Lambda^{\prime}$.  However, note that the $\Phi$-dependence
of $\Gamma_{\Lambda}$ and $\Gamma_{\Lambda^{\prime}}$ is in general
different.  At the same time, the choice~\eqref{eq:gammainiL} allows
us to
ensure $\Lambda \partial_{\Lambda}\Gamma_{\Lambda}\to
0$ for $m_{\text{ext},i}/\Lambda\to 0$, see also
below. Indeed, the condition $m_{\text{ext},i}/\Lambda \to 0$
ensures that Eq.~\eqref{eq:dm_ext0} is fulfilled
for $\Gamma_{\Lambda}$. Eq.~\eqref{eq:gammainiL} also offers
a practical way to compute the dependence of $\Gamma_{\Lambda^{\prime}}$ on the
external parameters. In other words, the chosen UV completion in form
  of ${\mathcal F}_{k>\Lambda^{\prime}}$ has to ensure the overall consistency
of the LEFT, and in particular the thermodynamical
consistency. Of course, this procedure is very general and also
  applies to the case of asymptotically free theories as well as to
  asymptotically safe theories where $\Lambda_{\text{UV}}$ is
  infinite.

For the construction of $\Gamma_{\Lambda}$ in case of LEFTs with $\Lambda_{\text{phys}}<\Lambda_\text{UV}$, it 
may even be required to choose $\Lambda >\Lambda_{\text{phys}}$.
At first glance, this appears to be in contradiction to the very definition of the scale $\Lambda_{\text{phys}}$.
Strictly speaking, this is
correct and an extension of LEFTs beyond $\Lambda_\text{phys}$
does not carry the physical fluctuation dynamics of the underlying fundamental theory for scales $\Lambda >\Lambda_{\text{phys}}$.
Nevertheless, we may have to choose $\Lambda >\Lambda_\text{phys}$ in order to suppress cutoff artefacts, i.e.\ 
the failure of Eq.~\eqref{eq:dm_ext0}. For the
generic flow equation~\eqref{eq:genflow}, the change of the initial
condition reads
\begin{align}\label{eq:flowdm_ext}
  \Lambda\0{\partial^2\Gamma_\Lambda[\Phi,m_{\text{ext}}]}{\partial m_{\text{ext},i}\,\partial{\Lambda}} =
 \0{\partial {
  \mathcal F}_{\Lambda}[\Phi,m_{\text{ext}}]}{\partial {m_{\text{ext},i}}} \,.
\end{align}
Integrating \eqref{eq:flowdm_ext} from $m^{(0)}_{\text{ext},i}$
to $m_\text{ext}$ leads to an even more convenient form,
\begin{align}\nonumber 
  &\Lambda\partial_\Lambda \Gamma_\Lambda[\Phi;m_\text{ext}]-
    \Lambda\partial_\Lambda 
    \Gamma_\Lambda[\Phi;m_\text{ext}^{(0)}]\\[2ex] &\hspace{2cm}= 
                                                    {\mathcal F}_{\Lambda}[\Phi;m_\text{ext}] - 
                                                    {\mathcal F}_{\Lambda}[\Phi;m_\text{ext}^{(0)}]\,.
\label{eq:Dflowdm_ext}\end{align}
If Eq.~\eqref{eq:dm_ext0} holds, then the initial effective action is not changed
apart from its explicit dependence on $m_\text{ext}$. The same
holds for the flow equation itself. Accordingly, if
Eqs.~\eqref{eq:flowdm_ext} and~\eqref{eq:Dflowdm_ext} are non-vanishing for a fixed initial effective
action $\Gamma_{\Lambda}$, then the (pre-)initial flow -- and hence the initial
effective action -- has to change for the RG-consistency
condition~\eq{eq:RGconsistency} to hold: with the representation of $\Gamma$ as the
integrated flow, see Eqs.~\eqref{eq:IntegratedFlow} and \eqref{eq:flowdm_ext}, we are
immediately led to the RG consistency condition~\eqref{eq:RGconsistency}. In turn, 
assuming~\eq{eq:dm_ext0}, and using the representation~\eq{eq:IntegratedFlow} of $\Gamma$ as the
integrated flow, we arrive at the important constraint
\begin{align}
  \Lambda\partial_{\Lambda} \Gamma[\Phi;m_\text{ext}]=
  -\left(  {\mathcal F}_{\Lambda}[\Phi;m_\text{ext}] - 
  {\mathcal F}_{\Lambda}[\Phi;m_\text{ext}^{(0)}]\right)\stackrel{!}{=}0 
  \,.
\label{eq:fdiff}
\end{align}
Here, the first term on the right-hand side arises from the
$\Lambda$-derivative of the integrated flow~\eqref{eq:IntegratedFlow},
whereas the second term originates from the $\Lambda$-derivative of
the initial effective action that is kept at its benchmark values for
the external parameters. Note that~Eq.~\eqref{eq:fdiff} has been used
in Ref.~\cite{Helmboldt:2014iya} for defining the ``thermal range"
$\Lambda_T[r]\equiv \Lambda[r;T]$ being the minimal cutoff value for
which Eq.~\eqref{eq:fdiff} holds to a given accuracy. From
our specific examples presented in Sec.~\ref{sec:rgcex}, it is
moreover possible to extract the ``density range" $\Lambda[r;\mu]$ and
the ``volume range" $\Lambda[r;V]$. Of course, the actual values of
these quantities depend on the regularization scheme specified by the
regulator shape function $r$.

More generally speaking, the external parameters
set the minimal value $\Lambda[r; m_\text{ext},m_\text{ext}^{(0)}]$ 
of the cutoff for which Eq.~\eqref{eq:fdiff} holds to a given accuracy. For the standard
benchmark defined by choosing the vacuum values for the external parameters, the third variable can be
dropped. For a given LEFT with a maximal physical UV range
$\Lambda_\text{phys}$, this entails that only results with
$m_\text{ext}$ in the set~${\mathcal M}_{\text{ext}}$, 
\begin{align}\label{eq:pyhsicsrange} 
\!\!\!\!{\mathcal M}_{\text{ext}}(m_\text{ext}^{(0)})\!=\! \left \{ m_\text{ext}\, |\, 
 \Lambda[r; m_\text{ext},m_\text{ext}^{(0)}]  \leq 
 \Lambda_\text{phys}[r]\,\right\}\,,
\end{align}
are fully trustworthy. We emphasize that all the above cutoff scales
naturally depend on the regularization scheme as defined by the choice for the 
regulator function $r$. In turn, the set ${\mathcal M}_{\text{ext}}$ should not depend on
$r$, but may be $r$-dependent in given low-level approximations. 

Provided that $\Lambda_{\text{phys}}$ is known for a LEFT at hand,
the set ${\mathcal M}_{\text{ext}}$ defines the physics range of
this LEFT. Interestingly, this discussion makes also clear that the
physics range for the external parameters depends on the chosen
benchmark value for the external parameters.  Of course, the latter
cannot be chosen freely, as the parameters in the initial effective
action $\Gamma_\Lambda[\Phi; m_\text{ext}^{(0)}]$ are fixed with the
aid of observables at $m_\text{ext}^{(0)}$. Only $m_\text{ext}^{(0)}$,
for which these observables are known, can be used as a
benchmark. Still, this suggests to use available first-principles result
from lattice or functional studies at finite temperature and small
chemical potential with $m_\text{ext}^{(0)}\neq 0$ as a benchmark
instead of the vacuum values. In case of QCD, this in principle allows for
more reliable LEFT computations of, e.g., finite-density effects, 
and is pursued within ``QCD-assisted" LEFTs.

Irrespective of the knowledge of $\Lambda_{\text{phys}}$ and the
corresponding physics range of the LEFT under consideration, it
is still crucial to use the strategy associated with Eq.~\eqref{eq:gammainiL} to remove or at
least suppress cutoff artefacts in the results for physical
observables within a given LEFT study. In Sec.~\ref{sec:rgcex}, we illustrate this strategy in detail 
with the aid of low-energy models of QCD.

In summary, the RG consistency condition~\eqref{eq:RGconsistency} of a
given theory is in general a non-trivial constraint on the initial
effective action at finite external parameters if
Eq.~\eqref{eq:dm_ext} applies to this theory. In the present FRG
framework, this is practically accessible via
Eqs.~\eqref{eq:flowdm_ext} and \eqref{eq:Dflowdm_ext}. Moreover, the formal discussion in
the present section leaves us with a practical toolbox for amending
computations of observables in the presence of finite external
parameters. In any case, we note that
the initial effective action is non-trivial if
Eq.~\eqref{eq:asUV} does not hold. 

\section{RG Consistency -- Examples}\label{sec:rgcex}
In this section, we apply our general discussion of RG consistency to
QCD low-energy models with $N_{\rm f}=2$ quark flavors
and $N_{\rm c}=3$ colors. In the past, low-energy
effective theories including part of the quantum, thermal and density
fluctuations have been studied to an increasing level of
sophistication. A rather large, but not complete, list of LEFTs ranges
from NJL- and QM-type models~\cite{Hatsuda:1985eb,*Asakawa:1989bq,*Klevansky:1992qe,Jungnickel:1995fp,*Berges:1997eu,*Berges:2000ew,Fukushima:2011jc,*Kamikado:2012bt,*Tripolt:2013jra,*Andersen:2014xxa,*Yokota:2017uzu,*vonSmekal:2012vx} over  
quark-meson-diquark 
models~\cite{Rajagopal:2000wf,*Buballa:2003qv,*Shovkovy:2004me,*Alford:2007xm,Strodthoff:2011tz} to models
including baryonic degrees of freedom~\cite{Floerchinger:2012xd,*Drews:2014wba,*Drews:2014spa,*Weyrich:2015hha}, and all of them may even be
augmented with statistical confinement in terms of a Polyakov loop
background and a corresponding Polyakov loop potential~\cite{Meisinger:1997jt,*Pisarski:2000eq,*Ratti:2005jh,*Fukushima:2003fw,*Roessner:2006xn,*Schaefer:2007pw,*Skokov:2010wb,*Skokov:2010uh,*Herbst:2010rf,*Strodthoff:2013cua,*Haas:2013qwp,*Pisarski:2016ixt,*Fukushima:2017csk}. Eventually,
these different models are nothing but different representations of 
the low-energy sector of QCD that emerges after the dynamical
decoupling of the gluonic degrees of freedom at cutoff scales
$\sim 0.4 \dots 1\,\text{GeV}$. For FRG investigations of this
decoupling phenomenon in fundamental QCD, see
Refs.~\cite{Gies:2002hq,Gies:2006wv,Braun:2009gm,Mitter:2014wpa,Braun:2014ata,Rennecke:2015eba,Cyrol:2017ewj};
for more detailed discussions of emergent LEFTS and further
investigations in this respect, see, e.g.,
Refs.~\cite{Pawlowski:2010ht,*Pawlowski:2014aha,Herbst:2010rf,Haas:2013qwp,Herbst:2013ufa,%
  Springer:2016cji}.

A detailed discussion of this interesting embedding of LEFTs in QCD
goes beyond the scope of the present work. Here, we rather aim at a
discussion of how cutoff artefacts can be removed consistently within
a given model study.  However, it is worth emphasizing that the
different LEFT representations of low-energy QCD discussed below can
be all mapped into each other within self-consistent and systematic
expansion schemes. Accordingly, the structural results obtained below
in one of these models extend straightforwardly to all representations
of low-energy QCD. In turn, the impact of truncation artefacts might
be limited to the specific model under investigation.

\subsection{Quark-meson model in the vacuum limit}\label{subsec:njlvac}
We start our discussion of specific examples with a variant of the
QM model as a representation of low-energy QCD in the vacuum limit. On the
mean-field level, its relation to NJL-type models~\cite{Nambu:1961tp,*Nambu:1961fr} 
is most apparent. Its classical or
UV action is nothing but an NJL-type model in its partially bosonized
form. In its instant form (no~$\sim\partial_i\phi$-terms),
the classical action of the QM-model reads
\begin{align}
S 
=\int d^4 x\,\Big\{ {\bar{q}} \Big( \partial\!\!\!\slash
+\frac{1}{2}\bar{h}(\sigma\!+\! {\rm i}\vec{\tau}\cdot\vec{\pi}\gamma_5) 
 \Big) q+\frac{1}{2}{\bar{m}}^2{\phi}^2  \Big\}\,,\label{Eq:HSTAction}
\end{align}
where $\bar{h}$ denotes the Yukawa coupling between quarks and the
scalar and pseudo-scalar mesons. The $\tau_i$'s are the Pauli matrices
which couple the quark spinors $q$ in flavor space.  The scalar
fields ${\phi}^{T}=(\sigma,\vec{\pi})$ do not carry an internal
charge, e.g.  color and flavor. Phenomenologically, these scalar
fields mediate the interaction between the quarks and carry the
quantum numbers of the $\sigma$-meson, $\sigma \sim (\bar{q}q)$, and
the pions, $\vec{\pi}\sim (\bar{q}\vec{\tau}\gamma_5\,q)$,
respectively.

Let us now compute the effective action of our model in a one-loop
approximation where we only take into account purely fermionic
loops. Of course, the effective action can be obtained in various
ways. We shall employ the Wetterich equation,
\eq{eq:genflow} with \eq{eq:wetterich}, which allows for a convenient
computation of the scale-dependent effective action. For
  simplicity, we shall drop terms of the
  form $\sim (\partial_{\mu}{\varphi})^2$ in our calculation 
  although they are generated by purely fermionic
  loops. Here,~$\varphi$ is the so-called classical scalar field associated 
  with the quantum field $\phi$ appearing in the action $S$.
  Our approximations imply that we neglect the RG running of
  the wavefunction renormalization of the scalar fields as well as the
  running of the Yukawa coupling, i.e. we keep $\bar{h}_k$
  constant, $\bar{h}_k\equiv \bar{h}$. By expanding
  the scalar fields~$\varphi$ about a homogeneous background $\bar{\varphi}$, we
  then arrive at the following result for the RG-scale dependent
  effective action:
\be
\!\!\!\!\! \frac{1}{V_4}\Gamma_k[\bar{\varphi}]  = 
\frac{1}{V_4}\Gamma_{\Lambda^{\prime}}[\bar{\varphi}] - 8N_{\rm c}L_k(\Lambda^{\prime},\tfrac{1}{4}\bar{h}^2\bar{\varphi}^2)\,,
\label{eq:Umfvac}
\ee
where $V_4$ is the four-dimensional volume of Euclidean
spacetime, the auxiliary function $L_k$ parametrizes the loop integral
(see below), and $\Lambda^{\prime}$ denotes the scale at which we
assume a simple form of the effective
action $\Gamma_{\Lambda^{\prime}}[\bar{\varphi}]$. Note that we do not indicate
the dependence of $\Gamma_k$ on the classical quark fields
corresponding to the quantum fields $q$ in the action $S$ here and in
the following as they are set to zero.

The initial condition employed to solve the differential equation~\eqref{eq:genflow} is given 
by $\Gamma_{\Lambda^{\prime}}[\bar{\varphi}]$. For example, 
as often done in conventional NJL/QM-type model studies, 
we choose
\be
\frac{1}{V_4}\Gamma_{\Lambda^{\prime}}[\bar{\varphi}]=
\frac{1}{2}m^2_{\Lambda^{\prime}}\bar{\varphi}^2\,.\label{eq:uvpot}
\ee
In this case, the parameters $\bar{h}$
and $\bar{m}^2_{\Lambda^{\prime}}$ are then fixed such that the
experimental/physical values of a given set of low-energy observables
are recovered in the long-range limit from the effective
action $\Gamma_{k\to 0}[\bar{\varphi}]$, e.g. the constituent quark
mass $m_{\rm q} = \tfrac{1}{2}\bar{h}|\bar{\varphi}_0|$ and the pion decay
constant $f_{\pi}=|\bar{\varphi}_0|$. In principle, the three
  parameters $\bar{h}$, $\bar{m}^2_{\Lambda^{\prime}}$,
  and $\Lambda^{\prime}$ can be used to fix the
  constituent quark mass, the pion decay constant, and the mass of the
  $\sigma$-meson. In our numerical studies below, we only
  use $\bar{m}^2_{\Lambda^{\prime}}$ and $\bar{h}$ to fix the
  constituent quark mass and the pion decay constituent. However, our
  line of arguments with respect to the RG consistency criterion can
  also be applied to the former case. The appearance of three
  parameters is related to the fact that the Yukawa coupling is
  marginally relevant with its RG flow being governed only by a Gau\ss
  ian fixed point~\cite{Braun:2011pp,Braun:2012zq}.  
  
  It is worth mentioning that it is not only conventional to parametrize the
effective action as a quadratic form as given in Eq.~\eqref{eq:uvpot}
at some scale $\Lambda^{\prime} \sim 0.4 \dots 1\,\text{GeV}$. It
rather mimics the form of the mesonic part of the effective action
in~QCD in this energy regime.  Indeed, it has been found in FRG
studies of fundamental QCD that mesonic self-interactions of higher
orders are
suppressed~\cite{Gies:2002hq,Gies:2006wv,Braun:2009gm,Mitter:2014wpa,Braun:2014ata,Rennecke:2015eba,Cyrol:2017ewj}. 

The auxiliary function $L_k$ parametrizing the loop integral in \eqref{eq:Umfvac} is defined as 
\be L_k(\Lambda,\chi)&=&\frac{1}{2}\int\frac{{\rm d}^4p}{(2\pi)^4}
\Big\{\ln(p^2(1+r_{\psi})^2 + \chi)\Big|_k \nn\\
&& \qquad - \ln(p^2(1+r_{\psi})^2 + \chi)\Big|_{\Lambda}\Big\}\,, \ee
where $p^2=p_0^2 + \dots + p_3^2$. For $k\to 0$, Eq.~\eqref{eq:Umfvac}
then corresponds to the standard mean-field result for the effective action
for a general (mass-like) regularization scheme as specified by the
regulator shape function $r_{\psi}$.

{The shape function $r_{\psi}$} is implicitly defined via the
definition of the regulator function $R_k\equiv R_k(p)$ appearing in
Eq.~\eqref{eq:wetterich}. In order to preserve chiral symmetry, we
choose the following general form for this
function~\cite{Jungnickel:1995fp,*Berges:1997eu,*Berges:2000ew}:
\be
R_k(p)=-p\!\!\!\slash\, r_{\psi}(\tfrac{p^2}{k^2})\,.
\ee
As also mentioned in 
Sec.~\ref{sec:RGconsistency}, the shape function $r_{\psi}$ is to a
large extent at our disposal~\cite{Wetterich:1992yh}.  For example,
using the Litim or flat  
regulator~\cite{Litim:2000ci,*Litim:2001up,*Litim:2001fd} for an
evaluation of the function $L_k$, we find
\be
L_k(\Lambda,\chi)&=& \frac{1}{2}\int\frac{{\rm d}^4p}{(2\pi)^4}\Bigl\{
\ln({p^2+\chi})\theta(k^2-p^2) \nn\\[2ex] 
&&\qquad + \ln({p^2+\chi})\theta(\Lambda^2-p^2)\theta(p^2-k^2)\nn\\[2ex] 
&&  \qquad\qquad - \ln({\Lambda^2 + \chi})\theta(\Lambda^2-p^2)\Bigr\}\,.
\label{eq:l0plr}
\ee
In the long-range limit $(k\to 0)$, this expression simplifies considerably (see e.g.\ \cite{Braun:2011pp}), 
\be
&& \!\!\!\!\!\!\! L_0(\Lambda,\chi)= \frac{1}{2}\int\frac{{\rm d}^4p}{(2\pi)^4}\theta(\Lambda^2\!-\! p^2)\left\{
\ln({p^2+\chi}) \right. \nn\\
&&\left. \qquad\qquad\qquad\qquad\qquad\qquad\quad - \ln({\Lambda^2 + \chi})
\right\}\,.
\ee
For comparison, we also give the result for $L_k$ as
obtained from a sharp regulator function which is often used in
mean-field studies, 
\be
&& L_k (\Lambda,\chi)\nn\\[2ex]
&& \quad=
\frac{1}{2}\int\frac{{\rm d}^4p}{(2\pi)^4}\theta(\Lambda^2\!-\! p^2)\theta(p^2\!-\!k^2)
\ln({p^2 \!+\! \chi})\,.\label{eq:lopsc}
\ee
As expected, this regulator function cuts off small as well as large momenta sharply.
For $k\to 0$, we then have
\be
L_0(\Lambda,\chi)\!=\! 
\frac{1}{2}\int\frac{{\rm d}^4p}{(2\pi)^4}\theta(\Lambda^2\!-\! p^2) \ln({p^2 + \chi})\,,
\label{eq:lopsc0}
\ee
which, together with Eq.~\eqref{eq:Umfvac}, yields indeed the standard result for the effective 
action $\Gamma[\bar{\varphi}]\equiv\Gamma_{k\to 0}[\bar{\varphi}]$ in the mean-field approximation.
Note also the difference in the expressions~\eqref{eq:l0plr} and~\eqref{eq:lopsc} 
for $L_k$ which can be traced back to the difference in the 
underlying regularization schemes. We add that the momentum integrations in Eqs.~\eqref{eq:l0plr}-\eqref{eq:lopsc0} 
can be performed analytically, if needed, see, e.g., Refs.~\cite{Braun:2011pp,Meyer:2001zp}.

Although our ansatz for the effective action at the
scale $\Lambda^{\prime}$ mimics the situation in QCD, the effective
action $\Gamma_{\Lambda^{\prime}}$ at the scale $\Lambda^{\prime}$
does not yet obey the RG-consistency condition~\eqref{eq:RGconsistency}, i.e. we
have $\Lambda^{\prime}\partial_{\Lambda^{\prime}}\Gamma_{\Lambda^{\prime}}\neq
0$.  Therefore, we now apply our general line of arguments detailed in
Sec.~\ref{sec:RGconsistency} to obtain an RG-consistent result for the
effective action of our present model in the mean-field approximation.
From our general discussion, we immediately conclude that the
effective action $\Gamma\equiv\Gamma_{k\to 0}$ does not depend on the
actual scale $\Lambda$ at which we fix $\Gamma_{\Lambda}$, provided that we
adapt $\Gamma_{\Lambda}$ accordingly, see Eq.~\eqref{eq:Gamma}.
Indeed, assuming $\Lambda>\Lambda^{\prime}$ and using
Eqs.~\eqref{eq:gammainiL} and~\eqref{eq:Gamma}, we obtain
\be \frac{1}{V_4}\Gamma[\bar{\varphi}] =
\frac{1}{V_4}\Gamma_{\Lambda}[\bar{\varphi}] - 8 N_{\rm c}
L_0(\Lambda,\tfrac{1}{4}\bar{h}^2\bar{\varphi}^2)\,,
\label{eq:upotl}
\ee
where
\be
\frac{1}{V_4}\Gamma_{\Lambda}[\bar{\varphi}] = \frac{1}{2}\bar{m}^2_{\Lambda^{\prime}}\bar{\varphi}^2 
+ 8 N_{\rm c} L_{\Lambda^{\prime}}(\Lambda,\tfrac{1}{4}\bar{h}^2\bar{\varphi}^2)\,.
\label{eq:ULprime}
\ee
Note that $\Gamma_{\Lambda^{\prime}}[\bar{\varphi}]$
and $\Gamma_{\Lambda}[\bar{\varphi}]$ obey a different dependence on
the field $\bar{\varphi}$.  This can be readily demonstrated for asymptotically large
scales $\Lambda$. In this case, the initial effective action $\Gamma_{\Lambda}$
receives $\Lambda$-dependent corrections only from terms up to fourth
order in the field $\bar \varphi$ as higher orders are suppressed by
powers of $\Lambda$:
\be \frac{1}{V_4}\Gamma_{\Lambda}[\bar{\varphi}] &=&\left(
  \frac{1}{2}\bar{m}^2_{\Lambda^{\prime}} +\frac{3 \bar h ^2
    \Lambda^2}{16 \pi^2}
\right) \bar \varphi^2 \nn \\[2ex]
&&\quad - \frac{3 \bar h ^4(1+ 4 \ln ( 2 \Lambda)) }{256 \pi^2}\bar\varphi^4\nn\\[2ex] 
&& \quad\quad- \frac{\bar h ^6 }{48\Lambda^2}\bar \varphi^6+
\CO\left(\frac{ \bar \varphi ^8}{\Lambda^4}\right)\,.
\label{eq:ULprimeExpansion}
\ee
Here, we have dropped field-independent constants and terms explicitly
depending on the scale $\Lambda^{\prime}<\Lambda$. In any case, in the long-range limit ($k\to 0$), the
effective action~\eqref{eq:upotl} agrees identically with the one
given in~\eqref{eq:Umfvac}, as it should be. In
particular, we find that the effective action $\Gamma$ obeys the RG
consistency condition~\eq{eq:RGconsistency}, that is 
$\Lambda\partial_{\Lambda}\Gamma[\bar{\varphi}]=0$.

For a study with finite external parameters, we can now
adjust $\Lambda$ such that cutoff artefacts are removed. The
latter may appear if $\Lambda > \Lambda^{\prime}$ has been chosen too
small initially for a specific range of the considered parameter
set. A priori, it may indeed be difficult to choose a suitable value
for $\Lambda$. However, our line of arguments given in
Sec.~\ref{sec:RGconsistency} shows how this issue can be resolved.
Even more, it allows us to investigate systematically cutoff effects
in the presence of external parameters since the vacuum physics is
left unchanged.\footnote{Of course, it is mandatory that the vacuum
  contributions to the effective action as well as those arising in
  the presence of finite external parameters are regularized
  consistently, i.e. in exactly the same way, as worked out in detail
  in Sec.~\ref{sec:RGconsistency}, see also Ref.~\cite{Braun:2015fva}
  for a discussion of this issue in terms of a {Polyakov}-loop
  extended NJL model.}  

Before we shall demonstrate this explicitly, we stress that our line
of arguments, which eventually led us to the RG consistency criterion in
Sec.~\ref{sec:RGconsistency}, goes qualitatively beyond what is sometimes called extended mean-field
theory in the literature. In fact, the
vacuum fermion loop associated with extended mean-field calculations
is naturally included in an RG treatment and should anyway not be
discarded in any other approach, see,
e.g.,~Refs.~\cite{Meyer:2001zp,Braun:2009si,Braun:2011pp} for detailed discussions of 
mean-field theory in the RG context and Refs.~\cite{Braun:2003ii,Skokov:2010sf} for approximative treatments 
of RG consistency in low-energy models of QCD.
Moreover, it is clear from our line of arguments that manifestation of 
RG consistency in general requires to include the fully field-dependent fermion loop and,
beyond the mean-field approximation, it even requires to include the fully
field-dependent contributions from {\it all} 
loop diagrams considered in a specific calculation of the
quantum effective action.
This also becomes apparent from the
right-hand side of the differential equation~\eqref{eq:genflow} which includes contributions
from all fields of a given theory, e.g. by means of the Wetterich equation.

\subsection{Diquarks -- equation of state}\label{subsec:dqeos}
As a second example, we consider the computation of the equation of state 
of a simple quark-diquark model as a function of the quark chemical 
potential at vanishing temperature. The model is defined by 
the following classical action (for reviews see, 
e.g., \cite{Rajagopal:2000wf,*Buballa:2003qv,*Shovkovy:2004me,*Alford:2007xm}): 
\begin{align}
\!\!\! S 
=&\int d^4 x\,\Big\{ {\bar{q}}\left(\partial\!\!\!\slash 
\!-\! \mu\gamma_0\right)q
 \!+\! \bar{\nu}^2\Delta^{\ast}_A\Delta_A  \nn\\[2ex]
 & \; + \bar{q}\gamma_5\tau_2\Delta_{A}^{\ast}T^{A}{\mathcal C}\bar{q}^{T}
 - q^{T}{\mathcal C}\gamma_5\tau_2\Delta_{A}T^{A}{q}\Big\}\,,
 \label{Eq:HSTActionDQ}
\end{align}
where ${\mathcal C}$ is the charge conjugation
operator and the sum over the color index $A$ runs only over
antisymmetric color generators $T^{A}$ in the fundamental
representation. The complex-valued scalar fields $\Delta_A$ carry the
quantum numbers of diquark
states, $\Delta_{A}\sim ( \bar{q}\gamma_5\tau_2 T^{A}{\mathcal
  C}\bar{q}^{T})$, with $J^{P}=0^{+}$. The parameter $\bar{\nu}$ is at
our disposal and can be used to determine the ground-state properties
of the vacuum in this model. From a general fixed-point
  analysis, see~\cite{Braun:2017srn,Braun:2018bik} and e.g.~\cite{Alford:1997zt} for a mean-field analysis, 
  it follows immediately that two qualitatively distinct scenarios are possible.
To be specific, we may choose $\bar{\nu}^2$ to be positive but
small such that already the ground state in the vacuum limit is
governed by the formation of a diquark condensate breaking the
$U_{\rm V}(1)$ symmetry of our model. Alternatively, we may choose a
sufficiently large value of $\bar{\nu}^2$ such that the $U_{\rm V}(1)$
symmetry is only broken at finite $\mu$ due to the existence of a 
  Cooper instability in the system but remains intact in the vacuum
limit. In the latter, we therefore conclude that a critical
value $\bar{\nu}_{\ast}$ (associated with a non-Gau\ss ian fixed
point) exists which separates these two distinct scenarios from each
other.

Let us now compute the effective action of this model in a
  one-loop approximation where we only take into account the purely
  fermionic loop again.  Moreover, we set the wavefunction
  renomalizations associated with the diquark fields to zero. In other
  words, we shall drop terms of the following form in our computation
  of the effective action:
\be
&& \int {\rm d}^4 x\,\Big\{\frac{1}{2}Z_{\perp}(|\Delta|^2)|\vec{\nabla} \Delta|^2 
+ \frac{1}{2}Z_{\parallel}(|\Delta|^2)|\partial_{\tau}\Delta|^2 \nn\\
 && \qquad\qquad\quad
 + \mu Z_{\mu}(|\Delta|^2) (\Delta \partial_{\tau}\Delta^{\ast} - \Delta^{\ast} \partial_{\tau}\Delta)\Big\}\,,
 \label{eq:dqzs}
\ee
where $\Delta^{\ast}{\mathcal O}\Delta\equiv
\sum_A\Delta^{\ast}_A{\mathcal O}\Delta_A$
and $|{\mathcal O}\Delta|^2 \equiv \sum_A |{\mathcal O}\Delta_A|^2$
with ${\mathcal O}$ being some operator acting on the diquark
fields. Note that, in general, such terms are dynamically generated due to 
quantum effects, even if only purely fermionic loops are taken into
account. As a consequence of the listed approximations, we also do not take
into account a scale dependence of the Yukawa-type quark-diquark
coupling but set it to be constant.  Using the Wetterich
equation~\eqref{eq:wetterich} and expanding the diquark fields about a
homogeneous background $\bar{\Delta}_A$, we then obtain the following
expression for the scale-dependent effective action:
\begin{align}
\frac{1}{V_4}\Gamma_k[\{\bar{\Delta}_A\}]\!=\! \frac{1}{V_4}\Gamma_{
  \Lambda^{\prime}}[\{\bar{\Delta}_A\}] \! -\! \frac{\mu^4}{6\pi^2}\!
-\!  8 M_k(\Lambda^{\prime},|\bar{\Delta}|^2)\,,
\label{eq:Umfdq}
\end{align}
where $|\bar{\Delta}|^2=\sum_A |\bar{\Delta}_A|^2$ and
$\Lambda^{\prime}$ denotes again the scale at which we know the form
of the effective action. The contribution $\sim\mu^4$ arises from
quark degrees of freedom which do not couple to the diquark fields and
therefore appear as non-interacting ``spectators".

The loop integral associated with the effective action~\eqref{eq:Umfdq} is 
parametrized by the function~$M_k$,
\be
\!\!\!\!\!\!\!\!\!\! M_k(\Lambda,\chi) 
=\frac{1}{2} \int \frac{{\rm d}^3p}{(2\pi)^3}
\sum_{\sigma=\pm 1} \left(\omega^{(\sigma)}\Big|_k
\!-\! \omega^{(\sigma)}\Big|_{\Lambda}\right)
\,,
\label{eq:Mdef}
\ee
where the auxiliary quantity $\omega_{\Delta}^{(\sigma)}$ may be
viewed as (infrared) regularized quasiparticle energy,
\be
\omega^{(\sigma)}=\sqrt{ ( |\vec{p}^{\,}|(1+r_{\psi}) +\sigma \mu)^2 + \chi}\,.
\label{eq:qpe}
\ee
Evidently, for $\chi=0$, we have
  $\omega^{(\sigma)}= |\vec{p}^{\,}|(1+r_{\psi}) +\sigma \mu$. In the
calculation of the loop integral~\eqref{eq:Mdef}, we have now employed
the class of so-called 3$d$ regulator functions which is
defined as
\be
R_k(p)=-\vec{p}\!\!\!\slash\, r_{\psi}(\tfrac{\vec{p}^{\,2}}{k^2})\,.
\ee
This class of regularization schemes is also frequently used in QCD
model studies since it allows to perform analytically the Matsubara
sums in at least some of the loop diagrams. However, it should also be noted 
that 3$d$ regularization schemes break the
{Poincar\'{e}} symmetry explicitly.  This explicit breaking is present
even in the limit of vanishing temperature and chemical potential,
see, e.g., Refs.~\cite{Braun:2009si,Helmboldt:2014iya,Pawlowski:2015mia,Pawlowski:2017gxj,Braun:2017srn}. The appearance of
this issue can be traced back to the fact that, by construction,
3$d$ regulators do not cut off the time-like momentum
modes, thereby treating the time-like and spatial modes
differently. We shall ignore this issue in our present study.

For convenience, we shall only consider the 3$d$ sharp
cutoff in our numerical studies below.  The function $M_k$ is then
given by
\begin{align}
  & M_k(\Lambda,\chi) 
    =\frac{1}{2} \int \frac{{\rm d}^3p}{(2\pi)^3} \theta(\Lambda^2 \!-\! 
    \vec{p}^{\,2})\theta(\vec{p}^{\,2}\!-\! k^2) \times \nn\\[2ex]
  & \qquad\;\; \times \Big\{ \sqrt{ ( |\vec{p}^{\,}| +\mu)^2 + \chi}  
    + \sqrt{ ( |\vec{p}^{\,}| -\mu)^2 + \chi}\Big\}
    \,,
\label{eq:Mk}
\end{align}
which reduces to the standard mean-field expression in the limit $k\to 0$. 
\begin{figure*}[t]
  \includegraphics[scale=0.7]{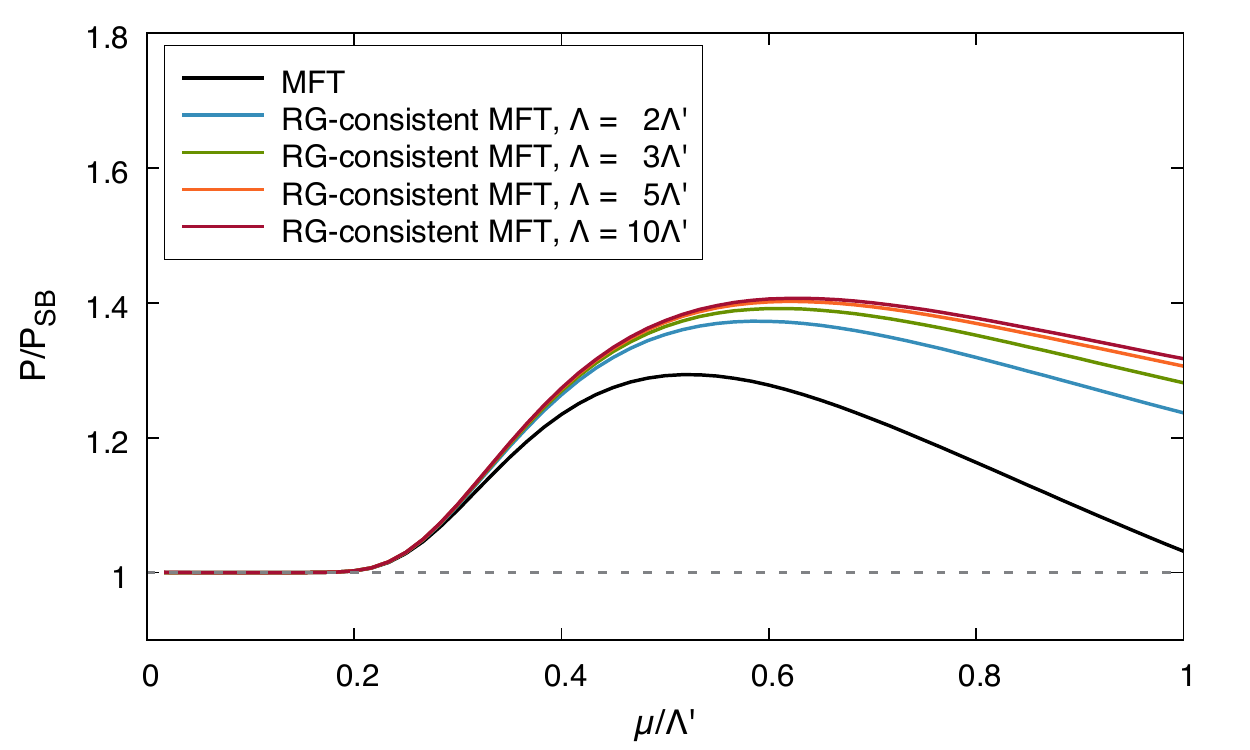}
  \includegraphics[scale=0.7]{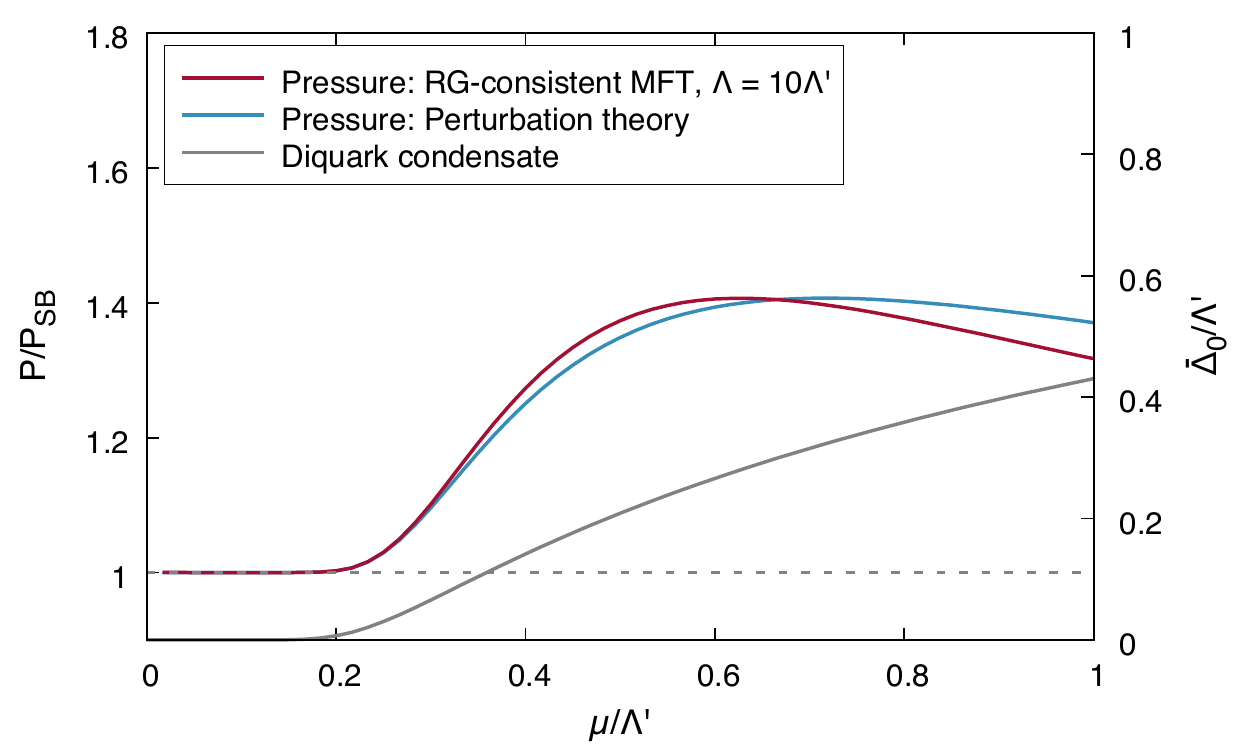}
  \caption{Left panel: Pressure $P/P_{\text{SB}}$ of our diquark model
    as a function of the chemical potential $\mu/\Lambda^{\prime}$ (with~$\Lambda^{\prime}=0.6\,\text{GeV}$) as
   obtained from conventional mean-field theory (MFT) with a UV
    cutoff $\Lambda=\Lambda^{\prime}$ (black line) as well as from
    RG-consistent MFT
    with $\Lambda/\Lambda^{\prime}=2,3,5,10$.  Right panel: Pressure
    $P/P_{\text{SB}}$ of our diquark model as a function of the
    chemical potential $\mu/\Lambda^{\prime}$ as obtained from 
    RG-consistent MFT
    with $\Lambda/\Lambda^{\prime}=10$ together with the perturbative
    expression for the pressure at leading order in the weak-coupling
    expansion, see Eq.~\eqref{eq:pressdqpt}.  Moreover, we also show
    the gap $\bar{\Delta}_0/\Lambda^{\prime}$ (gray line) as extracted
    from RG-consistent MFT
    with $\Lambda/\Lambda^{\prime}=10$.  }
\label{fig:pressdq}
\end{figure*}

Before we present our results for the equation of state of our diquark model, 
we would like to discuss first a subtlety in our
  calculation: In contrast to a possible renormalization of the quark
chemical potential driven by diagrams with internal bosonic and
fermionic lines, the renormalization of the chemical potential of the
diquarks associated with a term $\sim\mu^2 |\Delta|^2$ is already
included in our present analysis. Indeed, the field-dependent
renomalization factor $Y\equiv Y_{k\to 0}$ of the diquark chemical
potential is given by
\be
Y_{k}(|\bar{\Delta}|^2)=-\frac{1}{4 V_4}\partial_{\mu}^2 \Gamma_k[\{\bar{\Delta}_A\}]\Big|_{\mu=0}\,.
\ee
Using Eq.~\eqref{eq:Umfdq} for the effective action, we 
find $Y \sim  |\Delta|^2 \ln\Lambda^{\prime} + \dots$. Thus, $Y$ exhibits the same dependence on $\Lambda$
as expected for the renormalization factors of kinetic terms, such as 
the ones for the diquark fields in Eq.~\eqref{eq:dqzs}. This coincidence in the $\Lambda$-dependence 
of $Y$ and, e.g., $Z_{\parallel}$ 
is by no means accidental. It is rather related to a more abstract symmetry of our model 
which is associated with the so-called Silver-Blaze property of
quantum field theories~\cite{Cohen:2003kd,Marko:2014hea,Khan:2015puu,*Fu:2015naa}. 
In general, this property is linked to the fact that the free energy should not exhibit a dependence 
on the baryon/quark chemical potential at {\it zero} temperature, provided that it
is smaller than some critical value. Then, the corresponding symmetry is not violated. 
The critical value is set by the gaps in the propagators of the fields 
associated with a finite baryon number.
Note that the gap is not necessarily given by the physical (pole) mass. In our RG study, 
for example, the gap may also arise for~$k>0$ from the IR regularization of the propagator, 
see Ref.~\cite{Khan:2015puu} for details.

In the presence of the symmetry associated with the Silver-Blaze property~\cite{Khan:2015puu}, 
a finite renormalization factor $Y$ of the diquark chemical potential implies
that the renormalization factors $Z_{\perp}$, $Z_{\parallel}$,
and $Z_{\mu}$ of the diquark fields are in principle finite as well.  
In mean-field calculations, these renormalization factors are usually set to zero. Therefore, 
the resulting effective action violates the
Silver-Blaze property.\footnote{Irrespective of the regularization
  scheme, the so-called {Silver}-{Blaze} property of the theory is
  already violated by the fact that the quasiparticle energies are
  only positive semi-definite in (standard) mean-field approximations,
  see Eq.~\eqref{eq:qpe}.}  As already stated above, we shall not
compute these renormalization factors in this work but also
set them to zero.  Since we shall fix
the couplings/parameters of our model at a scale $k=\Lambda^{\prime} > \mu$,
i.e. at a point where the model is expected to respect the symmetry
associated with the Silver-Blaze property, we set the initial
condition for the renormalization factor $Y$ to zero as well.  This
ensures that this property is at least manifestly present at the
scale $\Lambda^{\prime}$ at which we fix the parameters of the model.
To be specific, we make the following ansatz for the effective action
at the scale $\Lambda^{\prime}$ in the vacuum limit,
\begin{align}
  \frac{1}{V_4}\Gamma_{\Lambda^{\prime}}[\{\bar{\Delta}_A\}]=
  \bar{\nu}^2_{\Lambda^{\prime}}|\bar{\Delta}|^2\,,
\label{eq:gamma0dq}
\end{align}
where $\bar{\nu}^2_{\Lambda^{\prime}}$ is at our disposal and
corresponds to the parameter $\bar{\nu}^2$ in the classical
action~\eqref{Eq:HSTActionDQ}.  However, our
choice~\eqref{eq:gamma0dq} for the effective action at the
scale $\Lambda^{\prime}$ does not imply that $Y$ remains zero at
scales $k\neq \Lambda^{\prime}$. 
Since we do not take into account
the running of the renormalization factors $Z_{\perp}$, $Z_{\parallel}$,
and $Z_{\mu}$, the symmetry associated
with the Silver-Blaze property is therefore in general violated
away from the scale $\Lambda^{\prime}$.  Still, the consideration of
the renormalization factor $Y$ is required to ensure RG consistency
within our model study, see below.

Along the lines of our discussion of the vacuum limit, we can now
construct an RG-consistent effective action $\Gamma_{k\to 0}$ from
\eqref{eq:Umfdq} by adapting the effective action at the
scale $\Lambda>\Lambda^{\prime}$ such that the effective action at 
scales $k\leq \Lambda^{\prime}$ remains unchanged:
\begin{align}
\frac{1}{V_4}\Gamma_k[\{\bar{\Delta}_A\}]\!=\!\frac{1}{V_4}
\Gamma_{\Lambda}[\{\bar{\Delta}_A\}]\!-\!\frac{\mu^4}{6\pi^2}
\!-\! 8 M_k(\Lambda,|\bar{\Delta}|^2)\,,
\label{eq:effacdqrgc}
\end{align}
where
\begin{align}
  \frac{1}{V_4}\Gamma_{\Lambda}[\{\bar{\Delta}_A\}]=&\,\frac{1}{V_4}\Gamma_{\Lambda^{\prime}}[\{\bar{\Delta}_A\}]
                                                      +8 M_{\Lambda^{\prime}}(\Lambda,|\bar{\Delta}|^2)\Big|_{\mu=0}\nn\\[2ex]
                                                    &\quad +\, 4\mu^2\Big(\partial_{\mu}^2 M_{\Lambda^{\prime}}(\Lambda,|\bar{\Delta}|^2)\Big|_{\mu=0}\Big)\,.
\end{align}
Here, the last term on the right-hand side accounts for the fact that
the diquark chemical potential is renormalized.  Using
Eq.~\eqref{eq:effacdqrgc}, we indeed find that $\Gamma$ is
RG-consistent in a strict sense in the limit $\Lambda\to\infty$ since
\begin{align}
 \Lambda \partial_{\Lambda} \Gamma[\{\bar{\Delta}_A\}]\!=\! 
- 2V_4|\bar{\Delta}|^2\mu^2 \left(\frac{\mu}{\pi\Lambda}\right)^2
\!+\! {\mathcal O}(1/\Lambda^{4})\,.
\end{align}
Moreover, we deduce from Eq.~\eqref{eq:effacdqrgc} that the
renormalization of the diquark chemical potential still vanishes
identically at the scale $\Lambda^{\prime}$,
\be
Y_{\Lambda^{\prime}}(|\bar{\Delta}|^2) 
=0\,,
\ee
as it should be. With our RG-consistent effective
action~\eqref{eq:effacdqrgc} at hand, we now compute the equation of
state of our diquark model. More specifically, we compute the
pressure $P$ which is directly obtained from the effective action:
\begin{align} 
P =-\frac{1}{V_4}\Gamma[\{\bar{\Delta}_A\}_{\text{gs}}]\Big|_{\mu}
+ \frac{1}{V_4}\Gamma[\{\bar{\Delta}_A\}_{\text{gs}}]\Big|_{\mu=0}\,.
\label{eq:pressdq}
\end{align}
Here, the subscript `gs' indicates that the effective action is
evaluated on the $\mu$-dependent minimum (i.e. on the ground-state
(gs) configuration of the fields $\{\bar{\Delta}_A\}$). Note that we
have normalized the pressure with respect to the pressure in the
vacuum limit. The latter is given by the second term on the right-hand
side of Eq.~\eqref{eq:pressdq}.

As an explicit example, we compute the pressure of our pure
diquark model
for $(\bar{\nu}_{\Lambda^{\prime}}/\bar{\nu}_{\ast})^2 = 4/3$
where $\bar{\nu}_{\ast}^2 \approx 0.036$. Moreover, we 
set $\Lambda^{\prime}=0.6\,\text{GeV}$ in the following.
Phenomenologically
speaking, our parameter choice implies that the $U_{\rm V}(1)$
symmetry is only broken at finite $\mu$ but remains intact in the
vacuum limit, see our discussion above. Thus, the ground state in the
vacuum limit is governed by ungapped quarks. 

In the left panel of Fig.~\ref{fig:pressdq}, we show our results for
the pressure $P/P_{\text{SB}}$ of our diquark model,
where $P_{\text{SB}}=\mu^4/(2\pi^2)$ denotes the Stefan-Boltzmann
limit of the pressure, i.e. the pressure of a free quark gas at zero
temperature.  We observe that cutoff effects become continuously
smaller when $\Lambda/\Lambda^{\prime}$ is increased. Recall that, in
our RG-consistent calculations, an increase of $\Lambda$ leaves the
model in the vacuum limit unchanged. Moreover, we find that the
corrections to the results from the conventional mean-field study are
significant. Indeed, the pressure obtained from the conventional
mean-field study underestimates the (effectively) cutoff-independent
result for the pressure obtained from our RG-consistent mean-field
study (with $\Lambda/\Lambda^{\prime} = 10$) by about $10\%$
at $\mu/\Lambda^{\prime}=1/2$. Thus, ``cutoff contaminations" are
clearly visible even at values of the chemical potential which seem to
be sufficiently small compared to the originally chosen
scale~$\Lambda^{\prime}$. At $\mu/\Lambda^{\prime}=1$, the results
from the conventional mean-field study and our RG-consistent
mean-field study (with $\Lambda/\Lambda^{\prime}=10$) then already
deviate by about $30\%$. Increasing $\mu$ even further, we observe
that the pressure approaches the {Stefan}-{Boltzmann} limit from
above, provided $\Lambda/\Lambda^{\prime}$ has been chosen
sufficiently large.

From \eqref{eq:pressdq}, we can also derive the perturbative
result for the pressure. At leading order
of $|\bar{\Delta}_0|^2/\mu^2$ in the weak-coupling expansion, we
indeed recover the well-known result~\cite{Rajagopal:2000ff,*Shovkovy:2002kv}:
\begin{align}
\frac{P}{P_{\text{SB}}}=1 + \frac{2|\bar{\Delta}_0|^2}{\mu^2}
+ \dots\,,
\label{eq:pressdqpt}
\end{align}
where $\bar{\Delta}_0$ denotes the gap as obtained from a minimization
of the effective action.  In the right panel of
Fig.~\ref{fig:pressdq}, we compare this perturbative result for the
pressure with the results from our RG-consistent mean-field
calculation with $\Lambda/\Lambda^{\prime}=10$.  Moreover, the gap as
obtained from the same RG-consistent calculation is shown. Plugging
this result for the gap into the perturbative
expression \eqref{eq:pressdqpt} for $P/P_{\text{SB}}$, we find very
good agreement with the RG-consistent results for the pressure in the
regime where~\mbox{$|\bar{\Delta}_0|/\mu \lesssim 0.5$}. For larger
values of $\mu$, the results from the perturbative approximation of
the pressure then exceed the results from the RG-consistent
calculation. Still, the perturbative expression for the pressure
appears to provide us with a reasonable estimate for the pressure over
a wide range of the chemical potential, at least for our present
choice for the model parameter $\bar{\nu}_{\Lambda^{\prime}}$.

\subsection{Quarks, mesons, and diquarks -- phase diagram}\label{subsec:qmdqpd}
Let us now turn to our third example, the computation of the phase
diagram and the zero-temperature equation of state of a
quark-meson-diquark model with two massless quark flavors and
$N_{\rm c}=3$ colors.  The classical action $S$ underlying our study
may be viewed as a combination of the actions already discussed in
Subsecs.~\ref{subsec:njlvac} and~\ref{subsec:dqeos} and reads
\begin{align}
S 
=&\,\int d^4 x\,\Big\{ {\bar{q}}\Big( \partial\!\!\!\slash 
\!-\! \mu\gamma_0
\!+\! \frac{1}{2}{\bar{h}}(\sigma\!+\! {\rm i}\vec{\tau}\cdot\vec{\pi}\gamma_5)
\Big)q\nn\\[2ex]
 &\qquad\qquad + \bar{q}\gamma_5\tau_2\Delta_{A}^{\ast}T^{A}{\mathcal C}\bar{q}^{T}
 - q^{T}{\mathcal C}\gamma_5\tau_2\Delta_{A}T^{A}{q}\nn\\[2ex]
  & \qquad\qquad\qquad \!+ \frac{1}{2}{\bar{m}}^2{\phi}^2  \!+\! \bar{\nu}^2\Delta^{\ast}_A\Delta_A  
 \Big\}\,,
 \label{Eq:HSTActionQMDQ}
\end{align}
where $\bar{h}$, $\bar{m}^2$ and $\bar{\nu}^2$ are parameters at our
disposal.  In the following, we compute the effective
action of this model in a one-loop approximation where we only take
into account purely fermionic loops. Moreover, the
wavefunction renormalization factors of the meson and diquark fields
are set to zero again. These approximations also imply that we neglect
the RG runnings of the Yukawa-type couplings of our model. 
As before, we moreover neglect
corrections to the wavefunction renormalization factors of the quark
fields (as well as to the quark chemical potential). 
Note that
our discussion in the previous subsection regarding the fate of the
{Silver}-{Blaze} property in mean-field-like calculations also holds
for the present study of a quark-meson-diquark model.
\begin{figure*}[t]
  \includegraphics[scale=0.7]{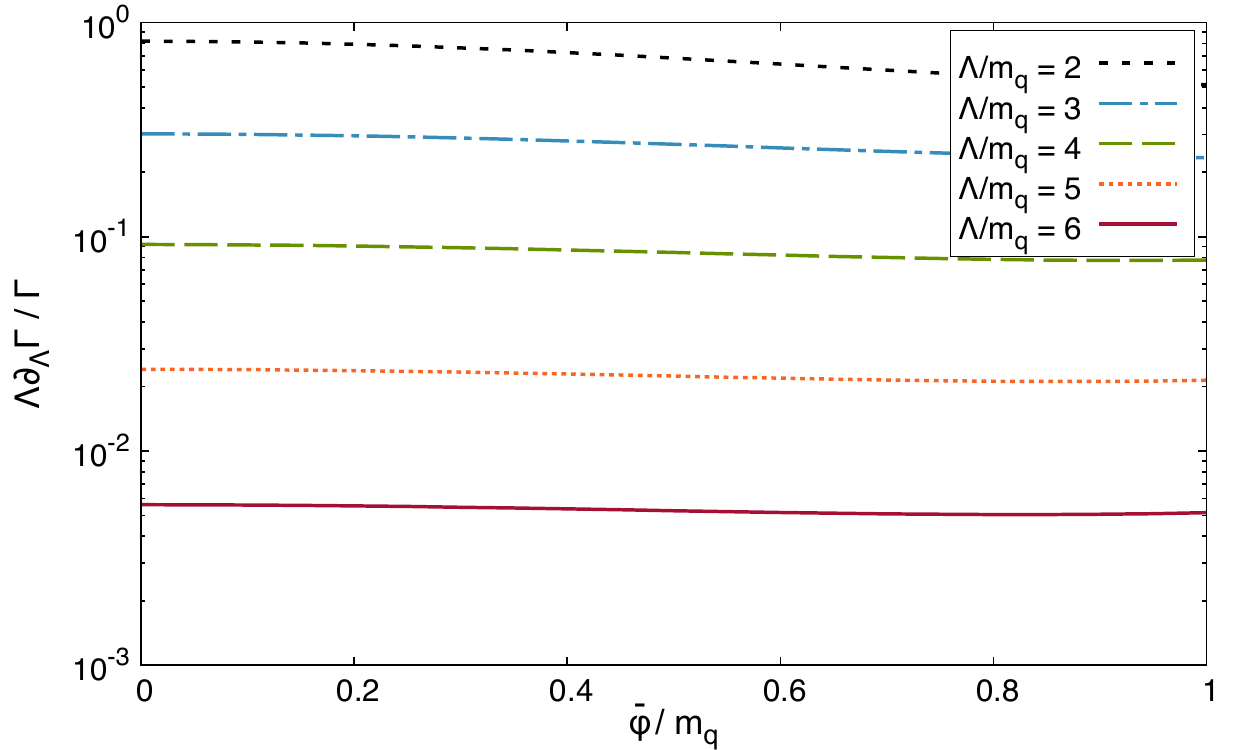}
  \includegraphics[scale=0.7]{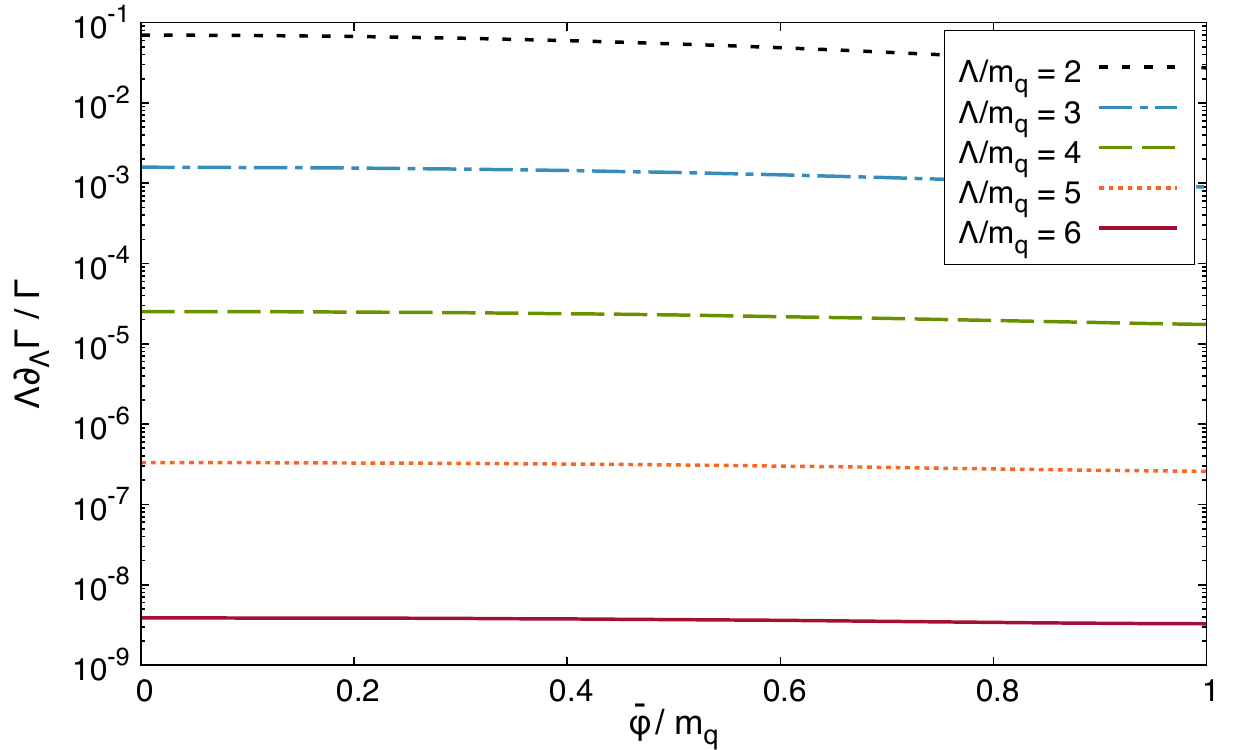}
\caption{Change of the effective action at $\bar{\Delta}=0$ 
under a variation of the UV scale $\Lambda$, i.e.  $\Lambda\partial_{\Lambda}\Gamma$, 
relative to the effective action $\Gamma$ itself 
 as a function of $\bar{\varphi}$
for $T/m_{\rm q}= 1/2$ and $\mu=0$ (left panel) as well as for $T/m_{\rm q}=1/5$ and $\mu/m_{\rm q}=1$ (right panel) 
 for various different values of $\Lambda/m_{\rm q}$, where~$m_{\rm q}\approx 0.300\,\text{GeV}$ is the vacuum quark mass.}
\label{fig:delG}
\end{figure*}

Using the {Wetterich} equation~\eqref{eq:wetterich} with the class of 3$d$ regulator functions and expanding 
the meson and diquark fields about homogeneous backgrounds $\bar{\varphi}$ and $\bar{\Delta}_A$, respectively, we obtain
the following result for the scale-dependent RG-consistent \mbox{effective action}:
\begin{align}
\frac{T}{V}\Gamma_k[\bar{\varphi},\{\bar{\Delta}_A\}] =&\, \frac{T}{V}\Gamma_{\Lambda}[\bar{\varphi},\{\bar{\Delta}_A\}] - 
4 L_k^{(T)}(\Lambda,\tfrac{1}{4}\bar{h}^2\bar{\varphi}^2)\nn\\[2ex]
& \qquad - 8M_k^{(T)}(\Lambda,\tfrac{1}{4}\bar{h}^2\bar{\varphi}^2,|\bar{\Delta}|^2)\,,
\label{eq:44}
\end{align}
where $T$ is the temperature, $V$ is the spatial volume of the system, 
and 
\begin{align}
& \frac{T}{V}\Gamma_{\Lambda}[\bar{\varphi},\{\bar{\Delta}_A\}]\nn\\[2ex]
& \qquad\quad =\frac{T}{V}\Gamma_{\Lambda^{\prime}}[\bar{\varphi},\{\bar{\Delta}_A\}]
+ 4L_{\Lambda^{\prime}}^{(T)}(\Lambda,\tfrac{1}{4}\bar{h}^2\bar{\varphi}^2)\Big|_{T=\mu=0}\nn\\[2ex]
&  \qquad\quad\quad + 4\mu^2\!\left(\!\partial_{\mu}^2 M_{\Lambda^{\prime}}^{(T)}(\Lambda,\tfrac{1}{4}
\bar{h}^2\!\bar{\varphi}^2,|\bar{\Delta}|^2)\Big|_{T=\mu=0}\!\right)\nn\\[2ex]
& \qquad\quad\quad\quad + 8 M_{\Lambda^{\prime}}^{(T)}(\Lambda,\tfrac{1}{4}\bar{h}^2
\bar{\varphi}^2,|\bar{\Delta}|^2)\Big|_{T=\mu=0}\,.
\label{eq:qmdqgl}
\end{align}
Here, the auxiliary functions $L_{k}^{(T)}$ and $M_{k}^{(T)}$
parametrize loop integrals in the presence of a heat bath with
temperature $T=1/\beta$, see below for their definitions.  The term
$\sim\mu^2$ in Eq.~\eqref{eq:qmdqgl} accounts for the renormalization
of the chemical potential of the diquarks. As done in the previous
subsections, we shall assume that the parameters of the model are
fixed at the scale $k=\Lambda^{\prime}<\Lambda$ by means of an ansatz
for $\Gamma_{\Lambda^{\prime}}$ in Eq.~\eqref{eq:qmdqgl}. For a study
of the effect of a finite temperature and/or quark chemical potential,
the scale $\Lambda$ then has to be chosen sufficiently large such that
cutoff artefacts are suppressed.  For $\Gamma_{\Lambda^{\prime}}$, to
be specific, we use the following ansatz (in the vacuum limit):
\begin{align}
\lim_{T\to 0}\frac{T}{V} \Gamma_{\Lambda^{\prime}}[\bar{\varphi},\{\bar{\Delta}_A\}]=
\frac{1}{2}\bar{m}^2_{\Lambda^{\prime}}\bar{\varphi}^2
+\bar{\nu}_{\Lambda^{\prime}}^2|\bar{\Delta}|^2\,.
\label{eq:glpqmdqm}
\end{align}
As often done in quark-meson-diquark model
studies~\cite{Rajagopal:2000wf,*Buballa:2003qv,*Shovkovy:2004me,*Alford:2007xm},
we shall relate the parameters appearing in Eq.~\eqref{eq:glpqmdqm}
via $2\bar{m}_{\Lambda^{\prime}}^2/\bar{h}^2=(3/4)\bar{\nu}^2_{\Lambda^{\prime}}$
and fix $\bar{h}$, $\bar{m}^2_{\Lambda^{\prime}}$ at
$\Lambda^{\prime}/m_{\rm q}=2$ in the vacuum limit such that we obtain
$m_{\rm q}=\frac{1}{2}\bar{h}\bar{\varphi}_0 \approx
0.300\,\text{GeV}$ for the quark mass and
$f_{\pi}=2m_{\rm q}/\bar{h}\approx 0.088\,\text{GeV}$ for the pion
decay constant.

For sufficiently large values of $\Lambda$, the effective action
$\Gamma_{\Lambda}$ receives corrections only from terms up to fourth
order in the fields $\bar{\varphi}$ and $\bar{\Delta}$,
respectively. Higher orders are suppressed when $\Lambda$ is
increased. This resembles the situation in
Subsec.~\ref{subsec:njlvac}.  Recall that we have
$\Lambda\partial_{\Lambda}\Gamma =0$ by construction at~$T=\mu=0$ and, for sufficiently 
large values of~$\Lambda$, also at~$T>0$ and/or~$\mu>0$.
From Eqs.~\eqref{eq:44} and~\eqref{eq:qmdqgl}, however, we deduce  that
$\Gamma_{\Lambda^{\prime}}$ depends on $T$ and $\mu$ and is no longer
only quadratic in the fields for $T>0$ and/or $\mu>0$. This implies
that RG consistency is in general violated in conventional QCD
low-energy model studies with fixed~$\Lambda^{\prime}=\Lambda$ 
since these modifications of
$\Gamma_{\Lambda^{\prime}}$ at~$T>0$ and/or~$\mu>0$ are not taken into account. There, the quadratic
form~\eqref{eq:glpqmdqm} is rather left unchanged for any value of the
external parameters.

For convenience, we shall restrict ourselves to the 3$d$
sharp regulator in our numerical calculations below. Then, the auxiliary
functions parametrizing the loop integrals in Eqs.~\eqref{eq:44}
and~\eqref{eq:qmdqgl}  read
\begin{align}
& L_{k}^{(T)}(\Lambda,\chi) \nn\\[2ex]
& \;= \frac{1}{2}\int\frac{{\rm d}^3p}{(2\pi)^3}\sum_{\sigma=\pm 1}\left\{ \left(\omega^{(\sigma)}_{\varphi}
 + 2T\ln \left( 1+{\rm e}^{-\beta\omega^{(\sigma)}_{\varphi}} \right) \right)\Big|_{k}\right.\nn\\[2ex]
& \qquad\qquad\qquad \left. -\left(\omega^{(\sigma)}_{\varphi} + 2T\ln \left( 1+{\rm e}^{
-\beta\omega^{(\sigma)}_{\varphi}}\right) \right)\Big|_{\Lambda}
\right\}
\end{align}
with
\begin{align}
\omega^{(\sigma)}_{\varphi}=\sqrt{\vec{p}^{\,2}(1+r_{\psi})^2 + \chi} +\sigma\mu
\end{align}
and
\begin{align}
& M_{k}^{(T)}(\Lambda,\chi,\xi) \nn\\[2ex]
& \;= \frac{1}{2}\int\frac{{\rm d}^3p}{(2\pi)^3}\sum_{\sigma=\pm 1}
\left\{ \left(\omega^{(\sigma)}_{\Delta} + 2T\ln \left( 1+{\rm e}^{
-\beta\omega^{(\sigma)}_{\Delta}} \right) \right)\Big|_{k}\right.\nn\\[2ex]
& \qquad\qquad\qquad \left. -\left(\omega^{(\sigma)}_{\Delta} + 2T\ln \left( 1+{\rm e}^{-\beta\omega^{(\sigma)}_{\Delta}}\right) \right)\Big|_{\Lambda}
\right\}
\end{align}
with 
\begin{align}
\omega^{(\sigma)}_{\Delta}=\sqrt{\left(\sqrt{\vec{p}^{\,2}(1+r_{\psi})^2 + \chi} +\sigma\mu\right)^2+ \xi}\,.
\end{align}
For $\xi=0$, we use
$\omega^{(\sigma)}_{\Delta}=\omega^{(\sigma)}_{\varphi}$ to preserve
the Silver-Blaze property along the axis associated with
$\bar{\Delta}=0$.  For the 3$d$ sharp regulator function,
for example, these functions are given by 
\begin{align}
& L_{k}^{(T)}(\Lambda,\chi)\nn\\[2ex] 
& \qquad =\frac{1}{2}\int\frac{{\rm d}^3p}{(2\pi)^3}
\theta(\Lambda^2\!-\! \vec{p}^{\,2})\theta(\vec{p}^{\,2}\!-\!k^2)
\sum_{\sigma=\pm 1} \left(\omega^{(\sigma)}_{\varphi} \right. \nn\\[2ex] 
& \qquad\qquad\qquad
\left. + 2T\ln \left( 1+{\rm e}^{-\beta\omega^{(\sigma)}_{\varphi}} \right) \right)\Big|_{(1+r_{\psi})\to 1}
\end{align}
and
\begin{align}
& M_{k}^{(T)}(\Lambda,\chi,\xi)\nn\\
& \qquad =\frac{1}{2}\int\frac{{\rm d}^3p}{(2\pi)^3}
\theta(\Lambda^2\!-\! \vec{p}^{\,2})\theta(\vec{p}^{\,2}\!-\!k^2)
\sum_{\sigma=\pm 1} \left(\omega^{(\sigma)}_{\Delta} \right. \nn\\
& \qquad\qquad\qquad
\left. + 2T\ln \left( 1+{\rm e}^{-\beta\omega^{(\sigma)}_{\Delta}} \right) \right)\Big|_{(1+r_{\psi})\to 1}\,.
\end{align}
Corresponding
  expressions for~$L_k^{(T)}$ for the 3$d$ Litim regulator can be
  found in Ref.~\cite{Braun:2011pp}. 
We note that the effective action~\eqref{eq:44} is identical to the effective 
action~\eqref{eq:effacdqrgc} for $\bar{\varphi}=0$ in the limit $T\to 0$. 

In Fig.~\ref{fig:delG}, we show
$(\Lambda\partial_{\Lambda}\Gamma) / \Gamma$,
i.e. the change of the effective action under variation of the scale
$\Lambda>\Lambda^{\prime}$ relative to $\Gamma$ itself,  at $\bar{\Delta}=0$ as a function
of $\bar{\varphi}$ for various different values of
$\Lambda/m_{\rm q}$.  We observe that $\Gamma$ exhibits 
a strong dependence on our choice for $\Lambda$ in the
phenomenologically most relevant regime
$\bar{\varphi} \lesssim f_{\pi}$. In particular, this is true close to
the critical temperature at $\mu=0$, see left panel of
Fig.~\ref{fig:delG}, where cutoff artefacts are still clearly present
in the effective action even for already seemingly large values of
$\Lambda>\Lambda^{\prime}$. At low temperature but large quark
chemical potential $\mu \gtrsim m_{\rm q}$, see right panel of
Fig.~\ref{fig:delG}, cutoff contaminations of the effective action are
also present but appear to be less strong compared to the case with
$\mu=0$. However, this is misleading as the minimum of the effective
action is pushed away from the axis with $\bar{\Delta}=0$ in this
regime. There, the dynamics is no longer governed by the pions and the
$\sigma$-meson but rather by the diquark degrees of freedom.  Indeed,
close to the physical minimum of the effective action in this regime,
cutoff effects even appear to be stronger as in the case with $\mu=0$.
This can be inferred from the phase diagram in the $(T,\mu)$ plane as
well as from the pressure at zero temperature. We emphasize
  that the value of $\Lambda$ associated with effectively converged
  results depends on the temperature, the quark chemical potential,
  and the employed regularization scheme. Note that the effective
actions associated with different values of $\Lambda>\Lambda^{\prime}$
agree identically in the vacuum limit, i.e. we have
$\Lambda\partial_{\Lambda}\Gamma=0$ in this~limit.
\begin{figure}[t]
  \includegraphics[width=1\linewidth]{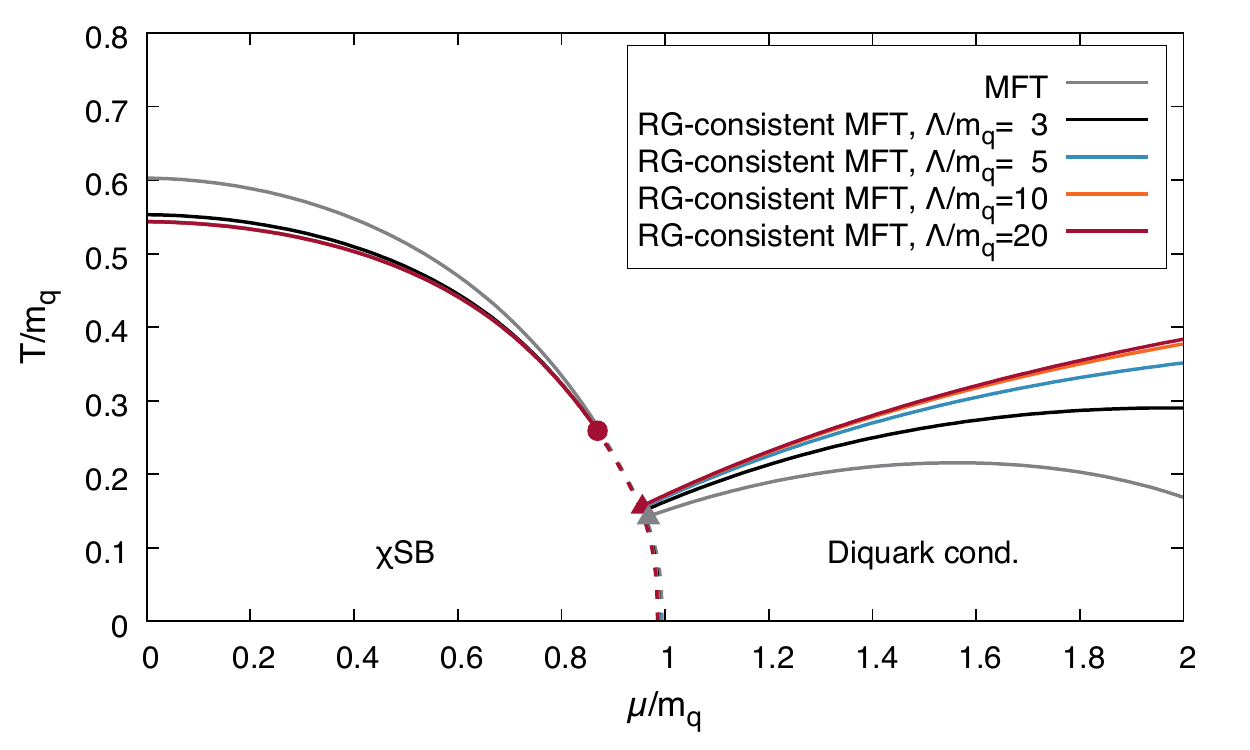}
\caption{Phase diagram of the quark-meson-diquark model in the plane spanned by the dimensionless temperature $T/m_{\rm q}$ 
and the dimensionless chemical potential $\mu/m_{\rm q}$ for various different values of $\Lambda/m_{\rm q}$ 
(with $m_{\rm q}\approx 0.300\,\text{GeV}$). Solid lines are associated with second-order phase transitions
whereas dashed lines are associated with first-order phase transitions.
Note that the effective actions obtained with different values of $\Lambda$ agree 
identically in the vacuum limit, i.e. the RG-consistency condition~\eqref{eq:RGconsistency} is strictly satisfied in 
this~limit.
}
\label{fig:pd}
\end{figure}

In Fig.~\ref{fig:pd}, we present the results for the $(T,\mu)$~phase diagram of our 
quark-meson-diquark model. Qualitatively, the structure of the phase diagram is determined by the emergence of 
three different phases: a phase governed by spontaneous chiral symmetry breaking at low temperature and 
small quark chemical potential, a phase governed by spontaneous $U_{\rm V}(1)$-symmetry breaking 
as associated with diquark condensation at low temperature and 
large chemical potential, and a symmetric high-temperature phase. Moreover, for our parameter choice, we observe the existence 
of a critical endpoint (depicted by the dot in Fig.~\ref{fig:pd}), at which the line of chiral second-order phase transitions meets a line of chiral first-order phase 
transitions, as well as a triple point (depicted by the triangle in Fig.~\ref{fig:pd}), at which the phase governed by 
chiral symmetry breaking meets the diquark phase and the symmetric high-temperature phase.
The general structure of the phase diagram suggests that a description of the dynamics in terms of only quarks, pions, 
and $\sigma$ mesons is insufficient for $T/\mu \lesssim 0.2$ and $\mu/m_{\rm q}\gtrsim 1$. Below this line, 
diquark degrees of freedom become relevant, as well-known from previous mean-field 
studies~\cite{Rajagopal:2000wf,*Buballa:2003qv,*Shovkovy:2004me,*Alford:2007xm}. 
Note that these general statements 
on the structure of the phase diagram are also 
in accordance with a recent Fierz-complete NJL model study beyond the mean-field limit~\cite{Braun:2018bik}.
Of course, in addition to the issue of an RG-consistent treatment of cutoff artefacts as discussed in our present work,
artefacts from specific truncations of the effective action may become relevant in the dense and/or
low-temperature regime, see, e.g., Ref.~\cite{Herbst:2013ail,*Fu:2016tey,*Tripolt:2017zgc}.

\begin{figure}[t]
  \includegraphics[width=1\linewidth]{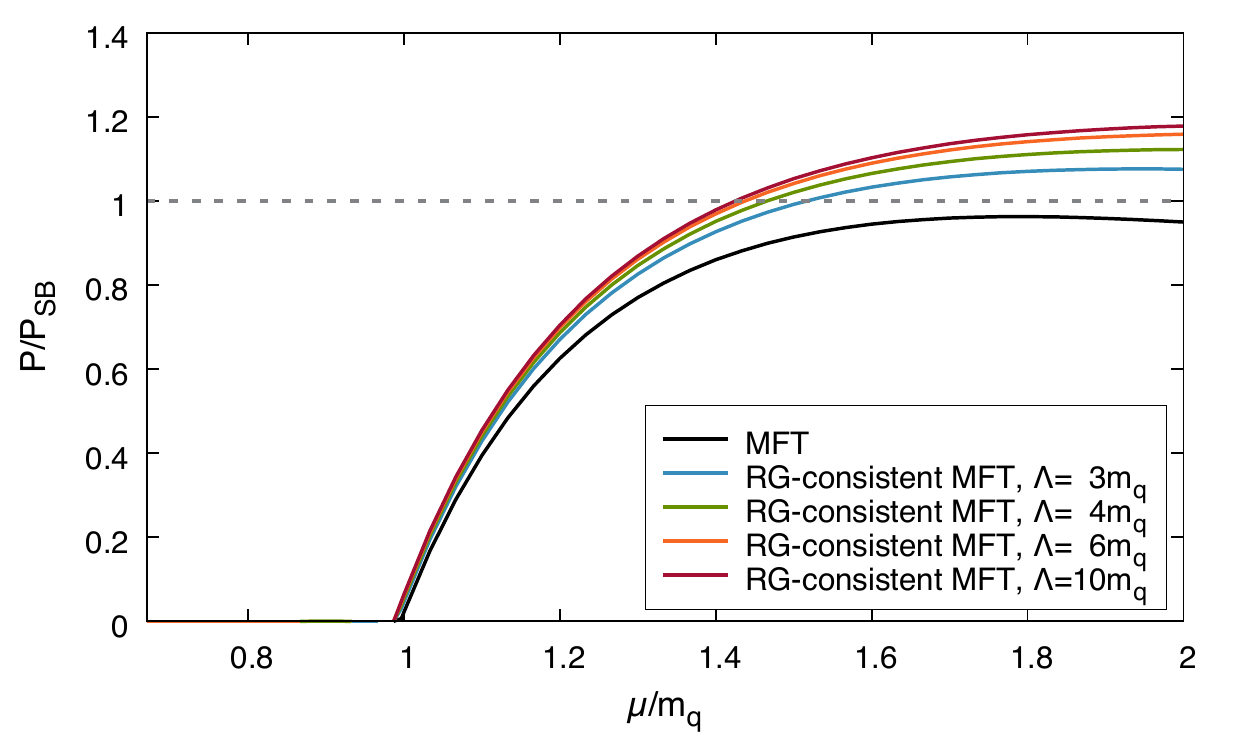}
\caption{Pressure $P/P_{\text{SB}}$ of our quark-meson-diquark model as a function 
of the chemical potential $\mu/m_{\rm q}$ (with $m_{\rm q}\approx 0.300\,\text{GeV}$) 
as obtained 
from conventional MFT associated with $\Lambda/m_{\rm q}\equiv\Lambda^{\prime}/m_{\rm q}=2$ (black line) 
as well as from RG-consistent MFT with $\Lambda/m_{\rm q}=3, 4, 6, 10$.
}
\label{fig:pressqmdq}
\end{figure}

The general structure of the phase diagram appears to be insensitive
with respect to an increase of the cutoff scale $\Lambda$, at least
for the values of the model parameters used in our numerical studies.
However, the positions of the two second-order phase transition lines
exhibit a strong dependence on $\Lambda$, meaning that they converge
only slowly when $\Lambda$ is increased, in particular at large
chemical potential, see Fig.~\ref{fig:pd}. To be more specific,
despite the fact that we employed a 3$d$ regulator,
the critical temperature at $\mu=0$ is lowered by about 10\% compared
to the conventional mean-field study when we take into account cutoff
corrections enforced by the RG-consistency condition~\eqref{eq:RGconsistency}.
In the regime governed by diquark dynamics, we observe that the
critical temperature is not decreased but rather increased
significantly when cutoff corrections are taken into account. 
Compared to the conventional mean-field study 
(associated with $\Lambda=\Lambda^{\prime}$), we indeed find a change of
about~30\% at $\mu/m_{\rm q}\approx 4/3$ 
  and about~100\%
at $\mu/m_{\rm q}\approx 2$.  
The strength of cutoff
artefacts in the high-density regime also becomes apparent in other
observables, such as the pressure of the system at zero temperature as
a function of the quark chemical potential, see
Fig.~\ref{fig:pressqmdq}.  Here, we find that the pressure now exceeds
the pressure $P_{\rm SB}$ of the free quark gas once cutoff artefacts
have been removed. Increasing the quark chemical potential further, we
eventually observe that the pressure approaches the pressure of the
free gas from above, as also observed for the pure diquark model, see
Fig.~\ref{fig:pressdq}.  Clearly, it appears crucial to enforce RG
consistency in the high-density regime. Note that our observations may
even be very relevant from a phenomenological point of view since the
associated corrections may significantly alter the equation of state
of dense strong-interaction matter as relevant for astrophysical
applications~\cite{BDHLPS}.
\subsection{Quarks and mesons -- finite-volume effects} 
As a fourth and final example, we demonstrate that our general line of
arguments detailed in Sec.~\ref{sec:RGconsistency} can also be applied
straightforwardly to studies beyond the mean-field approximation as
well as to studies with an external parameter other than the
temperature or the quark chemical potential. To this end, we employ again a variant of
the quark-meson model with two quark flavors and $N_{\rm c}=3$ colors
but we now take fluctuation effects into account to analyze the effect
of a finite cubic periodic box on the dynamics of this
model. To be specific, the classical action underlying our studies may
be viewed as an extension of the action~\eqref{Eq:HSTAction} and
reads
\begin{align}
&& S= \int d^4 x\,\Big\{ {\bar{q}}\Big( \partial\!\!\!\slash 
 \!+\! \frac{1}{2}{\bar{h}}(\sigma\!+\! {\rm i}\vec{\tau}\cdot\vec{\pi}\gamma_5)\Big)q\nn\\
 && + \frac{1}{2}(\partial_{\mu}\phi)^2+ U(\phi^2) - \bar{c}\sigma \Big\}\,.
\label{eq:qmmodel}
\end{align}
Compared to our previous studies, we allow for a term linear in
the $\sigma$-field. The latter breaks explicitly the chiral
symmetry. The associated parameter $\bar{c}$ is related to the quark
mass through a combination of the couplings of this model, see below. 
The inclusion of an explicit quark mass is now essential  
as we aim at a study of the effect of a finite cubic periodic box on
the dynamics of the model~\cite{Braun:2004yk,*Braun:2005fj}.

In the following we shall compute the effective action in the local potential approximation
where a possible space dependence of the expectation value of the scalar
fields is not taken into account and the wave-function renormalizations of the fields
are considered to be constant. Moreover, as also done in the studies presented in the previous subsections, 
we shall assume that the Yukawa coupling $\bar{h}$
does not depend on the RG scale $k$, i.e. $\bar{h}_k\equiv \bar{h}_{\Lambda^{\prime}}=\bar{h}$, with $\Lambda^{\prime}$ 
being the scale at which we fix the couplings of the model by means of 
an ansatz for the effective action $\Gamma_{\Lambda^{\prime}}$.
Still, we include effects beyond the mean-field approximation even 
within such a setting,
see, e.g., Refs.~\cite{Braun:2009si,Braun:2011pp} 
for a detailed discussion of the relation of the local potential approximation and the mean-field approximation. 
In any case, we only use this setting here to demonstrate how 
RG consistency can be ensured in approximations
which are more involved than the mean-field approximation. Of course, the line 
of arguments detailed in Sec.~\ref{sec:RGconsistency} is very general anyhow
and therefore does not depend on the underlying approximations by any means. 

A differential equation for the scale-dependent effective action $\Gamma_k$ can be derived
with the aid of the flow equation~\eqref{eq:wetterich}. Expanding the 
fields about a homogeneous background~$\bar{\varphi}$ and using the 3$d$ Litim regulator 
for both the quark and meson fields~\cite{Litim:2006ag,Blaizot:2006rj},  
we obtain~\cite{Braun:2010vd,Braun:2011iz,Tripolt:2013zfa}:
\be
k\partial_k U_k(\bar{\varphi}^2)= k\partial_k\left(\lim_{T\to 0}\frac{T}{V}\Gamma_k[\bar{\varphi}]\right) 
\ee
with $V=L^3$ and
\be
 k\partial_k U_k(\bar{\varphi}^2)
     &=& \frac{k^4}{2}
     \left( -\frac{24}{\sqrt{k^2 + \frac{1}{4}\bar{h}^2\bar{\varphi}^2}} + \frac{3}{\sqrt{k^2 + 2U^{\prime}_k}} \right.\nn\\
 && \left. +\frac{1}{\sqrt{k^2 + 2 U^{\prime}_k + 4\bar{\varphi}^2U^{\prime\prime}_k }} \right){\mathcal B}(kL)\,.
   \label{eq:rgvfloweq}
\ee
Here, $\bar{\varphi}$ denotes the homogeneous background of the scalar fields
and the primes denote derivatives with respect to $\bar{\varphi}^2$.
Note that the parameter $\bar{c}$ measuring the explicit breaking of the chiral symmetry 
does not depend on the RG scale $k$. Thus, we have
\be
U_k(\bar{\varphi}^2) - \bar{c}\bar{\sigma} = \lim_{T\to 0}\frac{T}{V}\Gamma_k[\bar{\varphi}]
\ee
in our present approximation 
where $\bar{\sigma}$ denotes the zeroth component of the field vector $\bar{\varphi}$. 

The first term on the right-hand side of the flow equation~\eqref{eq:rgvfloweq} is associated 
with the quark degrees of freedom. 
The second and the third term represent
contributions from the mesonic modes. By dropping the latter two contributions
to the RG flow of the effective action, we simply recover the mean-field effective action as already discussed above
for the 3$d$ Litim regulator in the infinite-volume limit.
The explicit dependence of the RG flow on the finite periodic cubic volume $V=L^3$ is encoded in the
momentum-modes counting function ${\mathcal B}$: 
\be 
  {\mathcal B}(kL)=\frac{1}{(kL)^3} \sum_{\vec{n} \in \mathbb{Z}^3} 
  \theta\!\left( (kL)^2    
    -(2\pi \vec{n})^2\right)\,,
    \label{eq:mmcount}
\ee
where $\vec n$ labels a three-dimensional vector of integers. In the
limit $L\to \infty$, we have ${\mathcal B}(kL)\to1/(6\pi^2)$. Thus,
the flow equation~\eqref{eq:rgvfloweq} agrees identically with the
known flow equation for the scale-dependent effective action in the
local potential approximation in the infinite-volume limit~\cite{Braun:2003ii,Schaefer:2004en}, 
as it should be. We note that, for finite~$L$, 
the right-hand side of the flow equation~\eqref{eq:rgvfloweq} is
discontinuous for the Litim regulator.  However, this does not
cause any conceptional problem. In fact, the resulting effective
action is still continuous as a function of $k$.  For more detailed
discussions of the properties of RG flows of finite systems with this
regulator, we refer the reader to
Refs.~\cite{Braun:2010vd,Braun:2011iz,Tripolt:2013zfa,Fister:2015eca,Klein:2017shl}.

In order to compute the scale-dependent effective action $\Gamma_k$ in
our present illustrational study, we shall now parametrize $U_k$ as
follows:
\be
U_k(\bar{\varphi}^2) &=&  \frac{1}{2}{\bar{m}^2_k}\left(\bar{\varphi}^2\!-\! \bar{\varphi}_{0,k}^2\right) 
 +\frac{1}{4}{\bar{\lambda}_k}\left(\bar{\varphi}^2\!-\! \bar{\varphi}_{0,k}^2\right)^2.
    \label{eq:potrg}
\ee
The condition
\be 
  \frac{\partial}{\partial
  \bar{\sigma}}\left(\lim_{T\to 0}\frac{T}{V}\Gamma_k[\bar{\varphi}]\right)
\Bigg|_{\bar{\varphi}=\bar{\varphi}_{0,k}} \stackrel{!}{=} 0
      \label{eq:minc}
\ee 
then ensures that the effective action is always expanded about the actual 
physical ground state by relating
the couplings $\bar{m}_k^2$ and $\bar{\varphi}_{0,k}$~\cite{Braun:2008sg}:
\be
  \bar{m}_k^2 \bar{\varphi}_{0,k} = \bar{c}\,. 
\ee
Thus, the RG flows of the two scale-dependent
couplings $\bar{\varphi}_{0,k}$ and $\bar{\lambda}_k$ parametrize the
RG flow of the effective action in our present approximation.  The 
flow equations for $\bar{\varphi}_{0,k}$ and $\bar{\lambda}_k$ can be
obtained by expanding the flow equation~\eqref{eq:rgvfloweq}
about $\bar{\varphi}_{0,k}$ and then projecting it onto the
logarithmic scale-derivative of the ansatz~\eqref{eq:potrg}. The
resulting flow equations can then be solved by specifying the values
of the two couplings at some scale $k=\Lambda^{\prime}$. For example,
the parameters $\bar{\varphi}_{0,\Lambda^{\prime}}$
and $\bar{\lambda}_{\Lambda^{\prime}}$ may be chosen such that the
physical values of a given set of low-energy observables are recovered
in the long-range limit $k\to 0$.

Let us now discuss how RG consistency can be ensured in the present
setting.  Following our general discussion in
Sec.~\ref{sec:RGconsistency}, this requires a suitable adaption of the
effective action $\Gamma_{\Lambda}$ at the
scale $k=\Lambda>\Lambda^{\prime}$.  In the infinite-volume limit, the
effective action $\Gamma_{\Lambda}$ at the scale $k=\Lambda$ is
obtained from the given effective action $\Gamma_{\Lambda^{\prime}}$
at the scale $\Lambda^{\prime}$ by solving the flow
equation~\eqref{eq:rgvfloweq} from $k=\Lambda^{\prime}$ up to the
scale $k=\Lambda>\Lambda^{\prime}$. This ensures RG consistency, 
see also Eq.~\eqref{eq:gammainiL} and the related
discussion.  In the present case, to be specific, this corresponds to
solving the set of coupled flow equations for $\bar{\varphi}_{0,k}$
and $\bar{\lambda}_k$ from $k=\Lambda^{\prime}$ up to the
scale $k=\Lambda>\Lambda^{\prime}$ with given values for the two
couplings at the scale $k=\Lambda^{\prime}$.  The effective
action $\Gamma=\Gamma_{k\to 0}$ then does not depend on our choice
for $\Lambda$, $\Lambda \partial_{\Lambda}\Gamma \to 0$, implying that
the values for the low-energy observables, such as the constituent
quark mass $m_{\rm q}$, the pion decay constant $f_{\pi}$, and the
pion mass $m_{\pi}$ do not depend on our actual choice
for $\Lambda$.

Of course, we also would like to ensure that the RG-consistency
condition is satisfied in our study of finite-volume effects.  To
discuss this issue further, we shall first assume that the parameters
of our model (i.e. the values of $\bar{\varphi}_{0,k}$
and $\bar{\lambda}_k$ at the scale $k=\Lambda^{\prime}$ as well
as $\bar{c}$) have been fixed in the infinite-volume limit as
discussed above.  RG consistency for $1/L>0$ can now be ensured by
fixing the two scale-dependent couplings $\bar{\varphi}_{0,k}$
and $\bar{\lambda}_k$ at a scale $\Lambda>\Lambda^{\prime}$ in such a
way that RG consistency is still ensured in the infinite-volume limit.
If $\Lambda$ has been chosen sufficiently large, i.e. $1/(\Lambda L)$ is
sufficiently small, then the effective action $\Gamma$ and therefore
also the physical observables become independent of $\Lambda$,
i.e. $\Lambda\partial_{\Lambda}\Gamma \to 0$, even for $1/L>0$.  Note
that the value for $\Lambda$ effectively ensuring RG consistency for a
given box size depends on the regularization scheme.
\begin{figure}[t]
  \includegraphics[width=1\linewidth]{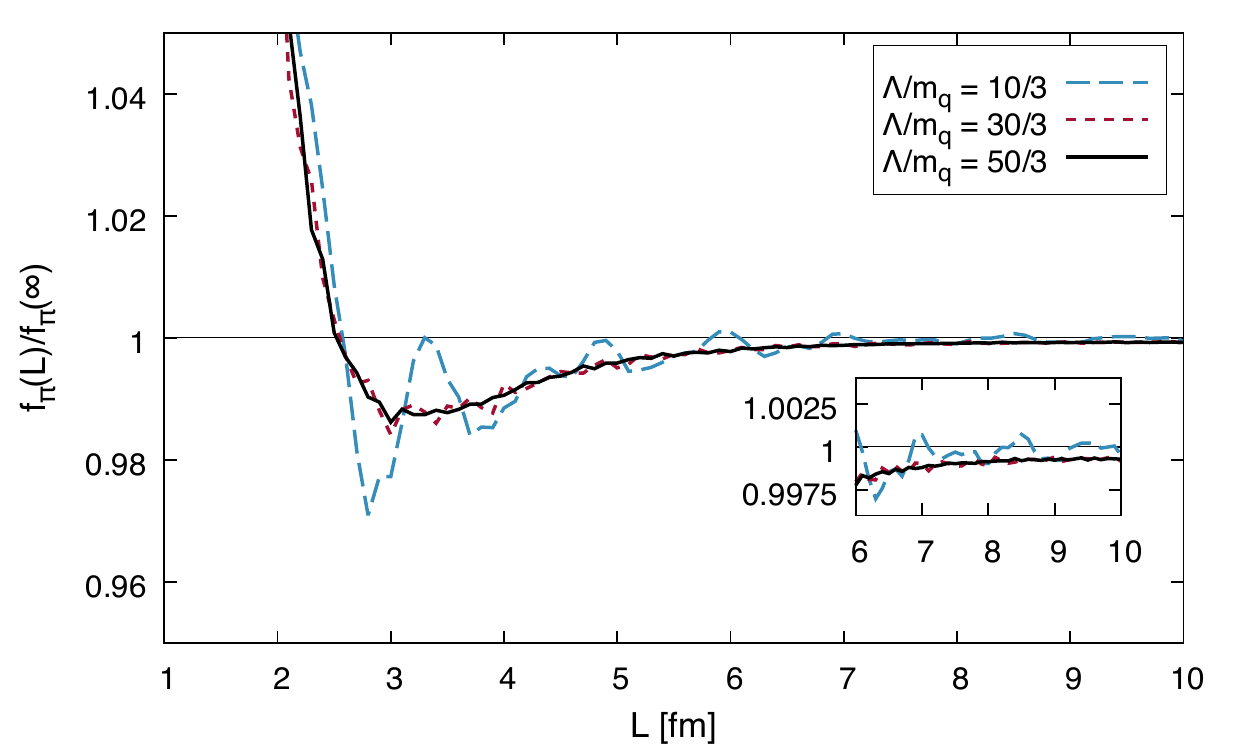}
\caption{Normalized pion decay constant $f_{\pi}(L)/f_{\pi}(\infty)$ as a function of the box size $L$  
computed with the 3$d$ Litim cutoff for $\Lambda/m_{\rm q}\approx 10/3,30/3,50/3$,
where $m_{\rm q}\equiv m_{\rm q}(\infty) \approx 0.300\,\text{GeV}$ and $f_{\pi}\equiv f_{\pi}(\infty)\approx 0.092\,\text{GeV}$.
For increasing $\Lambda L$, the pion decay constant $f_{\pi}(L)/f_{\pi}(\infty)$ is continuously ``smoothened". 
In any case, deviations of the results for $\Lambda/m_{\rm q}=10/3$ from those for $\Lambda/m_{\rm q}=50/3$ 
are found to be on the $1\%$ level at most.
Note that the effective actions associated with the different values of $\Lambda$ 
agree identically in the infinite-volume limit by construction, i.e. the RG-consistency condition~\eqref{eq:RGconsistency}
is exactly satisfied in this limit.
}
\label{fig:es}
\end{figure}

With an RG-consistent effective action at hand, we may now compute physical observables, such as 
the pressure $P$ of the system,
\be
P=-\frac{\partial}{\partial V}\left(\lim_{T\to 0}T\,\Gamma[\bar{\varphi}_0]\right)\,,
\ee
as we have done it in the previous subsections.\footnote{Note that the derivative of the effective action with respect 
to the volume is trivial in case of infinite-volume studies, see, e.g., Eq.~\eqref{eq:pressdq}. In the presence of a finite volume, the computation 
is more involved as the volume dependence of the couplings and $\bar{\varphi}_0$ has to be taken into account as well.}  
However, since we only aim at an illustration 
of how RG consistency is ensured in a study beyond the mean-field approximation, we shall only discuss 
the volume dependence of the simplest physical observable that can be extracted from the effective 
action in our present setting, namely the position of its minimum in the long-range limit, i.e. $f_{\pi}\equiv |\bar{\varphi}_0|=\lim_{k\to 0}\bar{\varphi}_{0,k}$.
To this end, we need to fix the initial conditions for the couplings $\bar{\varphi}_{0,k}$ and $\bar{\lambda}_k$ 
at some scale $\Lambda^{\prime}$ (as well as the parameter $\bar{c}$)
which corresponds to fixing the effective action at this scale. Here, we choose the parameters such that
the physical values of a given set of low-energy observables are recovered 
from our ansatz for the effective action $\Gamma$ in the infinite-volume limit.
To be specific, the parameters are determined such that we 
have $m_{\rm q}=\frac{1}{2}\bar{h}\bar{\varphi}_{0}\approx 0.300\,\text{GeV}$, 
$f_{\pi}=2 m_{\rm q}/\bar{h}\approx 0.092\,\text{GeV}$, and $m_{\pi}\approx 0.138\,\text{GeV}$,
in accordance with chiral perturbation theory~\cite{Colangelo:2003hf}. These values are obtained 
by choosing $\bar{\varphi}_{0,\Lambda^{\prime}}/m_{\rm q}\approx 1.08\cdot 10^{-2}$,
$\bar{\lambda}_{\Lambda^{\prime}}\approx 50.8$, $\bar{h}\approx 6.52$, $\bar{c}/m_{\rm q}^3\approx 6.48\cdot 10^{-2}$
at the scale $\Lambda^{\prime} /m_{\rm q}\approx 10/3$.  

In Fig.~\ref{fig:es}, we show our results for the pion decay constant
as a function of $L$ as obtained from calculations with
$\Lambda/m_{\rm q}\equiv \Lambda^{\prime}/m_{\rm q}\approx 10/3$ as
well as for $\Lambda/m_{\rm q}\approx 30/3$ and
$\Lambda/m_{\rm q}\approx 50/3$.  For $\Lambda=\Lambda^{\prime}$, the
pion decay constant seems to exhibit a pathological behavior when the
volume is decreased. For increasing box size, however, the dependence
of $f_{\pi}$ on $L$ is continuously ``smoothened" and $f_{\pi}$ eventually approaches its
value in the infinite-volume limit. This overall behavior of the pion decay
constant does not come unexpected and can be traced back to
the behavior of the momentum-modes counting
function~\eqref{eq:mmcount}. The non-analytic form of the latter
originates from the use of a non-analytic regulator function in our
study. With such regulators, the presence of a momentum cutoff becomes
very evident.  Indeed, we observe that the seemingly pathological behavior
is continuously ``smoothened" when $\Lambda$ is increased in an
RG-consistent manner.  The latter corresponds to increasing the
dimensionless quantity $\Lambda L$ and therefore also explains why
this behavior of $f_{\pi}$ observed for, e.g., $\Lambda/m_{\rm q}\approx 10/3$ and
small box sizes goes away when the box size is increased. In
particular, we observe that the results for the pion decay constant as
a function of $L$ converge when $\Lambda$ is increased. In any case,
we find that the deviations of the results for
$\Lambda/m_{\rm q}\approx 10/3$ from those for
$\Lambda/m_{\rm q}\approx 50/3$ are on the $1\%$ level at most for the
range of box sizes shown in Fig.~\ref{fig:es}. 
For a discussion of the general behavior of the pion decay constant as a
function of the box size, we refer the reader to
Ref.~\cite{Braun:2005gy} where also qualitative comparisons to
lattice QCD calculations~\cite{Guagnelli:2004ww,*Orth:2005kq} of
related quantities can be found.

Note that in practice it may be advantageous -- although not necessarily required --
  to employ analytic regulators which lead to an exponential
  suppression of cutoff effects when $\Lambda L$ is increased rather
  than a polynomial suppression as in the case of non-analytic
  regulators. Here, the strength of the exponential suppression is
  related to the first pole or cut in the complex plane introduced by
  the regulator. It is not necessarily the typical thermal damping in
  the presence of a mass. The latter decay is indeed only seen for 3$d$
  regulators, see Ref.~\cite{Fister:2015eca} for a detailed discussion. In this context, we emphasize again that the
  meaning of the actual value of the cutoff~$\Lambda$ is ambiguous
  without referring to the employed regularization scheme, see our
  discussion in Sec.~\ref{sec:RGconsistency}. Indeed, the values of the momentum cutoff~$\Lambda$ associated
  with analytic regulators used in the past to study fluctuation effects in finite
volumes~\cite{Braun:2004yk,*Braun:2005fj,Braun:2007td,*Klein:2010tk,Braun:2005gy,Fister:2015eca,Almasi:2016zqf} 
 effectively correspond to larger values of~$\Lambda$ in case of typically
 employed non-analytic regulators, such as the sharp or Litim regulator.

We would like to close our discussion with a word of caution regarding
the construction of UV completions of low-energy models. In
contradistinction to the mean-field studies in
Subsecs.~\ref{subsec:njlvac}-\ref{subsec:qmdqpd}, the UV cutoff scale
$\Lambda$ cannot be pushed to arbitrarily large values in the present
case where we have taken into account fluctuation effects. This can be
traced back to the fact that the mesonic fluctuations induce a
Landau-pole instability at large momentum scales, irrespective of the
employed regulator.  For the simple quartic approximation of the
effective action considered here, this instability occurs at
comparatively large scales $\Lambda > 5\,\text{GeV}$ (for the
3$d$ Litim regulator). However, the position of this
instability is shifted to smaller momentum scales when corrections of
higher order are included.  It may then not be possible anymore to
construct an RG-consistent effective action for the range of
parameters of interest. In such a situation, a suitable workaround to
preserve RG consistency at least approximately may be obtained by
simply dropping the mesonic fluctuations in the construction of the UV
completion, i.e. by only employing the corresponding mean-field UV
completion simply constructed from the fully field-dependent quark
loop as detailed in
Subsecs.~\ref{subsec:njlvac}-\ref{subsec:qmdqpd}. In practice, this may already
suffice to reduce cutoff artefacts to a large extent in studies beyond
the mean-field approximation. 

\section{Conclusions}\label{sec:conc}

In the present work we have discussed the concept of RG consistency in
the context of model studies, with an emphasis on studies in the
presence of external parameters.  In general, RG consistency requires
that the effective action $\Gamma$ of a given theory does not depend
on the cutoff scale $\Lambda$,
i.e. $\Lambda \partial_{\Lambda}\Gamma=0$,
also in the presence of external parameters.  After a detailed general
discussion of RG consistency in Sec.~\ref{sec:RGconsistency}, we have
given an illustrative discussion of RG consistency in mean-field studies
of a quark-meson model in the vacuum
limit, a diquark model at finite density, and a
quark-meson-diquark model at finite temperature
and density. We note that in the latter two cases, we had to take
into account the renormalization of the diquark chemical potential to
ensure RG consistency.  Moreover, we discussed RG consistency in
studies of finite-volume effects by considering the quark-meson model
beyond the mean-field approximation.

For regularization schemes and values of the cutoff scale $\Lambda$ as
widely employed in mean-field studies of QCD models, our
illustrational studies already suggest that ``cutoff contaminations"
of physical observables can be significant. Indeed, for the
zero-temperature pressure of our quark-meson-diquark model, we found
corrections of up to 30\% in the considered range for the quark
chemical potential. However, it is not only the computation of the
pressure that suffers from ``cutoff contaminations". For example, the
critical temperature of our quark-meson-diquark model at $\mu=0$ is
lowered by about 10\% when we take into account cutoff corrections
enforced by the RG-consistency condition~\eqref{eq:RGconsistency}.  In
general, such corrections do not necessarily only lead to a decrease
of the critical temperature. In fact, in the regime governed by
diquark condensation, cutoff corrections rather tend to increase
it. To be specific, the critical temperature is increased by
about~30\% at $\mu/m_{\rm q}= 4/3$ (with
$m_{\rm q}\approx 0.300\,\text{GeV}$) and already by more than~100\%
at $\mu/m_{\rm q}=2$ compared to the results from a conventional
mean-field study.  Thus, the implementation of RG consistency appears
to be very relevant in the high-density regime of our QCD models. 
For example, the associated corrections may significantly alter the
presently available equations of state of dense strong-interaction
matter as relevant for astrophysical applications~\cite{BDHLPS}. In any
case, our illustrational studies show clearly that it is of
phenomenological relevance to ensure RG consistency in general model
studies, even in the mean-field approximation.

{\it Acknowledgments.--~} The authors as members of the fQCD collaboration~\cite{fQCD} 
would like to thank the other members of this collaboration for discussions.
J.B. acknowledges support by HIC for FAIR within the LOEWE program of the State of Hesse. 
Moreover, this work is supported by the DFG through grant SFB~1245.


%
\bibliography{qcd}

\begin{thebibliography}{102}%
\makeatletter
\providecommand \@ifxundefined [1]{%
 \@ifx{#1\undefined}
}%
\providecommand \@ifnum [1]{%
 \ifnum #1\expandafter \@firstoftwo
 \else \expandafter \@secondoftwo
 \fi
}%
\providecommand \@ifx [1]{%
 \ifx #1\expandafter \@firstoftwo
 \else \expandafter \@secondoftwo
 \fi
}%
\providecommand \natexlab [1]{#1}%
\providecommand \enquote  [1]{``#1''}%
\providecommand \bibnamefont  [1]{#1}%
\providecommand \bibfnamefont [1]{#1}%
\providecommand \citenamefont [1]{#1}%
\providecommand \href@noop [0]{\@secondoftwo}%
\providecommand \href [0]{\begingroup \@sanitize@url \@href}%
\providecommand \@href[1]{\@@startlink{#1}\@@href}%
\providecommand \@@href[1]{\endgroup#1\@@endlink}%
\providecommand \@sanitize@url [0]{\catcode `\\12\catcode `\$12\catcode
  `\&12\catcode `\#12\catcode `\^12\catcode `\_12\catcode `\%12\relax}%
\providecommand \@@startlink[1]{}%
\providecommand \@@endlink[0]{}%
\providecommand \url  [0]{\begingroup\@sanitize@url \@url }%
\providecommand \@url [1]{\endgroup\@href {#1}{\urlprefix }}%
\providecommand \urlprefix  [0]{URL }%
\providecommand \Eprint [0]{\href }%
\providecommand \doibase [0]{http://dx.doi.org/}%
\providecommand \selectlanguage [0]{\@gobble}%
\providecommand \bibinfo  [0]{\@secondoftwo}%
\providecommand \bibfield  [0]{\@secondoftwo}%
\providecommand \translation [1]{[#1]}%
\providecommand \BibitemOpen [0]{}%
\providecommand \bibitemStop [0]{}%
\providecommand \bibitemNoStop [0]{.\EOS\space}%
\providecommand \EOS [0]{\spacefactor3000\relax}%
\providecommand \BibitemShut  [1]{\csname bibitem#1\endcsname}%
\let\auto@bib@innerbib\@empty
\bibitem [{\citenamefont {Hatsuda}\ and\ \citenamefont
  {Kunihiro}(1985)}]{Hatsuda:1985eb}%
  \BibitemOpen
  \bibfield  {author} {\bibinfo {author} {\bibfnamefont {T.}~\bibnamefont
  {Hatsuda}}\ and\ \bibinfo {author} {\bibfnamefont {T.}~\bibnamefont
  {Kunihiro}},\ }\href {\doibase 10.1103/PhysRevLett.55.158} {\bibfield
  {journal} {\bibinfo  {journal} {Phys. Rev. Lett.}\ }\textbf {\bibinfo
  {volume} {55}},\ \bibinfo {pages} {158} (\bibinfo {year} {1985})}\BibitemShut
  {NoStop}%
\bibitem [{\citenamefont {Asakawa}\ and\ \citenamefont
  {Yazaki}(1989)}]{Asakawa:1989bq}%
  \BibitemOpen
  \bibfield  {author} {\bibinfo {author} {\bibfnamefont {M.}~\bibnamefont
  {Asakawa}}\ and\ \bibinfo {author} {\bibfnamefont {K.}~\bibnamefont
  {Yazaki}},\ }\href {\doibase 10.1016/0375-9474(89)90002-X} {\bibfield
  {journal} {\bibinfo  {journal} {Nucl. Phys.}\ }\textbf {\bibinfo {volume}
  {A504}},\ \bibinfo {pages} {668} (\bibinfo {year} {1989})}\BibitemShut
  {NoStop}%
\bibitem [{\citenamefont {Klevansky}(1992)}]{Klevansky:1992qe}%
  \BibitemOpen
  \bibfield  {author} {\bibinfo {author} {\bibfnamefont {S.~P.}\ \bibnamefont
  {Klevansky}},\ }\href {\doibase 10.1103/RevModPhys.64.649} {\bibfield
  {journal} {\bibinfo  {journal} {Rev. Mod. Phys.}\ }\textbf {\bibinfo {volume}
  {64}},\ \bibinfo {pages} {649} (\bibinfo {year} {1992})}\BibitemShut
  {NoStop}%
\bibitem [{\citenamefont {Jungnickel}\ and\ \citenamefont
  {Wetterich}(1996)}]{Jungnickel:1995fp}%
  \BibitemOpen
  \bibfield  {author} {\bibinfo {author} {\bibfnamefont {D.~U.}\ \bibnamefont
  {Jungnickel}}\ and\ \bibinfo {author} {\bibfnamefont {C.}~\bibnamefont
  {Wetterich}},\ }\href@noop {} {\bibfield  {journal} {\bibinfo  {journal}
  {Phys. Rev.}\ }\textbf {\bibinfo {volume} {D53}},\ \bibinfo {pages} {5142}
  (\bibinfo {year} {1996})},\ \Eprint {http://arxiv.org/abs/hep-ph/9505267}
  {hep-ph/9505267} \BibitemShut {NoStop}%
\bibitem [{\citenamefont {Berges}\ \emph {et~al.}(1999)\citenamefont {Berges},
  \citenamefont {Jungnickel},\ and\ \citenamefont {Wetterich}}]{Berges:1997eu}%
  \BibitemOpen
  \bibfield  {author} {\bibinfo {author} {\bibfnamefont {J.}~\bibnamefont
  {Berges}}, \bibinfo {author} {\bibfnamefont {D.~U.}\ \bibnamefont
  {Jungnickel}}, \ and\ \bibinfo {author} {\bibfnamefont {C.}~\bibnamefont
  {Wetterich}},\ }\href@noop {} {\bibfield  {journal} {\bibinfo  {journal}
  {Phys. Rev.}\ }\textbf {\bibinfo {volume} {D59}},\ \bibinfo {pages} {034010}
  (\bibinfo {year} {1999})},\ \Eprint {http://arxiv.org/abs/hep-ph/9705474}
  {hep-ph/9705474} \BibitemShut {NoStop}%
\bibitem [{\citenamefont {Berges}\ \emph {et~al.}(2002)\citenamefont {Berges},
  \citenamefont {Tetradis},\ and\ \citenamefont {Wetterich}}]{Berges:2000ew}%
  \BibitemOpen
  \bibfield  {author} {\bibinfo {author} {\bibfnamefont {J.}~\bibnamefont
  {Berges}}, \bibinfo {author} {\bibfnamefont {N.}~\bibnamefont {Tetradis}}, \
  and\ \bibinfo {author} {\bibfnamefont {C.}~\bibnamefont {Wetterich}},\
  }\href@noop {} {\bibfield  {journal} {\bibinfo  {journal} {Phys. Rept.}\
  }\textbf {\bibinfo {volume} {363}},\ \bibinfo {pages} {223} (\bibinfo {year}
  {2002})},\ \Eprint {http://arxiv.org/abs/hep-ph/0005122} {hep-ph/0005122}
  \BibitemShut {NoStop}%
\bibitem [{\citenamefont {Fukushima}(2012)}]{Fukushima:2011jc}%
  \BibitemOpen
  \bibfield  {author} {\bibinfo {author} {\bibfnamefont {K.}~\bibnamefont
  {Fukushima}},\ }\href {\doibase 10.1088/0954-3899/39/1/013101} {\bibfield
  {journal} {\bibinfo  {journal} {J. Phys.}\ }\textbf {\bibinfo {volume}
  {G39}},\ \bibinfo {pages} {013101} (\bibinfo {year} {2012})},\ \Eprint
  {http://arxiv.org/abs/1108.2939} {arXiv:1108.2939 [hep-ph]} \BibitemShut
  {NoStop}%
\bibitem [{\citenamefont {Kamikado}\ \emph {et~al.}(2013)\citenamefont
  {Kamikado}, \citenamefont {Strodthoff}, \citenamefont {von Smekal},\ and\
  \citenamefont {Wambach}}]{Kamikado:2012bt}%
  \BibitemOpen
  \bibfield  {author} {\bibinfo {author} {\bibfnamefont {K.}~\bibnamefont
  {Kamikado}}, \bibinfo {author} {\bibfnamefont {N.}~\bibnamefont
  {Strodthoff}}, \bibinfo {author} {\bibfnamefont {L.}~\bibnamefont {von
  Smekal}}, \ and\ \bibinfo {author} {\bibfnamefont {J.}~\bibnamefont
  {Wambach}},\ }\href {\doibase 10.1016/j.physletb.2012.11.055} {\bibfield
  {journal} {\bibinfo  {journal} {Phys. Lett.}\ }\textbf {\bibinfo {volume}
  {B718}},\ \bibinfo {pages} {1044} (\bibinfo {year} {2013})},\ \Eprint
  {http://arxiv.org/abs/1207.0400} {arXiv:1207.0400 [hep-ph]} \BibitemShut
  {NoStop}%
\bibitem [{\citenamefont {Tripolt}\ \emph
  {et~al.}(2014{\natexlab{a}})\citenamefont {Tripolt}, \citenamefont
  {Strodthoff}, \citenamefont {von Smekal},\ and\ \citenamefont
  {Wambach}}]{Tripolt:2013jra}%
  \BibitemOpen
  \bibfield  {author} {\bibinfo {author} {\bibfnamefont {R.-A.}\ \bibnamefont
  {Tripolt}}, \bibinfo {author} {\bibfnamefont {N.}~\bibnamefont {Strodthoff}},
  \bibinfo {author} {\bibfnamefont {L.}~\bibnamefont {von Smekal}}, \ and\
  \bibinfo {author} {\bibfnamefont {J.}~\bibnamefont {Wambach}},\ }\href
  {\doibase 10.1103/PhysRevD.89.034010} {\bibfield  {journal} {\bibinfo
  {journal} {Phys. Rev.}\ }\textbf {\bibinfo {volume} {D89}},\ \bibinfo {pages}
  {034010} (\bibinfo {year} {2014}{\natexlab{a}})},\ \Eprint
  {http://arxiv.org/abs/1311.0630} {arXiv:1311.0630 [hep-ph]} \BibitemShut
  {NoStop}%
\bibitem [{\citenamefont {Andersen}\ \emph {et~al.}(2016)\citenamefont
  {Andersen}, \citenamefont {Naylor},\ and\ \citenamefont
  {Tranberg}}]{Andersen:2014xxa}%
  \BibitemOpen
  \bibfield  {author} {\bibinfo {author} {\bibfnamefont {J.~O.}\ \bibnamefont
  {Andersen}}, \bibinfo {author} {\bibfnamefont {W.~R.}\ \bibnamefont
  {Naylor}}, \ and\ \bibinfo {author} {\bibfnamefont {A.}~\bibnamefont
  {Tranberg}},\ }\href {\doibase 10.1103/RevModPhys.88.025001} {\bibfield
  {journal} {\bibinfo  {journal} {Rev. Mod. Phys.}\ }\textbf {\bibinfo {volume}
  {88}},\ \bibinfo {pages} {025001} (\bibinfo {year} {2016})},\ \Eprint
  {http://arxiv.org/abs/1411.7176} {arXiv:1411.7176 [hep-ph]} \BibitemShut
  {NoStop}%
\bibitem [{\citenamefont {Yokota}\ \emph {et~al.}(2017)\citenamefont {Yokota},
  \citenamefont {Kunihiro},\ and\ \citenamefont {Morita}}]{Yokota:2017uzu}%
  \BibitemOpen
  \bibfield  {author} {\bibinfo {author} {\bibfnamefont {T.}~\bibnamefont
  {Yokota}}, \bibinfo {author} {\bibfnamefont {T.}~\bibnamefont {Kunihiro}}, \
  and\ \bibinfo {author} {\bibfnamefont {K.}~\bibnamefont {Morita}},\ }\href
  {\doibase 10.1103/PhysRevD.96.074028} {\bibfield  {journal} {\bibinfo
  {journal} {Phys. Rev.}\ }\textbf {\bibinfo {volume} {D96}},\ \bibinfo {pages}
  {074028} (\bibinfo {year} {2017})},\ \Eprint
  {http://arxiv.org/abs/1707.05520} {arXiv:1707.05520 [hep-ph]} \BibitemShut
  {NoStop}%
\bibitem [{\citenamefont {von Smekal}(2012)}]{vonSmekal:2012vx}%
  \BibitemOpen
  \bibfield  {author} {\bibinfo {author} {\bibfnamefont {L.}~\bibnamefont {von
  Smekal}},\ }\href {\doibase 10.1016/j.nuclphysbps.2012.06.006} {\bibfield
  {journal} {\bibinfo  {journal} {Nucl. Phys. Proc. Suppl.}\ }\textbf {\bibinfo
  {volume} {228}},\ \bibinfo {pages} {179} (\bibinfo {year} {2012})},\ \Eprint
  {http://arxiv.org/abs/1205.4205} {arXiv:1205.4205 [hep-ph]} \BibitemShut
  {NoStop}%
\bibitem [{\citenamefont {Rajagopal}\ and\ \citenamefont
  {Wilczek}(2000)}]{Rajagopal:2000wf}%
  \BibitemOpen
  \bibfield  {author} {\bibinfo {author} {\bibfnamefont {K.}~\bibnamefont
  {Rajagopal}}\ and\ \bibinfo {author} {\bibfnamefont {F.}~\bibnamefont
  {Wilczek}},\ }in\ \href {\doibase 10.1142/9789812810458_0043} {\emph
  {\bibinfo {booktitle} {At the frontier of particle physics. Handbook of QCD.
  Vol. 1-3}}},\ \bibinfo {editor} {edited by\ \bibinfo {editor} {\bibfnamefont
  {M.}~\bibnamefont {Shifman}}\ and\ \bibinfo {editor} {\bibfnamefont
  {B.}~\bibnamefont {Ioffe}}}\ (\bibinfo {year} {2000})\ pp.\ \bibinfo {pages}
  {2061--2151},\ \Eprint {http://arxiv.org/abs/hep-ph/0011333}
  {arXiv:hep-ph/0011333 [hep-ph]} \BibitemShut {NoStop}%
\bibitem [{\citenamefont {Buballa}(2005)}]{Buballa:2003qv}%
  \BibitemOpen
  \bibfield  {author} {\bibinfo {author} {\bibfnamefont {M.}~\bibnamefont
  {Buballa}},\ }\href {\doibase 10.1016/j.physrep.2004.11.004} {\bibfield
  {journal} {\bibinfo  {journal} {Phys. Rept.}\ }\textbf {\bibinfo {volume}
  {407}},\ \bibinfo {pages} {205} (\bibinfo {year} {2005})},\ \Eprint
  {http://arxiv.org/abs/hep-ph/0402234} {arXiv:hep-ph/0402234 [hep-ph]}
  \BibitemShut {NoStop}%
\bibitem [{\citenamefont {Shovkovy}(2005)}]{Shovkovy:2004me}%
  \BibitemOpen
  \bibfield  {author} {\bibinfo {author} {\bibfnamefont {I.~A.}\ \bibnamefont
  {Shovkovy}},\ }\href {\doibase 10.1007/s10701-005-6440-x} {\bibfield
  {journal} {\bibinfo  {journal} {Found. Phys.}\ }\textbf {\bibinfo {volume}
  {35}},\ \bibinfo {pages} {1309} (\bibinfo {year} {2005})},\ \Eprint
  {http://arxiv.org/abs/nucl-th/0410091} {arXiv:nucl-th/0410091 [nucl-th]}
  \BibitemShut {NoStop}%
\bibitem [{\citenamefont {Alford}\ \emph {et~al.}(2008)\citenamefont {Alford},
  \citenamefont {Schmitt}, \citenamefont {Rajagopal},\ and\ \citenamefont
  {Schäfer}}]{Alford:2007xm}%
  \BibitemOpen
  \bibfield  {author} {\bibinfo {author} {\bibfnamefont {M.~G.}\ \bibnamefont
  {Alford}}, \bibinfo {author} {\bibfnamefont {A.}~\bibnamefont {Schmitt}},
  \bibinfo {author} {\bibfnamefont {K.}~\bibnamefont {Rajagopal}}, \ and\
  \bibinfo {author} {\bibfnamefont {T.}~\bibnamefont {Schäfer}},\ }\href
  {\doibase 10.1103/RevModPhys.80.1455} {\bibfield  {journal} {\bibinfo
  {journal} {Rev. Mod. Phys.}\ }\textbf {\bibinfo {volume} {80}},\ \bibinfo
  {pages} {1455} (\bibinfo {year} {2008})},\ \Eprint
  {http://arxiv.org/abs/0709.4635} {arXiv:0709.4635 [hep-ph]} \BibitemShut
  {NoStop}%
\bibitem [{\citenamefont {Strodthoff}\ \emph {et~al.}(2012)\citenamefont
  {Strodthoff}, \citenamefont {Schaefer},\ and\ \citenamefont {von
  Smekal}}]{Strodthoff:2011tz}%
  \BibitemOpen
  \bibfield  {author} {\bibinfo {author} {\bibfnamefont {N.}~\bibnamefont
  {Strodthoff}}, \bibinfo {author} {\bibfnamefont {B.-J.}\ \bibnamefont
  {Schaefer}}, \ and\ \bibinfo {author} {\bibfnamefont {L.}~\bibnamefont {von
  Smekal}},\ }\href {\doibase 10.1103/PhysRevD.85.074007} {\bibfield  {journal}
  {\bibinfo  {journal} {Phys. Rev.}\ }\textbf {\bibinfo {volume} {D85}},\
  \bibinfo {pages} {074007} (\bibinfo {year} {2012})},\ \Eprint
  {http://arxiv.org/abs/1112.5401} {arXiv:1112.5401 [hep-ph]} \BibitemShut
  {NoStop}%
\bibitem [{\citenamefont {Floerchinger}\ and\ \citenamefont
  {Wetterich}(2012)}]{Floerchinger:2012xd}%
  \BibitemOpen
  \bibfield  {author} {\bibinfo {author} {\bibfnamefont {S.}~\bibnamefont
  {Floerchinger}}\ and\ \bibinfo {author} {\bibfnamefont {C.}~\bibnamefont
  {Wetterich}},\ }\href {\doibase 10.1016/j.nuclphysa.2012.07.009} {\bibfield
  {journal} {\bibinfo  {journal} {Nucl. Phys.}\ }\textbf {\bibinfo {volume}
  {A890-891}},\ \bibinfo {pages} {11} (\bibinfo {year} {2012})},\ \Eprint
  {http://arxiv.org/abs/1202.1671} {arXiv:1202.1671 [nucl-th]} \BibitemShut
  {NoStop}%
\bibitem [{\citenamefont {Drews}\ and\ \citenamefont
  {Weise}(2014)}]{Drews:2014wba}%
  \BibitemOpen
  \bibfield  {author} {\bibinfo {author} {\bibfnamefont {M.}~\bibnamefont
  {Drews}}\ and\ \bibinfo {author} {\bibfnamefont {W.}~\bibnamefont {Weise}},\
  }\href {\doibase 10.1016/j.physletb.2014.09.051} {\bibfield  {journal}
  {\bibinfo  {journal} {Phys. Lett.}\ }\textbf {\bibinfo {volume} {B738}},\
  \bibinfo {pages} {187} (\bibinfo {year} {2014})},\ \Eprint
  {http://arxiv.org/abs/1404.0882} {arXiv:1404.0882 [nucl-th]} \BibitemShut
  {NoStop}%
\bibitem [{\citenamefont {Drews}\ and\ \citenamefont
  {Weise}(2015)}]{Drews:2014spa}%
  \BibitemOpen
  \bibfield  {author} {\bibinfo {author} {\bibfnamefont {M.}~\bibnamefont
  {Drews}}\ and\ \bibinfo {author} {\bibfnamefont {W.}~\bibnamefont {Weise}},\
  }\href {\doibase 10.1103/PhysRevC.91.035802} {\bibfield  {journal} {\bibinfo
  {journal} {Phys. Rev.}\ }\textbf {\bibinfo {volume} {C91}},\ \bibinfo {pages}
  {035802} (\bibinfo {year} {2015})},\ \Eprint {http://arxiv.org/abs/1412.7655}
  {arXiv:1412.7655 [nucl-th]} \BibitemShut {NoStop}%
\bibitem [{\citenamefont {Weyrich}\ \emph {et~al.}(2015)\citenamefont
  {Weyrich}, \citenamefont {Strodthoff},\ and\ \citenamefont {von
  Smekal}}]{Weyrich:2015hha}%
  \BibitemOpen
  \bibfield  {author} {\bibinfo {author} {\bibfnamefont {J.}~\bibnamefont
  {Weyrich}}, \bibinfo {author} {\bibfnamefont {N.}~\bibnamefont {Strodthoff}},
  \ and\ \bibinfo {author} {\bibfnamefont {L.}~\bibnamefont {von Smekal}},\
  }\href {\doibase 10.1103/PhysRevC.92.015214} {\bibfield  {journal} {\bibinfo
  {journal} {Phys. Rev.}\ }\textbf {\bibinfo {volume} {C92}},\ \bibinfo {pages}
  {015214} (\bibinfo {year} {2015})},\ \Eprint
  {http://arxiv.org/abs/1504.02697} {arXiv:1504.02697 [nucl-th]} \BibitemShut
  {NoStop}%
\bibitem [{\citenamefont {Meisinger}\ and\ \citenamefont
  {Ogilvie}(1997)}]{Meisinger:1997jt}%
  \BibitemOpen
  \bibfield  {author} {\bibinfo {author} {\bibfnamefont {P.~N.}\ \bibnamefont
  {Meisinger}}\ and\ \bibinfo {author} {\bibfnamefont {M.~C.}\ \bibnamefont
  {Ogilvie}},\ }\href@noop {} {\bibfield  {journal} {\bibinfo  {journal} {Phys.
  Lett.}\ }\textbf {\bibinfo {volume} {B407}},\ \bibinfo {pages} {297}
  (\bibinfo {year} {1997})},\ \Eprint {http://arxiv.org/abs/hep-lat/9703009}
  {hep-lat/9703009} \BibitemShut {NoStop}%
\bibitem [{\citenamefont {Pisarski}(2000)}]{Pisarski:2000eq}%
  \BibitemOpen
  \bibfield  {author} {\bibinfo {author} {\bibfnamefont {R.~D.}\ \bibnamefont
  {Pisarski}},\ }\href {\doibase 10.1103/PhysRevD.62.111501} {\bibfield
  {journal} {\bibinfo  {journal} {Phys. Rev.}\ }\textbf {\bibinfo {volume}
  {D62}},\ \bibinfo {pages} {111501} (\bibinfo {year} {2000})},\ \Eprint
  {http://arxiv.org/abs/hep-ph/0006205} {arXiv:hep-ph/0006205 [hep-ph]}
  \BibitemShut {NoStop}%
\bibitem [{\citenamefont {Ratti}\ \emph {et~al.}(2006)\citenamefont {Ratti},
  \citenamefont {Thaler},\ and\ \citenamefont {Weise}}]{Ratti:2005jh}%
  \BibitemOpen
  \bibfield  {author} {\bibinfo {author} {\bibfnamefont {C.}~\bibnamefont
  {Ratti}}, \bibinfo {author} {\bibfnamefont {M.~A.}\ \bibnamefont {Thaler}}, \
  and\ \bibinfo {author} {\bibfnamefont {W.}~\bibnamefont {Weise}},\ }\href
  {\doibase 10.1103/PhysRevD.73.014019} {\bibfield  {journal} {\bibinfo
  {journal} {Phys. Rev.}\ }\textbf {\bibinfo {volume} {D73}},\ \bibinfo {pages}
  {014019} (\bibinfo {year} {2006})},\ \Eprint
  {http://arxiv.org/abs/hep-ph/0506234} {arXiv:hep-ph/0506234} \BibitemShut
  {NoStop}%
\bibitem [{\citenamefont {Fukushima}(2004)}]{Fukushima:2003fw}%
  \BibitemOpen
  \bibfield  {author} {\bibinfo {author} {\bibfnamefont {K.}~\bibnamefont
  {Fukushima}},\ }\href@noop {} {\bibfield  {journal} {\bibinfo  {journal}
  {Phys. Lett.}\ }\textbf {\bibinfo {volume} {B591}},\ \bibinfo {pages} {277}
  (\bibinfo {year} {2004})},\ \Eprint {http://arxiv.org/abs/hep-ph/0310121}
  {hep-ph/0310121} \BibitemShut {NoStop}%
\bibitem [{\citenamefont {Roessner}\ \emph {et~al.}(2007)\citenamefont
  {Roessner}, \citenamefont {Ratti},\ and\ \citenamefont
  {Weise}}]{Roessner:2006xn}%
  \BibitemOpen
  \bibfield  {author} {\bibinfo {author} {\bibfnamefont {S.}~\bibnamefont
  {Roessner}}, \bibinfo {author} {\bibfnamefont {C.}~\bibnamefont {Ratti}}, \
  and\ \bibinfo {author} {\bibfnamefont {W.}~\bibnamefont {Weise}},\ }\href
  {\doibase 10.1103/PhysRevD.75.034007} {\bibfield  {journal} {\bibinfo
  {journal} {Phys. Rev.}\ }\textbf {\bibinfo {volume} {D75}},\ \bibinfo {pages}
  {034007} (\bibinfo {year} {2007})},\ \Eprint
  {http://arxiv.org/abs/hep-ph/0609281} {arXiv:hep-ph/0609281 [hep-ph]}
  \BibitemShut {NoStop}%
\bibitem [{\citenamefont {Schaefer}\ \emph {et~al.}(2007)\citenamefont
  {Schaefer}, \citenamefont {Pawlowski},\ and\ \citenamefont
  {Wambach}}]{Schaefer:2007pw}%
  \BibitemOpen
  \bibfield  {author} {\bibinfo {author} {\bibfnamefont {B.-J.}\ \bibnamefont
  {Schaefer}}, \bibinfo {author} {\bibfnamefont {J.~M.}\ \bibnamefont
  {Pawlowski}}, \ and\ \bibinfo {author} {\bibfnamefont {J.}~\bibnamefont
  {Wambach}},\ }\href {\doibase 10.1103/PhysRevD.76.074023} {\bibfield
  {journal} {\bibinfo  {journal} {Phys. Rev.}\ }\textbf {\bibinfo {volume}
  {D76}},\ \bibinfo {pages} {074023} (\bibinfo {year} {2007})},\ \Eprint
  {http://arxiv.org/abs/0704.3234} {arXiv:0704.3234 [hep-ph]} \BibitemShut
  {NoStop}%
\bibitem [{\citenamefont {Skokov}\ \emph
  {et~al.}(2010{\natexlab{a}})\citenamefont {Skokov}, \citenamefont {Stokic},
  \citenamefont {Friman},\ and\ \citenamefont {Redlich}}]{Skokov:2010wb}%
  \BibitemOpen
  \bibfield  {author} {\bibinfo {author} {\bibfnamefont {V.}~\bibnamefont
  {Skokov}}, \bibinfo {author} {\bibfnamefont {B.}~\bibnamefont {Stokic}},
  \bibinfo {author} {\bibfnamefont {B.}~\bibnamefont {Friman}}, \ and\ \bibinfo
  {author} {\bibfnamefont {K.}~\bibnamefont {Redlich}},\ }\href {\doibase
  10.1103/PhysRevC.82.015206} {\bibfield  {journal} {\bibinfo  {journal} {Phys.
  Rev.}\ }\textbf {\bibinfo {volume} {C82}},\ \bibinfo {pages} {015206}
  (\bibinfo {year} {2010}{\natexlab{a}})},\ \Eprint
  {http://arxiv.org/abs/1004.2665} {arXiv:1004.2665 [hep-ph]} \BibitemShut
  {NoStop}%
\bibitem [{\citenamefont {Skokov}\ \emph {et~al.}(2011)\citenamefont {Skokov},
  \citenamefont {Friman},\ and\ \citenamefont {Redlich}}]{Skokov:2010uh}%
  \BibitemOpen
  \bibfield  {author} {\bibinfo {author} {\bibfnamefont {V.}~\bibnamefont
  {Skokov}}, \bibinfo {author} {\bibfnamefont {B.}~\bibnamefont {Friman}}, \
  and\ \bibinfo {author} {\bibfnamefont {K.}~\bibnamefont {Redlich}},\ }\href
  {\doibase 10.1103/PhysRevC.83.054904} {\bibfield  {journal} {\bibinfo
  {journal} {Phys. Rev.}\ }\textbf {\bibinfo {volume} {C83}},\ \bibinfo {pages}
  {054904} (\bibinfo {year} {2011})},\ \Eprint {http://arxiv.org/abs/1008.4570}
  {arXiv:1008.4570 [hep-ph]} \BibitemShut {NoStop}%
\bibitem [{\citenamefont {Herbst}\ \emph {et~al.}(2011)\citenamefont {Herbst},
  \citenamefont {Pawlowski},\ and\ \citenamefont {Schaefer}}]{Herbst:2010rf}%
  \BibitemOpen
  \bibfield  {author} {\bibinfo {author} {\bibfnamefont {T.~K.}\ \bibnamefont
  {Herbst}}, \bibinfo {author} {\bibfnamefont {J.~M.}\ \bibnamefont
  {Pawlowski}}, \ and\ \bibinfo {author} {\bibfnamefont {B.-J.}\ \bibnamefont
  {Schaefer}},\ }\href {\doibase 10.1016/j.physletb.2010.12.003} {\bibfield
  {journal} {\bibinfo  {journal} {Phys. Lett.}\ }\textbf {\bibinfo {volume}
  {B696}},\ \bibinfo {pages} {58} (\bibinfo {year} {2011})},\ \Eprint
  {http://arxiv.org/abs/1008.0081} {arXiv:1008.0081 [hep-ph]} \BibitemShut
  {NoStop}%
\bibitem [{\citenamefont {Strodthoff}\ and\ \citenamefont {von
  Smekal}(2014)}]{Strodthoff:2013cua}%
  \BibitemOpen
  \bibfield  {author} {\bibinfo {author} {\bibfnamefont {N.}~\bibnamefont
  {Strodthoff}}\ and\ \bibinfo {author} {\bibfnamefont {L.}~\bibnamefont {von
  Smekal}},\ }\href {\doibase 10.1016/j.physletb.2014.03.008} {\bibfield
  {journal} {\bibinfo  {journal} {Phys. Lett.}\ }\textbf {\bibinfo {volume}
  {B731}},\ \bibinfo {pages} {350} (\bibinfo {year} {2014})},\ \Eprint
  {http://arxiv.org/abs/1306.2897} {arXiv:1306.2897 [hep-ph]} \BibitemShut
  {NoStop}%
\bibitem [{\citenamefont {Haas}\ \emph {et~al.}(2013)\citenamefont {Haas},
  \citenamefont {Stiele}, \citenamefont {Braun}, \citenamefont {Pawlowski},\
  and\ \citenamefont {Schaffner-Bielich}}]{Haas:2013qwp}%
  \BibitemOpen
  \bibfield  {author} {\bibinfo {author} {\bibfnamefont {L.~M.}\ \bibnamefont
  {Haas}}, \bibinfo {author} {\bibfnamefont {R.}~\bibnamefont {Stiele}},
  \bibinfo {author} {\bibfnamefont {J.}~\bibnamefont {Braun}}, \bibinfo
  {author} {\bibfnamefont {J.~M.}\ \bibnamefont {Pawlowski}}, \ and\ \bibinfo
  {author} {\bibfnamefont {J.}~\bibnamefont {Schaffner-Bielich}},\ }\href
  {\doibase 10.1103/PhysRevD.87.076004} {\bibfield  {journal} {\bibinfo
  {journal} {Phys.Rev.}\ }\textbf {\bibinfo {volume} {D87}},\ \bibinfo {pages}
  {076004} (\bibinfo {year} {2013})},\ \Eprint {http://arxiv.org/abs/1302.1993}
  {arXiv:1302.1993 [hep-ph]} \BibitemShut {NoStop}%
\bibitem [{\citenamefont {Pisarski}\ and\ \citenamefont
  {Skokov}(2016)}]{Pisarski:2016ixt}%
  \BibitemOpen
  \bibfield  {author} {\bibinfo {author} {\bibfnamefont {R.~D.}\ \bibnamefont
  {Pisarski}}\ and\ \bibinfo {author} {\bibfnamefont {V.~V.}\ \bibnamefont
  {Skokov}},\ }\href {\doibase 10.1103/PhysRevD.94.034015} {\bibfield
  {journal} {\bibinfo  {journal} {Phys. Rev.}\ }\textbf {\bibinfo {volume}
  {D94}},\ \bibinfo {pages} {034015} (\bibinfo {year} {2016})},\ \Eprint
  {http://arxiv.org/abs/1604.00022} {arXiv:1604.00022 [hep-ph]} \BibitemShut
  {NoStop}%
\bibitem [{\citenamefont {Fukushima}\ and\ \citenamefont
  {Skokov}(2017)}]{Fukushima:2017csk}%
  \BibitemOpen
  \bibfield  {author} {\bibinfo {author} {\bibfnamefont {K.}~\bibnamefont
  {Fukushima}}\ and\ \bibinfo {author} {\bibfnamefont {V.}~\bibnamefont
  {Skokov}},\ }\href {\doibase 10.1016/j.ppnp.2017.05.002} {\bibfield
  {journal} {\bibinfo  {journal} {Prog. Part. Nucl. Phys.}\ }\textbf {\bibinfo
  {volume} {96}},\ \bibinfo {pages} {154} (\bibinfo {year} {2017})},\ \Eprint
  {http://arxiv.org/abs/1705.00718} {arXiv:1705.00718 [hep-ph]} \BibitemShut
  {NoStop}%
\bibitem [{\citenamefont {Gies}\ and\ \citenamefont
  {Wetterich}(2004)}]{Gies:2002hq}%
  \BibitemOpen
  \bibfield  {author} {\bibinfo {author} {\bibfnamefont {H.}~\bibnamefont
  {Gies}}\ and\ \bibinfo {author} {\bibfnamefont {C.}~\bibnamefont
  {Wetterich}},\ }\href@noop {} {\bibfield  {journal} {\bibinfo  {journal}
  {Phys. Rev.}\ }\textbf {\bibinfo {volume} {D69}},\ \bibinfo {pages} {025001}
  (\bibinfo {year} {2004})},\ \Eprint {http://arxiv.org/abs/hep-th/0209183}
  {hep-th/0209183} \BibitemShut {NoStop}%
\bibitem [{\citenamefont {Gies}(2012)}]{Gies:2006wv}%
  \BibitemOpen
  \bibfield  {author} {\bibinfo {author} {\bibfnamefont {H.}~\bibnamefont
  {Gies}},\ }\href {\doibase 10.1007/978-3-642-27320-9_6} {\bibfield  {journal}
  {\bibinfo  {journal} {Lect. Notes Phys.}\ }\textbf {\bibinfo {volume}
  {852}},\ \bibinfo {pages} {287} (\bibinfo {year} {2012})},\ \Eprint
  {http://arxiv.org/abs/hep-ph/0611146} {arXiv:hep-ph/0611146 [hep-ph]}
  \BibitemShut {NoStop}%
\bibitem [{\citenamefont {Braun}\ \emph
  {et~al.}(2011{\natexlab{a}})\citenamefont {Braun}, \citenamefont {Haas},
  \citenamefont {Marhauser},\ and\ \citenamefont {Pawlowski}}]{Braun:2009gm}%
  \BibitemOpen
  \bibfield  {author} {\bibinfo {author} {\bibfnamefont {J.}~\bibnamefont
  {Braun}}, \bibinfo {author} {\bibfnamefont {L.~M.}\ \bibnamefont {Haas}},
  \bibinfo {author} {\bibfnamefont {F.}~\bibnamefont {Marhauser}}, \ and\
  \bibinfo {author} {\bibfnamefont {J.~M.}\ \bibnamefont {Pawlowski}},\ }\href
  {\doibase 10.1103/PhysRevLett.106.022002} {\bibfield  {journal} {\bibinfo
  {journal} {Phys. Rev. Lett.}\ }\textbf {\bibinfo {volume} {106}},\ \bibinfo
  {pages} {022002} (\bibinfo {year} {2011}{\natexlab{a}})},\ \Eprint
  {http://arxiv.org/abs/0908.0008} {arXiv:0908.0008 [hep-ph]} \BibitemShut
  {NoStop}%
\bibitem [{\citenamefont {Mitter}\ \emph {et~al.}(2015)\citenamefont {Mitter},
  \citenamefont {Pawlowski},\ and\ \citenamefont
  {Strodthoff}}]{Mitter:2014wpa}%
  \BibitemOpen
  \bibfield  {author} {\bibinfo {author} {\bibfnamefont {M.}~\bibnamefont
  {Mitter}}, \bibinfo {author} {\bibfnamefont {J.~M.}\ \bibnamefont
  {Pawlowski}}, \ and\ \bibinfo {author} {\bibfnamefont {N.}~\bibnamefont
  {Strodthoff}},\ }\href {\doibase 10.1103/PhysRevD.91.054035} {\bibfield
  {journal} {\bibinfo  {journal} {Phys. Rev.}\ }\textbf {\bibinfo {volume}
  {D91}},\ \bibinfo {pages} {054035} (\bibinfo {year} {2015})},\ \Eprint
  {http://arxiv.org/abs/1411.7978} {arXiv:1411.7978 [hep-ph]} \BibitemShut
  {NoStop}%
\bibitem [{\citenamefont {Braun}\ \emph
  {et~al.}(2016{\natexlab{a}})\citenamefont {Braun}, \citenamefont {Fister},
  \citenamefont {Pawlowski},\ and\ \citenamefont {Rennecke}}]{Braun:2014ata}%
  \BibitemOpen
  \bibfield  {author} {\bibinfo {author} {\bibfnamefont {J.}~\bibnamefont
  {Braun}}, \bibinfo {author} {\bibfnamefont {L.}~\bibnamefont {Fister}},
  \bibinfo {author} {\bibfnamefont {J.~M.}\ \bibnamefont {Pawlowski}}, \ and\
  \bibinfo {author} {\bibfnamefont {F.}~\bibnamefont {Rennecke}},\ }\href
  {\doibase 10.1103/PhysRevD.94.034016} {\bibfield  {journal} {\bibinfo
  {journal} {Phys. Rev.}\ }\textbf {\bibinfo {volume} {D94}},\ \bibinfo {pages}
  {034016} (\bibinfo {year} {2016}{\natexlab{a}})},\ \Eprint
  {http://arxiv.org/abs/1412.1045} {arXiv:1412.1045 [hep-ph]} \BibitemShut
  {NoStop}%
\bibitem [{\citenamefont {Rennecke}(2015)}]{Rennecke:2015eba}%
  \BibitemOpen
  \bibfield  {author} {\bibinfo {author} {\bibfnamefont {F.}~\bibnamefont
  {Rennecke}},\ }\href {\doibase 10.1103/PhysRevD.92.076012} {\bibfield
  {journal} {\bibinfo  {journal} {Phys. Rev.}\ }\textbf {\bibinfo {volume}
  {D92}},\ \bibinfo {pages} {076012} (\bibinfo {year} {2015})},\ \Eprint
  {http://arxiv.org/abs/1504.03585} {arXiv:1504.03585 [hep-ph]} \BibitemShut
  {NoStop}%
\bibitem [{\citenamefont {Cyrol}\ \emph {et~al.}(2018)\citenamefont {Cyrol},
  \citenamefont {Mitter}, \citenamefont {Pawlowski},\ and\ \citenamefont
  {Strodthoff}}]{Cyrol:2017ewj}%
  \BibitemOpen
  \bibfield  {author} {\bibinfo {author} {\bibfnamefont {A.~K.}\ \bibnamefont
  {Cyrol}}, \bibinfo {author} {\bibfnamefont {M.}~\bibnamefont {Mitter}},
  \bibinfo {author} {\bibfnamefont {J.~M.}\ \bibnamefont {Pawlowski}}, \ and\
  \bibinfo {author} {\bibfnamefont {N.}~\bibnamefont {Strodthoff}},\ }\href
  {\doibase 10.1103/PhysRevD.97.054006} {\bibfield  {journal} {\bibinfo
  {journal} {Phys. Rev.}\ }\textbf {\bibinfo {volume} {D97}},\ \bibinfo {pages}
  {054006} (\bibinfo {year} {2018})},\ \Eprint
  {http://arxiv.org/abs/1706.06326} {arXiv:1706.06326 [hep-ph]} \BibitemShut
  {NoStop}%
\bibitem [{\citenamefont {Pawlowski}(2011)}]{Pawlowski:2010ht}%
  \BibitemOpen
  \bibfield  {author} {\bibinfo {author} {\bibfnamefont {J.~M.}\ \bibnamefont
  {Pawlowski}},\ }\href {\doibase 10.1063/1.3574945} {\bibfield  {journal}
  {\bibinfo  {journal} {AIP Conf. Proc.}\ }\textbf {\bibinfo {volume} {1343}},\
  \bibinfo {pages} {75} (\bibinfo {year} {2011})},\ \Eprint
  {http://arxiv.org/abs/1012.5075} {arXiv:1012.5075 [hep-ph]} \BibitemShut
  {NoStop}%
\bibitem [{\citenamefont {Pawlowski}(2014)}]{Pawlowski:2014aha}%
  \BibitemOpen
  \bibfield  {author} {\bibinfo {author} {\bibfnamefont {J.~M.}\ \bibnamefont
  {Pawlowski}},\ }\href {\doibase 10.1016/j.nuclphysa.2014.09.074} {\bibfield
  {journal} {\bibinfo  {journal} {Nucl. Phys.}\ }\textbf {\bibinfo {volume}
  {A931}},\ \bibinfo {pages} {113} (\bibinfo {year} {2014})}\BibitemShut
  {NoStop}%
\bibitem [{\citenamefont {Herbst}\ \emph {et~al.}(2014)\citenamefont {Herbst},
  \citenamefont {Mitter}, \citenamefont {Pawlowski}, \citenamefont {Schaefer},\
  and\ \citenamefont {Stiele}}]{Herbst:2013ufa}%
  \BibitemOpen
  \bibfield  {author} {\bibinfo {author} {\bibfnamefont {T.~K.}\ \bibnamefont
  {Herbst}}, \bibinfo {author} {\bibfnamefont {M.}~\bibnamefont {Mitter}},
  \bibinfo {author} {\bibfnamefont {J.~M.}\ \bibnamefont {Pawlowski}}, \bibinfo
  {author} {\bibfnamefont {B.-J.}\ \bibnamefont {Schaefer}}, \ and\ \bibinfo
  {author} {\bibfnamefont {R.}~\bibnamefont {Stiele}},\ }\href {\doibase
  10.1016/j.physletb.2014.02.045} {\bibfield  {journal} {\bibinfo  {journal}
  {Phys. Lett.}\ }\textbf {\bibinfo {volume} {B731}},\ \bibinfo {pages} {248}
  (\bibinfo {year} {2014})},\ \Eprint {http://arxiv.org/abs/1308.3621}
  {arXiv:1308.3621 [hep-ph]} \BibitemShut {NoStop}%
\bibitem [{\citenamefont {Springer}\ \emph {et~al.}(2017)\citenamefont
  {Springer}, \citenamefont {Braun}, \citenamefont {Rechenberger},\ and\
  \citenamefont {Rennecke}}]{Springer:2016cji}%
  \BibitemOpen
  \bibfield  {author} {\bibinfo {author} {\bibfnamefont {P.}~\bibnamefont
  {Springer}}, \bibinfo {author} {\bibfnamefont {J.}~\bibnamefont {Braun}},
  \bibinfo {author} {\bibfnamefont {S.}~\bibnamefont {Rechenberger}}, \ and\
  \bibinfo {author} {\bibfnamefont {F.}~\bibnamefont {Rennecke}},\ }\href
  {\doibase 10.1051/epjconf/201713703022} {\bibfield  {journal} {\bibinfo
  {journal} {EPJ Web Conf.}\ }\textbf {\bibinfo {volume} {137}},\ \bibinfo
  {pages} {03022} (\bibinfo {year} {2017})},\ \Eprint
  {http://arxiv.org/abs/1611.06020} {arXiv:1611.06020 [hep-ph]} \BibitemShut
  {NoStop}%
\bibitem [{\citenamefont {Wetterich}(1993)}]{Wetterich:1992yh}%
  \BibitemOpen
  \bibfield  {author} {\bibinfo {author} {\bibfnamefont {C.}~\bibnamefont
  {Wetterich}},\ }\href@noop {} {\bibfield  {journal} {\bibinfo  {journal}
  {Phys. Lett.}\ }\textbf {\bibinfo {volume} {B301}},\ \bibinfo {pages} {90}
  (\bibinfo {year} {1993})}\BibitemShut {NoStop}%
\bibitem [{\citenamefont {Reuter}\ and\ \citenamefont
  {Wetterich}(1994)}]{Reuter:1993kw}%
  \BibitemOpen
  \bibfield  {author} {\bibinfo {author} {\bibfnamefont {M.}~\bibnamefont
  {Reuter}}\ and\ \bibinfo {author} {\bibfnamefont {C.}~\bibnamefont
  {Wetterich}},\ }\href@noop {} {\bibfield  {journal} {\bibinfo  {journal}
  {Nucl. Phys.}\ }\textbf {\bibinfo {volume} {B417}},\ \bibinfo {pages} {181}
  (\bibinfo {year} {1994})}\BibitemShut {NoStop}%
\bibitem [{\citenamefont {Pawlowski}(2001)}]{Pawlowski:2001df}%
  \BibitemOpen
  \bibfield  {author} {\bibinfo {author} {\bibfnamefont {J.~M.}\ \bibnamefont
  {Pawlowski}},\ }\href {\doibase 10.1142/S0217751X01004785} {\bibfield
  {journal} {\bibinfo  {journal} {Int. J. Mod. Phys.}\ }\textbf {\bibinfo
  {volume} {A16}},\ \bibinfo {pages} {2105} (\bibinfo {year}
  {2001})}\BibitemShut {NoStop}%
\bibitem [{\citenamefont {Litim}\ and\ \citenamefont
  {Pawlowski}(2002{\natexlab{a}})}]{Litim:2002hj}%
  \BibitemOpen
  \bibfield  {author} {\bibinfo {author} {\bibfnamefont {D.~F.}\ \bibnamefont
  {Litim}}\ and\ \bibinfo {author} {\bibfnamefont {J.~M.}\ \bibnamefont
  {Pawlowski}},\ }\href {\doibase 10.1016/S0370-2693(02)02693-X} {\bibfield
  {journal} {\bibinfo  {journal} {Phys. Lett.}\ }\textbf {\bibinfo {volume}
  {B546}},\ \bibinfo {pages} {279} (\bibinfo {year} {2002}{\natexlab{a}})},\
  \Eprint {http://arxiv.org/abs/hep-th/0208216} {arXiv:hep-th/0208216 [hep-th]}
  \BibitemShut {NoStop}%
\bibitem [{\citenamefont {Gies}(2002)}]{Gies:2002af}%
  \BibitemOpen
  \bibfield  {author} {\bibinfo {author} {\bibfnamefont {H.}~\bibnamefont
  {Gies}},\ }\href {\doibase 10.1103/PhysRevD.66.025006} {\bibfield  {journal}
  {\bibinfo  {journal} {Phys. Rev.}\ }\textbf {\bibinfo {volume} {D66}},\
  \bibinfo {pages} {025006} (\bibinfo {year} {2002})},\ \Eprint
  {http://arxiv.org/abs/hep-th/0202207} {arXiv:hep-th/0202207} \BibitemShut
  {NoStop}%
\bibitem [{\citenamefont {Pawlowski}(2007)}]{Pawlowski:2005xe}%
  \BibitemOpen
  \bibfield  {author} {\bibinfo {author} {\bibfnamefont {J.~M.}\ \bibnamefont
  {Pawlowski}},\ }\href {\doibase 10.1016/j.aop.2007.01.007} {\bibfield
  {journal} {\bibinfo  {journal} {Annals Phys.}\ }\textbf {\bibinfo {volume}
  {322}},\ \bibinfo {pages} {2831} (\bibinfo {year} {2007})},\ \Eprint
  {http://arxiv.org/abs/hep-th/0512261} {arXiv:hep-th/0512261} \BibitemShut
  {NoStop}%
\bibitem [{\citenamefont {Pawlowski}\ \emph
  {et~al.}(2017{\natexlab{a}})\citenamefont {Pawlowski}, \citenamefont
  {Scherer}, \citenamefont {Schmidt},\ and\ \citenamefont
  {Wetzel}}]{Pawlowski:2015mlf}%
  \BibitemOpen
  \bibfield  {author} {\bibinfo {author} {\bibfnamefont {J.~M.}\ \bibnamefont
  {Pawlowski}}, \bibinfo {author} {\bibfnamefont {M.~M.}\ \bibnamefont
  {Scherer}}, \bibinfo {author} {\bibfnamefont {R.}~\bibnamefont {Schmidt}}, \
  and\ \bibinfo {author} {\bibfnamefont {S.~J.}\ \bibnamefont {Wetzel}},\
  }\href {\doibase 10.1016/j.aop.2017.06.017} {\bibfield  {journal} {\bibinfo
  {journal} {Annals Phys.}\ }\textbf {\bibinfo {volume} {384}},\ \bibinfo
  {pages} {165} (\bibinfo {year} {2017}{\natexlab{a}})},\ \Eprint
  {http://arxiv.org/abs/1512.03598} {arXiv:1512.03598 [hep-th]} \BibitemShut
  {NoStop}%
\bibitem [{\citenamefont {Litim}\ and\ \citenamefont
  {Pawlowski}(2002{\natexlab{b}})}]{Litim:2001ky}%
  \BibitemOpen
  \bibfield  {author} {\bibinfo {author} {\bibfnamefont {D.~F.}\ \bibnamefont
  {Litim}}\ and\ \bibinfo {author} {\bibfnamefont {J.~M.}\ \bibnamefont
  {Pawlowski}},\ }\href {\doibase 10.1103/PhysRevD.65.081701} {\bibfield
  {journal} {\bibinfo  {journal} {Phys. Rev.}\ }\textbf {\bibinfo {volume}
  {D65}},\ \bibinfo {pages} {081701} (\bibinfo {year} {2002}{\natexlab{b}})},\
  \Eprint {http://arxiv.org/abs/hep-th/0111191} {arXiv:hep-th/0111191 [hep-th]}
  \BibitemShut {NoStop}%
\bibitem [{\citenamefont {Litim}\ and\ \citenamefont
  {Pawlowski}(2002{\natexlab{c}})}]{Litim:2002xm}%
  \BibitemOpen
  \bibfield  {author} {\bibinfo {author} {\bibfnamefont {D.~F.}\ \bibnamefont
  {Litim}}\ and\ \bibinfo {author} {\bibfnamefont {J.~M.}\ \bibnamefont
  {Pawlowski}},\ }\href@noop {} {\bibfield  {journal} {\bibinfo  {journal}
  {Phys. Rev.}\ }\textbf {\bibinfo {volume} {D66}},\ \bibinfo {pages} {025030}
  (\bibinfo {year} {2002}{\natexlab{c}})},\ \Eprint
  {http://arxiv.org/abs/hep-th/0202188} {hep-th/0202188} \BibitemShut {NoStop}%
\bibitem [{\citenamefont {Rosten}(2012)}]{Rosten:2010vm}%
  \BibitemOpen
  \bibfield  {author} {\bibinfo {author} {\bibfnamefont {O.~J.}\ \bibnamefont
  {Rosten}},\ }\href {\doibase 10.1016/j.physrep.2011.12.003} {\bibfield
  {journal} {\bibinfo  {journal} {Phys. Rept.}\ }\textbf {\bibinfo {volume}
  {511}},\ \bibinfo {pages} {177} (\bibinfo {year} {2012})},\ \Eprint
  {http://arxiv.org/abs/1003.1366} {arXiv:1003.1366 [hep-th]} \BibitemShut
  {NoStop}%
\bibitem [{\citenamefont {Braun}\ \emph {et~al.}(2004)\citenamefont {Braun},
  \citenamefont {Schwenzer},\ and\ \citenamefont {Pirner}}]{Braun:2003ii}%
  \BibitemOpen
  \bibfield  {author} {\bibinfo {author} {\bibfnamefont {J.}~\bibnamefont
  {Braun}}, \bibinfo {author} {\bibfnamefont {K.}~\bibnamefont {Schwenzer}}, \
  and\ \bibinfo {author} {\bibfnamefont {H.-J.}\ \bibnamefont {Pirner}},\
  }\href@noop {} {\bibfield  {journal} {\bibinfo  {journal} {Phys. Rev.}\
  }\textbf {\bibinfo {volume} {D70}},\ \bibinfo {pages} {085016} (\bibinfo
  {year} {2004})},\ \Eprint {http://arxiv.org/abs/hep-ph/0312277}
  {hep-ph/0312277} \BibitemShut {NoStop}%
\bibitem [{\citenamefont {Helmboldt}\ \emph {et~al.}(2015)\citenamefont
  {Helmboldt}, \citenamefont {Pawlowski},\ and\ \citenamefont
  {Strodthoff}}]{Helmboldt:2014iya}%
  \BibitemOpen
  \bibfield  {author} {\bibinfo {author} {\bibfnamefont {A.~J.}\ \bibnamefont
  {Helmboldt}}, \bibinfo {author} {\bibfnamefont {J.~M.}\ \bibnamefont
  {Pawlowski}}, \ and\ \bibinfo {author} {\bibfnamefont {N.}~\bibnamefont
  {Strodthoff}},\ }\href {\doibase 10.1103/PhysRevD.91.054010} {\bibfield
  {journal} {\bibinfo  {journal} {Phys.Rev.}\ }\textbf {\bibinfo {volume}
  {D91}},\ \bibinfo {pages} {054010} (\bibinfo {year} {2015})},\ \Eprint
  {http://arxiv.org/abs/1409.8414} {arXiv:1409.8414 [hep-ph]} \BibitemShut
  {NoStop}%
\bibitem [{\citenamefont {Nambu}\ and\ \citenamefont
  {Jona-Lasinio}(1961{\natexlab{a}})}]{Nambu:1961tp}%
  \BibitemOpen
  \bibfield  {author} {\bibinfo {author} {\bibfnamefont {Y.}~\bibnamefont
  {Nambu}}\ and\ \bibinfo {author} {\bibfnamefont {G.}~\bibnamefont
  {Jona-Lasinio}},\ }\href {\doibase 10.1103/PhysRev.122.345} {\bibfield
  {journal} {\bibinfo  {journal} {Phys. Rev.}\ }\textbf {\bibinfo {volume}
  {122}},\ \bibinfo {pages} {345} (\bibinfo {year}
  {1961}{\natexlab{a}})}\BibitemShut {NoStop}%
\bibitem [{\citenamefont {Nambu}\ and\ \citenamefont
  {Jona-Lasinio}(1961{\natexlab{b}})}]{Nambu:1961fr}%
  \BibitemOpen
  \bibfield  {author} {\bibinfo {author} {\bibfnamefont {Y.}~\bibnamefont
  {Nambu}}\ and\ \bibinfo {author} {\bibfnamefont {G.}~\bibnamefont
  {Jona-Lasinio}},\ }\href {\doibase 10.1103/PhysRev.124.246} {\bibfield
  {journal} {\bibinfo  {journal} {Phys. Rev.}\ }\textbf {\bibinfo {volume}
  {124}},\ \bibinfo {pages} {246} (\bibinfo {year}
  {1961}{\natexlab{b}})}\BibitemShut {NoStop}%
\bibitem [{\citenamefont {Braun}(2012)}]{Braun:2011pp}%
  \BibitemOpen
  \bibfield  {author} {\bibinfo {author} {\bibfnamefont {J.}~\bibnamefont
  {Braun}},\ }\href {\doibase 10.1088/0954-3899/39/3/033001} {\bibfield
  {journal} {\bibinfo  {journal} {J. Phys.}\ }\textbf {\bibinfo {volume}
  {G39}},\ \bibinfo {pages} {033001} (\bibinfo {year} {2012})},\ \Eprint
  {http://arxiv.org/abs/1108.4449} {arXiv:1108.4449 [hep-ph]} \BibitemShut
  {NoStop}%
\bibitem [{\citenamefont {Braun}\ and\ \citenamefont
  {Herbst}(2012)}]{Braun:2012zq}%
  \BibitemOpen
  \bibfield  {author} {\bibinfo {author} {\bibfnamefont {J.}~\bibnamefont
  {Braun}}\ and\ \bibinfo {author} {\bibfnamefont {T.~K.}\ \bibnamefont
  {Herbst}},\ }\href@noop {} {\  (\bibinfo {year} {2012})},\ \Eprint
  {http://arxiv.org/abs/1205.0779} {arXiv:1205.0779 [hep-ph]} \BibitemShut
  {NoStop}%
\bibitem [{\citenamefont {Litim}(2000)}]{Litim:2000ci}%
  \BibitemOpen
  \bibfield  {author} {\bibinfo {author} {\bibfnamefont {D.~F.}\ \bibnamefont
  {Litim}},\ }\href@noop {} {\bibfield  {journal} {\bibinfo  {journal} {Phys.
  Lett.}\ }\textbf {\bibinfo {volume} {B486}},\ \bibinfo {pages} {92} (\bibinfo
  {year} {2000})},\ \Eprint {http://arxiv.org/abs/hep-th/0005245}
  {hep-th/0005245} \BibitemShut {NoStop}%
\bibitem [{\citenamefont {Litim}(2001{\natexlab{a}})}]{Litim:2001up}%
  \BibitemOpen
  \bibfield  {author} {\bibinfo {author} {\bibfnamefont {D.~F.}\ \bibnamefont
  {Litim}},\ }\href@noop {} {\bibfield  {journal} {\bibinfo  {journal} {Phys.
  Rev.}\ }\textbf {\bibinfo {volume} {D64}},\ \bibinfo {pages} {105007}
  (\bibinfo {year} {2001}{\natexlab{a}})},\ \Eprint
  {http://arxiv.org/abs/hep-th/0103195} {hep-th/0103195} \BibitemShut {NoStop}%
\bibitem [{\citenamefont {Litim}(2001{\natexlab{b}})}]{Litim:2001fd}%
  \BibitemOpen
  \bibfield  {author} {\bibinfo {author} {\bibfnamefont {D.~F.}\ \bibnamefont
  {Litim}},\ }\href {\doibase 10.1142/S0217751X01004748} {\bibfield  {journal}
  {\bibinfo  {journal} {Int. J. Mod. Phys.}\ }\textbf {\bibinfo {volume}
  {A16}},\ \bibinfo {pages} {2081} (\bibinfo {year} {2001}{\natexlab{b}})},\
  \Eprint {http://arxiv.org/abs/hep-th/0104221} {arXiv:hep-th/0104221}
  \BibitemShut {NoStop}%
\bibitem [{\citenamefont {Meyer}\ \emph {et~al.}(2002)\citenamefont {Meyer},
  \citenamefont {Schwenzer}, \citenamefont {Pirner},\ and\ \citenamefont
  {Deandrea}}]{Meyer:2001zp}%
  \BibitemOpen
  \bibfield  {author} {\bibinfo {author} {\bibfnamefont {J.}~\bibnamefont
  {Meyer}}, \bibinfo {author} {\bibfnamefont {K.}~\bibnamefont {Schwenzer}},
  \bibinfo {author} {\bibfnamefont {H.-J.}\ \bibnamefont {Pirner}}, \ and\
  \bibinfo {author} {\bibfnamefont {A.}~\bibnamefont {Deandrea}},\ }\href@noop
  {} {\bibfield  {journal} {\bibinfo  {journal} {Phys. Lett.}\ }\textbf
  {\bibinfo {volume} {B526}},\ \bibinfo {pages} {79} (\bibinfo {year}
  {2002})},\ \Eprint {http://arxiv.org/abs/hep-ph/0110279} {hep-ph/0110279}
  \BibitemShut {NoStop}%
\bibitem [{\citenamefont {Braun}\ \emph
  {et~al.}(2016{\natexlab{b}})\citenamefont {Braun}, \citenamefont {Karbstein},
  \citenamefont {Rechenberger},\ and\ \citenamefont {Roscher}}]{Braun:2015fva}%
  \BibitemOpen
  \bibfield  {author} {\bibinfo {author} {\bibfnamefont {J.}~\bibnamefont
  {Braun}}, \bibinfo {author} {\bibfnamefont {F.}~\bibnamefont {Karbstein}},
  \bibinfo {author} {\bibfnamefont {S.}~\bibnamefont {Rechenberger}}, \ and\
  \bibinfo {author} {\bibfnamefont {D.}~\bibnamefont {Roscher}},\ }\href
  {\doibase 10.1103/PhysRevD.93.014032} {\bibfield  {journal} {\bibinfo
  {journal} {Phys. Rev.}\ }\textbf {\bibinfo {volume} {D93}},\ \bibinfo {pages}
  {014032} (\bibinfo {year} {2016}{\natexlab{b}})},\ \Eprint
  {http://arxiv.org/abs/1510.04012} {arXiv:1510.04012 [hep-ph]} \BibitemShut
  {NoStop}%
\bibitem [{\citenamefont {Braun}(2010)}]{Braun:2009si}%
  \BibitemOpen
  \bibfield  {author} {\bibinfo {author} {\bibfnamefont {J.}~\bibnamefont
  {Braun}},\ }\href {\doibase 10.1103/PhysRevD.81.016008} {\bibfield  {journal}
  {\bibinfo  {journal} {Phys. Rev.}\ }\textbf {\bibinfo {volume} {D81}},\
  \bibinfo {pages} {016008} (\bibinfo {year} {2010})},\ \Eprint
  {http://arxiv.org/abs/0908.1543} {arXiv:0908.1543 [hep-ph]} \BibitemShut
  {NoStop}%
\bibitem [{\citenamefont {Skokov}\ \emph
  {et~al.}(2010{\natexlab{b}})\citenamefont {Skokov}, \citenamefont {Friman},
  \citenamefont {Nakano}, \citenamefont {Redlich},\ and\ \citenamefont
  {Schaefer}}]{Skokov:2010sf}%
  \BibitemOpen
  \bibfield  {author} {\bibinfo {author} {\bibfnamefont {V.}~\bibnamefont
  {Skokov}}, \bibinfo {author} {\bibfnamefont {B.}~\bibnamefont {Friman}},
  \bibinfo {author} {\bibfnamefont {E.}~\bibnamefont {Nakano}}, \bibinfo
  {author} {\bibfnamefont {K.}~\bibnamefont {Redlich}}, \ and\ \bibinfo
  {author} {\bibfnamefont {B.~J.}\ \bibnamefont {Schaefer}},\ }\href {\doibase
  10.1103/PhysRevD.82.034029} {\bibfield  {journal} {\bibinfo  {journal} {Phys.
  Rev.}\ }\textbf {\bibinfo {volume} {D82}},\ \bibinfo {pages} {034029}
  (\bibinfo {year} {2010}{\natexlab{b}})},\ \Eprint
  {http://arxiv.org/abs/1005.3166} {arXiv:1005.3166 [hep-ph]} \BibitemShut
  {NoStop}%
\bibitem [{\citenamefont {Braun}\ \emph {et~al.}(2017)\citenamefont {Braun},
  \citenamefont {Leonhardt},\ and\ \citenamefont {Pospiech}}]{Braun:2017srn}%
  \BibitemOpen
  \bibfield  {author} {\bibinfo {author} {\bibfnamefont {J.}~\bibnamefont
  {Braun}}, \bibinfo {author} {\bibfnamefont {M.}~\bibnamefont {Leonhardt}}, \
  and\ \bibinfo {author} {\bibfnamefont {M.}~\bibnamefont {Pospiech}},\ }\href
  {\doibase 10.1103/PhysRevD.96.076003} {\bibfield  {journal} {\bibinfo
  {journal} {Phys. Rev.}\ }\textbf {\bibinfo {volume} {D96}},\ \bibinfo {pages}
  {076003} (\bibinfo {year} {2017})},\ \Eprint
  {http://arxiv.org/abs/1705.00074} {arXiv:1705.00074 [hep-ph]} \BibitemShut
  {NoStop}%
\bibitem [{\citenamefont {Braun}\ \emph {et~al.}(2018)\citenamefont {Braun},
  \citenamefont {Leonhardt},\ and\ \citenamefont {Pospiech}}]{Braun:2018bik}%
  \BibitemOpen
  \bibfield  {author} {\bibinfo {author} {\bibfnamefont {J.}~\bibnamefont
  {Braun}}, \bibinfo {author} {\bibfnamefont {M.}~\bibnamefont {Leonhardt}}, \
  and\ \bibinfo {author} {\bibfnamefont {M.}~\bibnamefont {Pospiech}},\ }\href
  {\doibase 10.1103/PhysRevD.97.076010} {\bibfield  {journal} {\bibinfo
  {journal} {Phys. Rev.}\ }\textbf {\bibinfo {volume} {D97}},\ \bibinfo {pages}
  {076010} (\bibinfo {year} {2018})},\ \Eprint
  {http://arxiv.org/abs/1801.08338} {arXiv:1801.08338 [hep-ph]} \BibitemShut
  {NoStop}%
\bibitem [{\citenamefont {Alford}\ \emph {et~al.}(1998)\citenamefont {Alford},
  \citenamefont {Rajagopal},\ and\ \citenamefont {Wilczek}}]{Alford:1997zt}%
  \BibitemOpen
  \bibfield  {author} {\bibinfo {author} {\bibfnamefont {M.~G.}\ \bibnamefont
  {Alford}}, \bibinfo {author} {\bibfnamefont {K.}~\bibnamefont {Rajagopal}}, \
  and\ \bibinfo {author} {\bibfnamefont {F.}~\bibnamefont {Wilczek}},\ }\href
  {\doibase 10.1016/S0370-2693(98)00051-3} {\bibfield  {journal} {\bibinfo
  {journal} {Phys. Lett.}\ }\textbf {\bibinfo {volume} {B422}},\ \bibinfo
  {pages} {247} (\bibinfo {year} {1998})},\ \Eprint
  {http://arxiv.org/abs/hep-ph/9711395} {arXiv:hep-ph/9711395 [hep-ph]}
  \BibitemShut {NoStop}%
\bibitem [{\citenamefont {Pawlowski}\ and\ \citenamefont
  {Strodthoff}(2015)}]{Pawlowski:2015mia}%
  \BibitemOpen
  \bibfield  {author} {\bibinfo {author} {\bibfnamefont {J.~M.}\ \bibnamefont
  {Pawlowski}}\ and\ \bibinfo {author} {\bibfnamefont {N.}~\bibnamefont
  {Strodthoff}},\ }\href {\doibase 10.1103/PhysRevD.92.094009} {\bibfield
  {journal} {\bibinfo  {journal} {Phys. Rev.}\ }\textbf {\bibinfo {volume}
  {D92}},\ \bibinfo {pages} {094009} (\bibinfo {year} {2015})},\ \Eprint
  {http://arxiv.org/abs/1508.01160} {arXiv:1508.01160 [hep-ph]} \BibitemShut
  {NoStop}%
\bibitem [{\citenamefont {Pawlowski}\ \emph
  {et~al.}(2017{\natexlab{b}})\citenamefont {Pawlowski}, \citenamefont
  {Strodthoff},\ and\ \citenamefont {Wink}}]{Pawlowski:2017gxj}%
  \BibitemOpen
  \bibfield  {author} {\bibinfo {author} {\bibfnamefont {J.~M.}\ \bibnamefont
  {Pawlowski}}, \bibinfo {author} {\bibfnamefont {N.}~\bibnamefont
  {Strodthoff}}, \ and\ \bibinfo {author} {\bibfnamefont {N.}~\bibnamefont
  {Wink}},\ }\href@noop {} {\  (\bibinfo {year} {2017}{\natexlab{b}})},\
  \Eprint {http://arxiv.org/abs/1711.07444} {arXiv:1711.07444 [hep-th]}
  \BibitemShut {NoStop}%
\bibitem [{\citenamefont {Cohen}(2003)}]{Cohen:2003kd}%
  \BibitemOpen
  \bibfield  {author} {\bibinfo {author} {\bibfnamefont {T.~D.}\ \bibnamefont
  {Cohen}},\ }\href {\doibase 10.1103/PhysRevLett.91.222001} {\bibfield
  {journal} {\bibinfo  {journal} {Phys. Rev. Lett.}\ }\textbf {\bibinfo
  {volume} {91}},\ \bibinfo {pages} {222001} (\bibinfo {year} {2003})},\
  \Eprint {http://arxiv.org/abs/hep-ph/0307089} {arXiv:hep-ph/0307089 [hep-ph]}
  \BibitemShut {NoStop}%
\bibitem [{\citenamefont {Mark{\'o}}\ \emph {et~al.}(2014)\citenamefont
  {Mark{\'o}}, \citenamefont {Reinosa},\ and\ \citenamefont
  {Szep}}]{Marko:2014hea}%
  \BibitemOpen
  \bibfield  {author} {\bibinfo {author} {\bibfnamefont {G.}~\bibnamefont
  {Mark{\'o}}}, \bibinfo {author} {\bibfnamefont {U.}~\bibnamefont {Reinosa}},
  \ and\ \bibinfo {author} {\bibfnamefont {Z.}~\bibnamefont {Szep}},\ }\href
  {\doibase 10.1103/PhysRevD.90.125021} {\bibfield  {journal} {\bibinfo
  {journal} {Phys. Rev.}\ }\textbf {\bibinfo {volume} {D90}},\ \bibinfo {pages}
  {125021} (\bibinfo {year} {2014})},\ \Eprint {http://arxiv.org/abs/1410.6998}
  {arXiv:1410.6998 [hep-ph]} \BibitemShut {NoStop}%
\bibitem [{\citenamefont {Khan}\ \emph {et~al.}(2015)\citenamefont {Khan},
  \citenamefont {Pawlowski}, \citenamefont {Rennecke},\ and\ \citenamefont
  {Scherer}}]{Khan:2015puu}%
  \BibitemOpen
  \bibfield  {author} {\bibinfo {author} {\bibfnamefont {N.}~\bibnamefont
  {Khan}}, \bibinfo {author} {\bibfnamefont {J.~M.}\ \bibnamefont {Pawlowski}},
  \bibinfo {author} {\bibfnamefont {F.}~\bibnamefont {Rennecke}}, \ and\
  \bibinfo {author} {\bibfnamefont {M.~M.}\ \bibnamefont {Scherer}},\
  }\href@noop {} {\  (\bibinfo {year} {2015})},\ \Eprint
  {http://arxiv.org/abs/1512.03673} {arXiv:1512.03673 [hep-ph]} \BibitemShut
  {NoStop}%
\bibitem [{\citenamefont {Fu}\ and\ \citenamefont
  {Pawlowski}(2015)}]{Fu:2015naa}%
  \BibitemOpen
  \bibfield  {author} {\bibinfo {author} {\bibfnamefont {W.-j.}\ \bibnamefont
  {Fu}}\ and\ \bibinfo {author} {\bibfnamefont {J.~M.}\ \bibnamefont
  {Pawlowski}},\ }\href {\doibase 10.1103/PhysRevD.92.116006} {\bibfield
  {journal} {\bibinfo  {journal} {Phys. Rev.}\ }\textbf {\bibinfo {volume}
  {D92}},\ \bibinfo {pages} {116006} (\bibinfo {year} {2015})},\ \Eprint
  {http://arxiv.org/abs/1508.06504} {arXiv:1508.06504 [hep-ph]} \BibitemShut
  {NoStop}%
\bibitem [{\citenamefont {Rajagopal}\ and\ \citenamefont
  {Wilczek}(2001)}]{Rajagopal:2000ff}%
  \BibitemOpen
  \bibfield  {author} {\bibinfo {author} {\bibfnamefont {K.}~\bibnamefont
  {Rajagopal}}\ and\ \bibinfo {author} {\bibfnamefont {F.}~\bibnamefont
  {Wilczek}},\ }\href {\doibase 10.1103/PhysRevLett.86.3492} {\bibfield
  {journal} {\bibinfo  {journal} {Phys. Rev. Lett.}\ }\textbf {\bibinfo
  {volume} {86}},\ \bibinfo {pages} {3492} (\bibinfo {year} {2001})},\ \Eprint
  {http://arxiv.org/abs/hep-ph/0012039} {arXiv:hep-ph/0012039 [hep-ph]}
  \BibitemShut {NoStop}%
\bibitem [{\citenamefont {Shovkovy}\ and\ \citenamefont
  {Ellis}(2002)}]{Shovkovy:2002kv}%
  \BibitemOpen
  \bibfield  {author} {\bibinfo {author} {\bibfnamefont {I.~A.}\ \bibnamefont
  {Shovkovy}}\ and\ \bibinfo {author} {\bibfnamefont {P.~J.}\ \bibnamefont
  {Ellis}},\ }\href {\doibase 10.1103/PhysRevC.66.015802} {\bibfield  {journal}
  {\bibinfo  {journal} {Phys. Rev.}\ }\textbf {\bibinfo {volume} {C66}},\
  \bibinfo {pages} {015802} (\bibinfo {year} {2002})},\ \Eprint
  {http://arxiv.org/abs/hep-ph/0204132} {arXiv:hep-ph/0204132 [hep-ph]}
  \BibitemShut {NoStop}%
\bibitem [{\citenamefont {Herbst}\ \emph {et~al.}(2013)\citenamefont {Herbst},
  \citenamefont {Pawlowski},\ and\ \citenamefont {Schaefer}}]{Herbst:2013ail}%
  \BibitemOpen
  \bibfield  {author} {\bibinfo {author} {\bibfnamefont {T.~K.}\ \bibnamefont
  {Herbst}}, \bibinfo {author} {\bibfnamefont {J.~M.}\ \bibnamefont
  {Pawlowski}}, \ and\ \bibinfo {author} {\bibfnamefont {B.-J.}\ \bibnamefont
  {Schaefer}},\ }\href {\doibase 10.1103/PhysRevD.88.014007} {\bibfield
  {journal} {\bibinfo  {journal} {Phys. Rev.}\ }\textbf {\bibinfo {volume}
  {D88}},\ \bibinfo {pages} {014007} (\bibinfo {year} {2013})},\ \Eprint
  {http://arxiv.org/abs/1302.1426} {arXiv:1302.1426 [hep-ph]} \BibitemShut
  {NoStop}%
\bibitem [{\citenamefont {Fu}\ \emph {et~al.}(2016)\citenamefont {Fu},
  \citenamefont {Pawlowski}, \citenamefont {Rennecke},\ and\ \citenamefont
  {Schaefer}}]{Fu:2016tey}%
  \BibitemOpen
  \bibfield  {author} {\bibinfo {author} {\bibfnamefont {W.-j.}\ \bibnamefont
  {Fu}}, \bibinfo {author} {\bibfnamefont {J.~M.}\ \bibnamefont {Pawlowski}},
  \bibinfo {author} {\bibfnamefont {F.}~\bibnamefont {Rennecke}}, \ and\
  \bibinfo {author} {\bibfnamefont {B.-J.}\ \bibnamefont {Schaefer}},\ }\href
  {\doibase 10.1103/PhysRevD.94.116020} {\bibfield  {journal} {\bibinfo
  {journal} {Phys. Rev.}\ }\textbf {\bibinfo {volume} {D94}},\ \bibinfo {pages}
  {116020} (\bibinfo {year} {2016})},\ \Eprint
  {http://arxiv.org/abs/1608.04302} {arXiv:1608.04302 [hep-ph]} \BibitemShut
  {NoStop}%
\bibitem [{\citenamefont {Tripolt}\ \emph {et~al.}(2018)\citenamefont
  {Tripolt}, \citenamefont {Schaefer}, \citenamefont {von Smekal},\ and\
  \citenamefont {Wambach}}]{Tripolt:2017zgc}%
  \BibitemOpen
  \bibfield  {author} {\bibinfo {author} {\bibfnamefont {R.-A.}\ \bibnamefont
  {Tripolt}}, \bibinfo {author} {\bibfnamefont {B.-J.}\ \bibnamefont
  {Schaefer}}, \bibinfo {author} {\bibfnamefont {L.}~\bibnamefont {von
  Smekal}}, \ and\ \bibinfo {author} {\bibfnamefont {J.}~\bibnamefont
  {Wambach}},\ }\href {\doibase 10.1103/PhysRevD.97.034022} {\bibfield
  {journal} {\bibinfo  {journal} {Phys. Rev.}\ }\textbf {\bibinfo {volume}
  {D97}},\ \bibinfo {pages} {034022} (\bibinfo {year} {2018})},\ \Eprint
  {http://arxiv.org/abs/1709.05991} {arXiv:1709.05991 [hep-ph]} \BibitemShut
  {NoStop}%
\bibitem [{\citenamefont {Braun}\ \emph {et~al.}(tion)\citenamefont {Braun},
  \citenamefont {Drischler}, \citenamefont {Hebeler}, \citenamefont
  {Leonhardt}, \citenamefont {Pospiech},\ and\ \citenamefont
  {Schwenk}}]{BDHLPS}%
  \BibitemOpen
  \bibfield  {author} {\bibinfo {author} {\bibfnamefont {J.}~\bibnamefont
  {Braun}}, \bibinfo {author} {\bibfnamefont {C.}~\bibnamefont {Drischler}},
  \bibinfo {author} {\bibfnamefont {K.}~\bibnamefont {Hebeler}}, \bibinfo
  {author} {\bibfnamefont {M.}~\bibnamefont {Leonhardt}}, \bibinfo {author}
  {\bibfnamefont {M.}~\bibnamefont {Pospiech}}, \ and\ \bibinfo {author}
  {\bibfnamefont {A.}~\bibnamefont {Schwenk}},\ }\href@noop {} {\  (\bibinfo
  {year} {in preparation})}\BibitemShut {NoStop}%
\bibitem [{\citenamefont {Braun}\ \emph
  {et~al.}(2005{\natexlab{a}})\citenamefont {Braun}, \citenamefont {Klein},\
  and\ \citenamefont {Pirner}}]{Braun:2004yk}%
  \BibitemOpen
  \bibfield  {author} {\bibinfo {author} {\bibfnamefont {J.}~\bibnamefont
  {Braun}}, \bibinfo {author} {\bibfnamefont {B.}~\bibnamefont {Klein}}, \ and\
  \bibinfo {author} {\bibfnamefont {H.~J.}\ \bibnamefont {Pirner}},\
  }\href@noop {} {\bibfield  {journal} {\bibinfo  {journal} {Phys. Rev.}\
  }\textbf {\bibinfo {volume} {D71}},\ \bibinfo {pages} {014032} (\bibinfo
  {year} {2005}{\natexlab{a}})},\ \Eprint {http://arxiv.org/abs/hep-ph/0408116}
  {hep-ph/0408116} \BibitemShut {NoStop}%
\bibitem [{\citenamefont {Braun}\ \emph {et~al.}(2006)\citenamefont {Braun},
  \citenamefont {Klein}, \citenamefont {Pirner},\ and\ \citenamefont
  {Rezaeian}}]{Braun:2005fj}%
  \BibitemOpen
  \bibfield  {author} {\bibinfo {author} {\bibfnamefont {J.}~\bibnamefont
  {Braun}}, \bibinfo {author} {\bibfnamefont {B.}~\bibnamefont {Klein}},
  \bibinfo {author} {\bibfnamefont {H.~J.}\ \bibnamefont {Pirner}}, \ and\
  \bibinfo {author} {\bibfnamefont {A.~H.}\ \bibnamefont {Rezaeian}},\
  }\href@noop {} {\bibfield  {journal} {\bibinfo  {journal} {Phys. Rev.}\
  }\textbf {\bibinfo {volume} {D73}},\ \bibinfo {pages} {074010} (\bibinfo
  {year} {2006})},\ \Eprint {http://arxiv.org/abs/hep-ph/0512274}
  {hep-ph/0512274} \BibitemShut {NoStop}%
\bibitem [{\citenamefont {Litim}\ and\ \citenamefont
  {Pawlowski}(2006)}]{Litim:2006ag}%
  \BibitemOpen
  \bibfield  {author} {\bibinfo {author} {\bibfnamefont {D.~F.}\ \bibnamefont
  {Litim}}\ and\ \bibinfo {author} {\bibfnamefont {J.~M.}\ \bibnamefont
  {Pawlowski}},\ }\href {\doibase 10.1088/1126-6708/2006/11/026} {\bibfield
  {journal} {\bibinfo  {journal} {JHEP}\ }\textbf {\bibinfo {volume} {11}},\
  \bibinfo {pages} {026} (\bibinfo {year} {2006})},\ \Eprint
  {http://arxiv.org/abs/hep-th/0609122} {arXiv:hep-th/0609122} \BibitemShut
  {NoStop}%
\bibitem [{\citenamefont {Blaizot}\ \emph {et~al.}(2007)\citenamefont
  {Blaizot}, \citenamefont {Ipp}, \citenamefont {Mendez-Galain},\ and\
  \citenamefont {Wschebor}}]{Blaizot:2006rj}%
  \BibitemOpen
  \bibfield  {author} {\bibinfo {author} {\bibfnamefont {J.-P.}\ \bibnamefont
  {Blaizot}}, \bibinfo {author} {\bibfnamefont {A.}~\bibnamefont {Ipp}},
  \bibinfo {author} {\bibfnamefont {R.}~\bibnamefont {Mendez-Galain}}, \ and\
  \bibinfo {author} {\bibfnamefont {N.}~\bibnamefont {Wschebor}},\ }\href
  {\doibase 10.1016/j.nuclphysa.2006.11.139} {\bibfield  {journal} {\bibinfo
  {journal} {Nucl. Phys.}\ }\textbf {\bibinfo {volume} {A784}},\ \bibinfo
  {pages} {376} (\bibinfo {year} {2007})},\ \Eprint
  {http://arxiv.org/abs/hep-ph/0610004} {arXiv:hep-ph/0610004} \BibitemShut
  {NoStop}%
\bibitem [{\citenamefont {Braun}\ \emph
  {et~al.}(2011{\natexlab{b}})\citenamefont {Braun}, \citenamefont {Klein},\
  and\ \citenamefont {Piasecki}}]{Braun:2010vd}%
  \BibitemOpen
  \bibfield  {author} {\bibinfo {author} {\bibfnamefont {J.}~\bibnamefont
  {Braun}}, \bibinfo {author} {\bibfnamefont {B.}~\bibnamefont {Klein}}, \ and\
  \bibinfo {author} {\bibfnamefont {P.}~\bibnamefont {Piasecki}},\ }\href
  {\doibase 10.1140/epjc/s10052-011-1576-7} {\bibfield  {journal} {\bibinfo
  {journal} {Eur. Phys. J.}\ }\textbf {\bibinfo {volume} {C71}},\ \bibinfo
  {pages} {1576} (\bibinfo {year} {2011}{\natexlab{b}})},\ \Eprint
  {http://arxiv.org/abs/1008.2155} {arXiv:1008.2155 [hep-ph]} \BibitemShut
  {NoStop}%
\bibitem [{\citenamefont {Braun}\ \emph {et~al.}(2012)\citenamefont {Braun},
  \citenamefont {Klein},\ and\ \citenamefont {Schaefer}}]{Braun:2011iz}%
  \BibitemOpen
  \bibfield  {author} {\bibinfo {author} {\bibfnamefont {J.}~\bibnamefont
  {Braun}}, \bibinfo {author} {\bibfnamefont {B.}~\bibnamefont {Klein}}, \ and\
  \bibinfo {author} {\bibfnamefont {B.-J.}\ \bibnamefont {Schaefer}},\ }\href
  {\doibase 10.1016/j.physletb.2012.05.053} {\bibfield  {journal} {\bibinfo
  {journal} {Phys. Lett.}\ }\textbf {\bibinfo {volume} {B713}},\ \bibinfo
  {pages} {216} (\bibinfo {year} {2012})},\ \Eprint
  {http://arxiv.org/abs/1110.0849} {arXiv:1110.0849 [hep-ph]} \BibitemShut
  {NoStop}%
\bibitem [{\citenamefont {Tripolt}\ \emph
  {et~al.}(2014{\natexlab{b}})\citenamefont {Tripolt}, \citenamefont {Braun},
  \citenamefont {Klein},\ and\ \citenamefont {Schaefer}}]{Tripolt:2013zfa}%
  \BibitemOpen
  \bibfield  {author} {\bibinfo {author} {\bibfnamefont {R.-A.}\ \bibnamefont
  {Tripolt}}, \bibinfo {author} {\bibfnamefont {J.}~\bibnamefont {Braun}},
  \bibinfo {author} {\bibfnamefont {B.}~\bibnamefont {Klein}}, \ and\ \bibinfo
  {author} {\bibfnamefont {B.-J.}\ \bibnamefont {Schaefer}},\ }\href {\doibase
  10.1103/PhysRevD.90.054012} {\bibfield  {journal} {\bibinfo  {journal} {Phys.
  Rev.}\ }\textbf {\bibinfo {volume} {D90}},\ \bibinfo {pages} {054012}
  (\bibinfo {year} {2014}{\natexlab{b}})},\ \Eprint
  {http://arxiv.org/abs/1308.0164} {arXiv:1308.0164 [hep-ph]} \BibitemShut
  {NoStop}%
\bibitem [{\citenamefont {Schaefer}\ and\ \citenamefont
  {Wambach}(2005)}]{Schaefer:2004en}%
  \BibitemOpen
  \bibfield  {author} {\bibinfo {author} {\bibfnamefont {B.-J.}\ \bibnamefont
  {Schaefer}}\ and\ \bibinfo {author} {\bibfnamefont {J.}~\bibnamefont
  {Wambach}},\ }\href@noop {} {\bibfield  {journal} {\bibinfo  {journal} {Nucl.
  Phys.}\ }\textbf {\bibinfo {volume} {A757}},\ \bibinfo {pages} {479}
  (\bibinfo {year} {2005})},\ \Eprint {http://arxiv.org/abs/nucl-th/0403039}
  {nucl-th/0403039} \BibitemShut {NoStop}%
\bibitem [{\citenamefont {Fister}\ and\ \citenamefont
  {Pawlowski}(2015)}]{Fister:2015eca}%
  \BibitemOpen
  \bibfield  {author} {\bibinfo {author} {\bibfnamefont {L.}~\bibnamefont
  {Fister}}\ and\ \bibinfo {author} {\bibfnamefont {J.~M.}\ \bibnamefont
  {Pawlowski}},\ }\href {\doibase 10.1103/PhysRevD.92.076009} {\bibfield
  {journal} {\bibinfo  {journal} {Phys. Rev.}\ }\textbf {\bibinfo {volume}
  {D92}},\ \bibinfo {pages} {076009} (\bibinfo {year} {2015})},\ \Eprint
  {http://arxiv.org/abs/1504.05166} {arXiv:1504.05166 [hep-ph]} \BibitemShut
  {NoStop}%
\bibitem [{\citenamefont {Klein}(2017)}]{Klein:2017shl}%
  \BibitemOpen
  \bibfield  {author} {\bibinfo {author} {\bibfnamefont {B.}~\bibnamefont
  {Klein}},\ }\href {\doibase 10.1016/j.physrep.2017.09.002} {\bibfield
  {journal} {\bibinfo  {journal} {Phys. Rept.}\ }\textbf {\bibinfo {volume}
  {707-708}},\ \bibinfo {pages} {1} (\bibinfo {year} {2017})},\ \Eprint
  {http://arxiv.org/abs/1710.05357} {arXiv:1710.05357 [hep-ph]} \BibitemShut
  {NoStop}%
\bibitem [{\citenamefont {Braun}\ and\ \citenamefont
  {Klein}(2009)}]{Braun:2008sg}%
  \BibitemOpen
  \bibfield  {author} {\bibinfo {author} {\bibfnamefont {J.}~\bibnamefont
  {Braun}}\ and\ \bibinfo {author} {\bibfnamefont {B.}~\bibnamefont {Klein}},\
  }\href {\doibase 10.1140/epjc/s10052-009-1098-8} {\bibfield  {journal}
  {\bibinfo  {journal} {Eur. Phys. J.}\ }\textbf {\bibinfo {volume} {C63}},\
  \bibinfo {pages} {443} (\bibinfo {year} {2009})},\ \Eprint
  {http://arxiv.org/abs/0810.0857} {arXiv:0810.0857 [hep-ph]} \BibitemShut
  {NoStop}%
\bibitem [{\citenamefont {Colangelo}\ and\ \citenamefont
  {D{\"u}rr}(2004)}]{Colangelo:2003hf}%
  \BibitemOpen
  \bibfield  {author} {\bibinfo {author} {\bibfnamefont {G.}~\bibnamefont
  {Colangelo}}\ and\ \bibinfo {author} {\bibfnamefont {S.}~\bibnamefont
  {D{\"u}rr}},\ }\href@noop {} {\bibfield  {journal} {\bibinfo  {journal} {Eur.
  Phys. J.}\ }\textbf {\bibinfo {volume} {C33}},\ \bibinfo {pages} {543}
  (\bibinfo {year} {2004})},\ \Eprint {http://arxiv.org/abs/hep-lat/0311023}
  {hep-lat/0311023} \BibitemShut {NoStop}%
\bibitem [{\citenamefont {Braun}\ \emph
  {et~al.}(2005{\natexlab{b}})\citenamefont {Braun}, \citenamefont {Klein},\
  and\ \citenamefont {Pirner}}]{Braun:2005gy}%
  \BibitemOpen
  \bibfield  {author} {\bibinfo {author} {\bibfnamefont {J.}~\bibnamefont
  {Braun}}, \bibinfo {author} {\bibfnamefont {B.}~\bibnamefont {Klein}}, \ and\
  \bibinfo {author} {\bibfnamefont {H.~J.}\ \bibnamefont {Pirner}},\
  }\href@noop {} {\bibfield  {journal} {\bibinfo  {journal} {Phys. Rev.}\
  }\textbf {\bibinfo {volume} {D72}},\ \bibinfo {pages} {034017} (\bibinfo
  {year} {2005}{\natexlab{b}})},\ \Eprint {http://arxiv.org/abs/hep-ph/0504127}
  {hep-ph/0504127} \BibitemShut {NoStop}%
\bibitem [{\citenamefont {Guagnelli}\ \emph {et~al.}(2004)\citenamefont
  {Guagnelli} \emph {et~al.}}]{Guagnelli:2004ww}%
  \BibitemOpen
  \bibfield  {author} {\bibinfo {author} {\bibfnamefont {M.}~\bibnamefont
  {Guagnelli}} \emph {et~al.} (\bibinfo {collaboration} {Zeuthen-Rome
  (ZeRo)}),\ }\href@noop {} {\bibfield  {journal} {\bibinfo  {journal} {Phys.
  Lett.}\ }\textbf {\bibinfo {volume} {B597}},\ \bibinfo {pages} {216}
  (\bibinfo {year} {2004})},\ \Eprint {http://arxiv.org/abs/hep-lat/0403009}
  {hep-lat/0403009} \BibitemShut {NoStop}%
\bibitem [{\citenamefont {Orth}\ \emph {et~al.}(2005)\citenamefont {Orth},
  \citenamefont {Lippert},\ and\ \citenamefont {Schilling}}]{Orth:2005kq}%
  \BibitemOpen
  \bibfield  {author} {\bibinfo {author} {\bibfnamefont {B.}~\bibnamefont
  {Orth}}, \bibinfo {author} {\bibfnamefont {T.}~\bibnamefont {Lippert}}, \
  and\ \bibinfo {author} {\bibfnamefont {K.}~\bibnamefont {Schilling}},\ }\href
  {\doibase 10.1103/PhysRevD.72.014503} {\bibfield  {journal} {\bibinfo
  {journal} {Phys. Rev.}\ }\textbf {\bibinfo {volume} {D72}},\ \bibinfo {pages}
  {014503} (\bibinfo {year} {2005})},\ \Eprint
  {http://arxiv.org/abs/hep-lat/0503016} {arXiv:hep-lat/0503016 [hep-lat]}
  \BibitemShut {NoStop}%
\bibitem [{\citenamefont {Braun}\ and\ \citenamefont
  {Klein}(2008)}]{Braun:2007td}%
  \BibitemOpen
  \bibfield  {author} {\bibinfo {author} {\bibfnamefont {J.}~\bibnamefont
  {Braun}}\ and\ \bibinfo {author} {\bibfnamefont {B.}~\bibnamefont {Klein}},\
  }\href {\doibase 10.1103/PhysRevD.77.096008} {\bibfield  {journal} {\bibinfo
  {journal} {Phys. Rev.}\ }\textbf {\bibinfo {volume} {D77}},\ \bibinfo {pages}
  {096008} (\bibinfo {year} {2008})},\ \Eprint {http://arxiv.org/abs/0712.3574}
  {arXiv:0712.3574 [hep-th]} \BibitemShut {NoStop}%
\bibitem [{\citenamefont {Klein}\ \emph {et~al.}(2010)\citenamefont {Klein},
  \citenamefont {Braun},\ and\ \citenamefont {Schaefer}}]{Klein:2010tk}%
  \BibitemOpen
  \bibfield  {author} {\bibinfo {author} {\bibfnamefont {B.}~\bibnamefont
  {Klein}}, \bibinfo {author} {\bibfnamefont {J.}~\bibnamefont {Braun}}, \ and\
  \bibinfo {author} {\bibfnamefont {B.-J.}\ \bibnamefont {Schaefer}},\
  }\href@noop {} {\bibfield  {journal} {\bibinfo  {journal} {PoS}\ }\textbf
  {\bibinfo {volume} {LATTICE2010}},\ \bibinfo {pages} {193} (\bibinfo {year}
  {2010})},\ \Eprint {http://arxiv.org/abs/1011.1435} {arXiv:1011.1435
  [hep-ph]} \BibitemShut {NoStop}%
\bibitem [{\citenamefont {Almasi}\ \emph {et~al.}(2017)\citenamefont {Almasi},
  \citenamefont {Pisarski},\ and\ \citenamefont {Skokov}}]{Almasi:2016zqf}%
  \BibitemOpen
  \bibfield  {author} {\bibinfo {author} {\bibfnamefont {G.}~\bibnamefont
  {Almasi}}, \bibinfo {author} {\bibfnamefont {R.}~\bibnamefont {Pisarski}}, \
  and\ \bibinfo {author} {\bibfnamefont {V.}~\bibnamefont {Skokov}},\ }\href
  {\doibase 10.1103/PhysRevD.95.056015} {\bibfield  {journal} {\bibinfo
  {journal} {Phys. Rev.}\ }\textbf {\bibinfo {volume} {D95}},\ \bibinfo {pages}
  {056015} (\bibinfo {year} {2017})},\ \Eprint
  {http://arxiv.org/abs/1612.04416} {arXiv:1612.04416 [hep-ph]} \BibitemShut
  {NoStop}%
\bibitem [{fQC()}]{fQCD}%
  \BibitemOpen
  \href@noop {} {}\bibinfo {note} {{\it fQCD Collaboration}, J. Braun, L.
  Corell, A.~K. Cyrol, W.-j. Fu, C. Huang, M. Leonhardt, M. Mitter, J.~M.
  Pawlowski, M. Pospiech, F. Rennecke, C. Schneider, R. Wen, N. Wink, S. Yin
  (members as of June 2018)}\BibitemShut {NoStop}%
\end{thebibliography}%

\end{document}